\documentclass[a4paper,fleqn]{cas-sc}

\usepackage[authoryear,longnamesfirst]{natbib}
\usepackage{bm}
\usepackage{relsize}
\usepackage{amsthm}

\def\tsc#1{\csdef{#1}{\textsc{\lowercase{#1}}\xspace}}
\tsc{WGM}
\tsc{QE}
\tsc{EP}
\tsc{PMS}
\tsc{BEC}
\tsc{DE}
\DeclareMathOperator*{\A}{ \mathlarger{\mathlarger{\mathlarger{\boldsymbol{\mathsf{A}}}}} }

\begin{document}
\let\WriteBookmarks\relax
\def\floatpagepagefraction{1}
\def\textpagefraction{.001}
\shorttitle{Leveraging social media news}
\shortauthors{Y. Xie et~al.}

\title [mode = title]{Revisit of Two-dimensional CEM on Crack Branching: from Single Crack-tip Tracking to Multiple Crack-tips Tracking }                      



\author[1]{Yuxi Xie}[type=editor,
                        auid=000,bioid=1,
                        orcid=0000-0001-8681-4053]
\cormark[1]
\ead{yuxi_xie2017@outlook.com}
\ead[url]{https://www.linkedin.com/in/yuxi-x-142917134/}

\credit{Conceptualization of this study, Methodology, Code Implementation, Writing - original draft preparation and revision}

\affiliation[1]{organization={Computational and Multiscale Mechanics Group, ANSYS Inc},
                addressline={7374 Las Positas Rd}, 
                city={Livermore},
                postcode={94551}, 
                state={C.A.},
                country={USA}}


\author[2]{Hongyou Cao}[%
  ]

\credit{Writing - revision}

\affiliation[2]{organization={School of Civil Engineering and Architecture, Wuhan University of Technology},
                postcode={430070}, 
                city={Wuhan},
                country={China}}

\author[3]{Miao Su}

\credit{Advices on revision}

\affiliation[3]{organization={School of Civil Engineering, Changsha University of Science and Technology},
                city={Changsha},
                postcode={410114}, 
                state={Hunan}, 
                country={China}}

\author[4]{Zhipeng Lai}

\credit{Advices on revision}

\affiliation[4]{organization={School of Civil Engineering, Central South University},
                city={Changsha},
                postcode={410075}, 
                state={Hunan}, 
                country={China}}

\author[1]{Xiaolong He}

\credit{Advices on revision}

\cortext[cor1]{Corresponding author}


\begin{abstract}
In this work, a Multiple Crack-tips Tracking algorithm in two-dimensional Crack Element Model (MCT-2D-CEM) is developed, aiming at modeling and predicting advanced and complicated crack patterns in two-dimensional dynamic fracturing problems, such as crack branching and fragmentation. Based on the developed fracture energy release rate formulation of split elementary topology, the Multiple Crack-tips Tracking algorithm is proposed and a series of benchmark examples are provided to validate effectiveness and efficiency in modeling crack branching and fragmentation. Besides, the proposed MCR-2D-CEM can still model single crack propagation but extra micro-cracks are introduced. GPU acceleration is employed in all two-dimensional simulations, providing high computational efficiency, consistency, and accuracy.
\end{abstract}

\begin{graphicalabstract}
\includegraphics[height=2.5in]{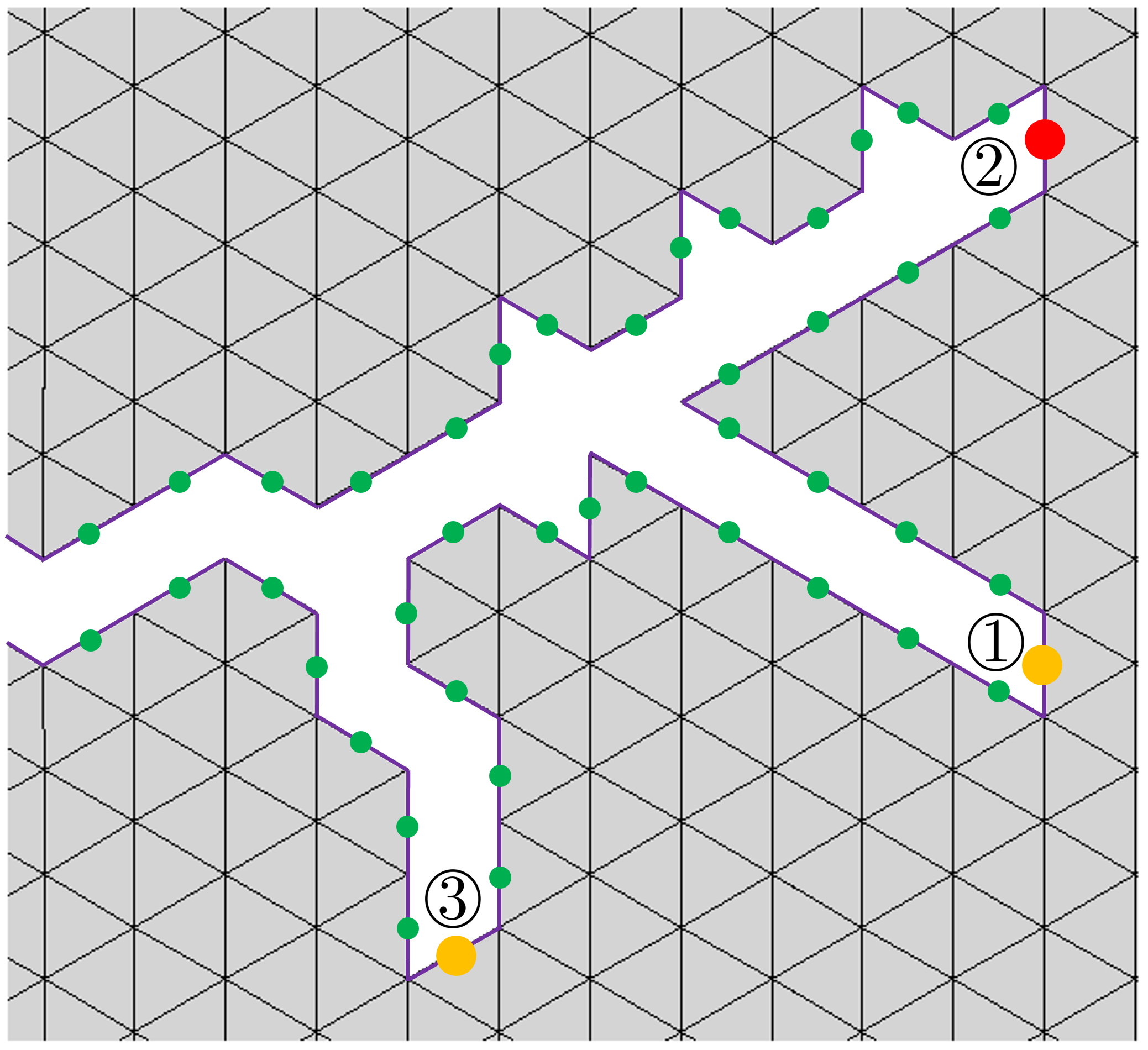} \\
\hfill \\
\includegraphics[height=1.5in]{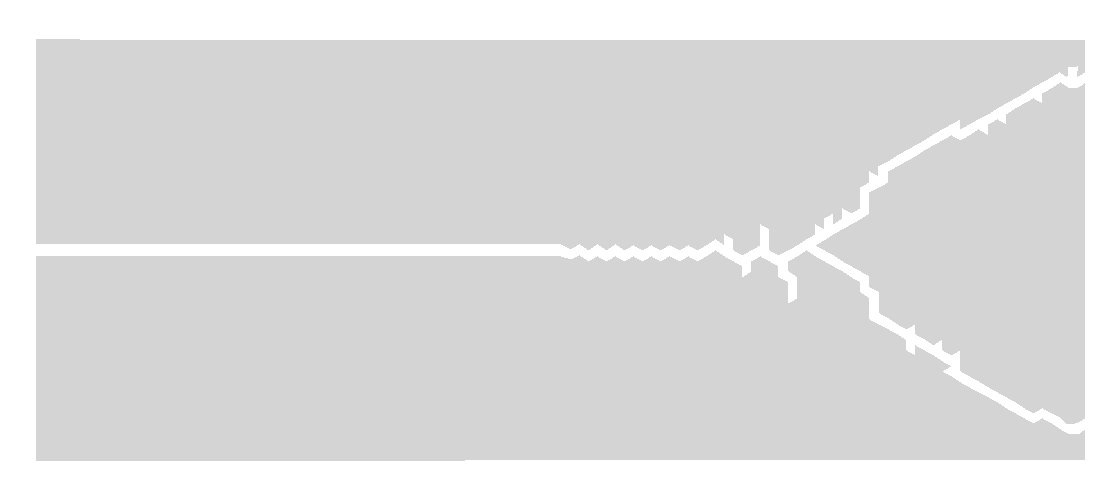} \\
\hfill \\
\includegraphics[height=3.0in]{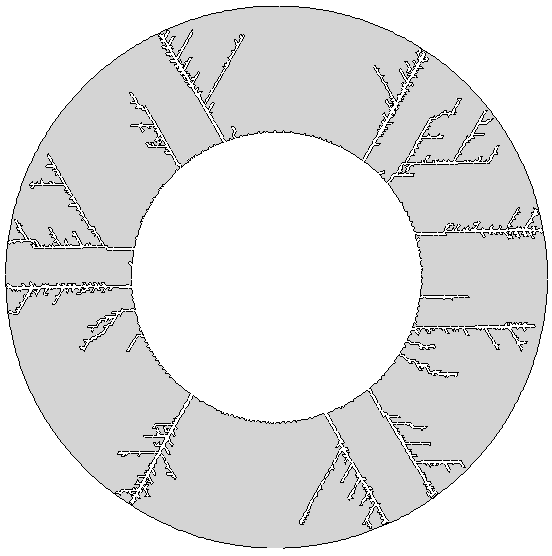}
\end{graphicalabstract}

\begin{highlights}
\item An improved Multiple Crack-tips Tracking algorithm in two-dimensional Crack Element Model (MCT-2D-CEM) is proposed, which is based on two-dimensional CEM with modified crack tracking algorithm, to overcome shortages of single crack-tip tracking algorithm.
\item The proposed MCT-2D-CEM satisfies accuracy and efficiency requirements in capturing two-dimensional crack branching with Neumann and Dirichlet boundary condition.
\item Advanced and complicated crack patterns such as two-dimensional fragmentation which has more than two crack-tips to be tracked,  can be successfully modeled. 
\item Employing NVIDIA GPU acceleration for 2D benchmark simulations illustrates the method’s readiness for integration into high-performance computing (HPC) workflows in industry.
\end{highlights}


\maketitle

\section{Introduction}
The study of dynamic crack propagation and bifurcation has remained a central topic in fracture mechanics, as it not only deepens our theoretical understanding of failure mechanism but also plays a crucial role in solving practical problems encountered in geophysics, materials science, and structural engineering (\cite{bouchbinder2014dynamics}). The development of modern fracture mechanics began in 1913, when Inglis (\cite{inglis1913stress}) examined how stress is distributed around elliptical holes, effectively treating these features as initial cracks. This work laid the groundwork for later contributions by Griffth (\cite{griffith1921vi}), who applied thermodynamic principles to explain that a crack grows when the energy released during propagation surpasses the material's resistance to fracture (\cite{anderson2005fracture}). Building on these ideas, Irwin (\cite{irwin1957analysis}) extended the theory by introducing the concept of stress intensity factors, providing a practical means to quantify crack behavior under various loading conditions. 

Fracture is inherently a dynamic process so that studying dynamic fracture behavior is not only of greater relevance to real-world engineering applications but also poses significantly higher computational challenges (\cite{cox2005modern}). The foundation of dynamic fracture mechanics was laid in 1948, when Mott (\cite{mott1948brittle}) extended Griffith’s energy criterion by incorporating an inertial term and introducing time dependence into the analysis. In the regime of small-scale yielding, the dynamic stress intensity factor serves as a key parameter for predicting crack initiation (\cite{sih1968some}): once it reaches the material's dynamic fracture toughness, crack propagation commences. Under extreme stress conditions, the advancing crack may undergo bifurcation, forming complex branching patterns resembling a river delta structure (\cite{ha2010studies}).

Crack branching is a common phenomenon encountered in numerous engineering applications, particularly in structures composed of brittle materials and in metal alloys affected by stress corrosion cracking (\cite{kalthoff1973propagation}, \cite{nishioka2002generation}). Branching can manifest in both symmetric and asymmetric forms, depending on the material properties, loading conditions, and stress distribution near the crack tip. Understanding the mechanisms that drive crack branching is crucial, as such behavior often marks a transition from stable to unstable crack growth, which can significantly compromise structural integrity. The investigation of crack branching encompasses several key aspects, including the identification of branching criteria, theoretical and empirical studies of the branching process, precise experimental measurements, and the development of advanced numerical models capable of capturing this complex phenomenon. Each of these components plays an important role in enabling accurate simulation and prediction of fracture behavior under dynamic or quasi-static loading conditions.

In the context of linear elastic fracture mechanics (LEFM), the propagation of a single, steady-state crack is typically governed by criteria based on the stress intensity factor or the energy release rate, both of which quantify the driving force for crack extension. In contrast, nonlinear fracture mechanics adopts the $\mathcal{J}$-integral as a fundamental parameter. The $\mathcal{J}$-integral serves as a generalized measure of the energy release rate under nonlinear material behavior and is often interpreted as an extension of the stress intensity factor concept to inelastic or large-deformation regimes. Despite significant advances in fracture mechanics, the mechanisms underlying crack branching remain complex and not yet fully understood. Early theories proposed that crack branching is primarily governed by the crack propagation velocity suggesting that once a crack exceeds a critical speed, the stress field ahead of the crack tip undergoes a qualitative change, prompting the crack to split. However, subsequent experimental observations revealed that the velocities at which cracks actually branch are often significantly lower than those predicted by theory (\cite{ramulu1985mechanics}), indicating that velocity alone may not be sufficient to explain the phenomenon. As a result, researchers have explored additional contributing factors, such as the interaction with microcracks near the crack front, local perturbations in the stress field caused by twisting or tilting of the crack path, and dynamic instabilities intrinsic to the fracture process. These factors introduce further complexity, particularly under dynamic loading conditions. Notably, it has been shown that when such near-tip instabilities are suppressed, the formation of supersonic cracks becomes possible, challenging conventional assumptions about crack speed limits (\cite{guozden2010supersonic}). Although numerous experimental (\cite{schardin1959velocity}, \cite{kerkhof1973general}, \cite{kobayashi1978dynamic}), theoretical (\cite{broberg1960propagation}, \cite{eshelby1970inelastic}, \cite{gao1993surface}) and numerical studies (\cite{xu1994numerical}, \cite{rabczuk2004cracking}, \cite{song2008comparative}) have examined various aspects of crack branching, a comprehensive and unified theory that accurately describes the onset of dynamic instabilities and the branching process remains elusive. This continues to be an active area of research, with implications for the predictive modeling of fracture in brittle solids and structural materials under high strain rates.

A novel two-dimensional and three-dimensional Crack Element Model was proposed by Xie et al. in \cite{xie2025practical} and in \cite{xie2025gpu}, respectively. The two-dimensional CEM (\cite{xie2025practical}) was initially proposed to deal with single crack propagation problem in two-dimension and it was extended to three-dimensional CEM (\cite{xie2025gpu}) naturally. Out of authors' expectation, the three-dimensional CEM demonstrates strong capability in dealing with crack branching under Neumann and Dirichlet boundary conditions. Therefore, to be a series of comprehensive studies on fracture problems, it is necessary to seek out an improved two-dimensional CEM to resolve crack branching and even fragmentation in two-dimension. Contrary to three-dimensional CEM which captures crack branching spontaneously without any artificial criteria such as velocity of crack-tip, the present study introduces an energy-based Multiple Crack-tips Tracking algorithm under the framework of two-dimensional CEM, aiming at solving problems of single-crack and multiple-cracks such as crack branching and fragmentation by using one set algorithm.  

To help readers catch up on the present study, a brief outline of the paper is summarized as follows: In Section.2 we briefly review fundamental ideas of two-dimensional Crack Element Model (CEM) and element-dependent computation of fracture energy release rate. Then the locally single crack-tip tracking algorithm is extended to Multiple Crack-tips Tracking (MCT) algorithm with two-dimensional CEM in Section.3. To validate the proposed MCT-2D-CEM, some classical benchmark examples of two-dimensional dynamic fracture, including two-dimensional crack branching and fragmentation, are introduced in Section.4. In Section.5, We provide a summary of the current research, highlight unresolved challenges in existing studies, and outline potential directions for future developments.

\section{Brief review of Single Crack-tip in 2D CEM}
In this section, we briefly review explicit formulation of two-dimensional Edge-based Smooth Finite Element Method (ES-FEM) and the two-dimensional Crack Element Model with single crack-tip tracking algorithm. The notations and indices utilized in the present work are introduced as well.

\subsection{2D formulation of Edge-based Smoothed FEM}
\begin{figure}[htp]
\begin{minipage}{0.45\linewidth}
   \begin{center}
      \includegraphics[height=2.2in]{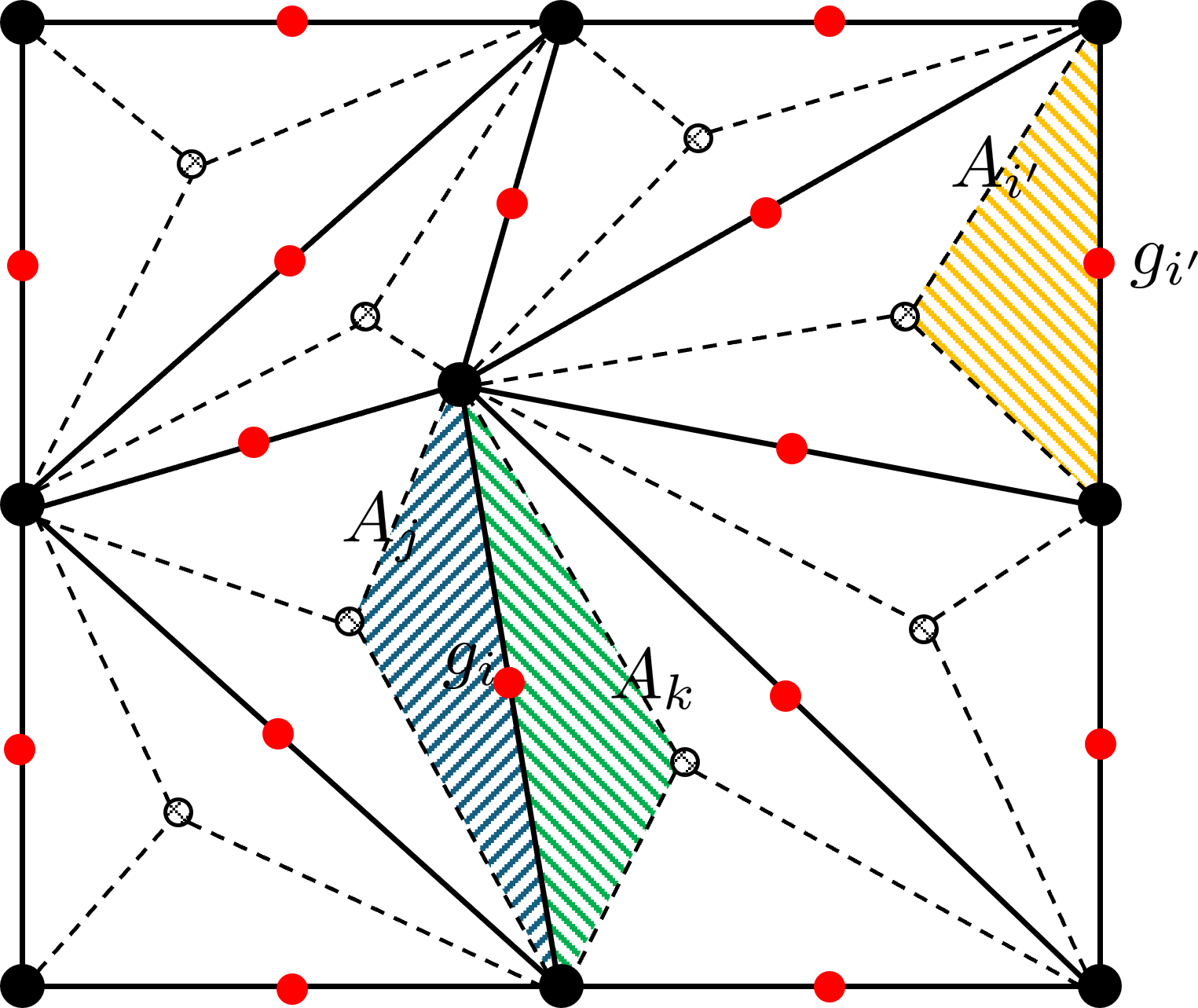}
  \end{center}
   \begin{center}
   (a)
   \end{center}
   \end{minipage}
   \hfill
\begin{minipage}{0.45\linewidth}
   \begin{center}
      \includegraphics[height=2.2in]{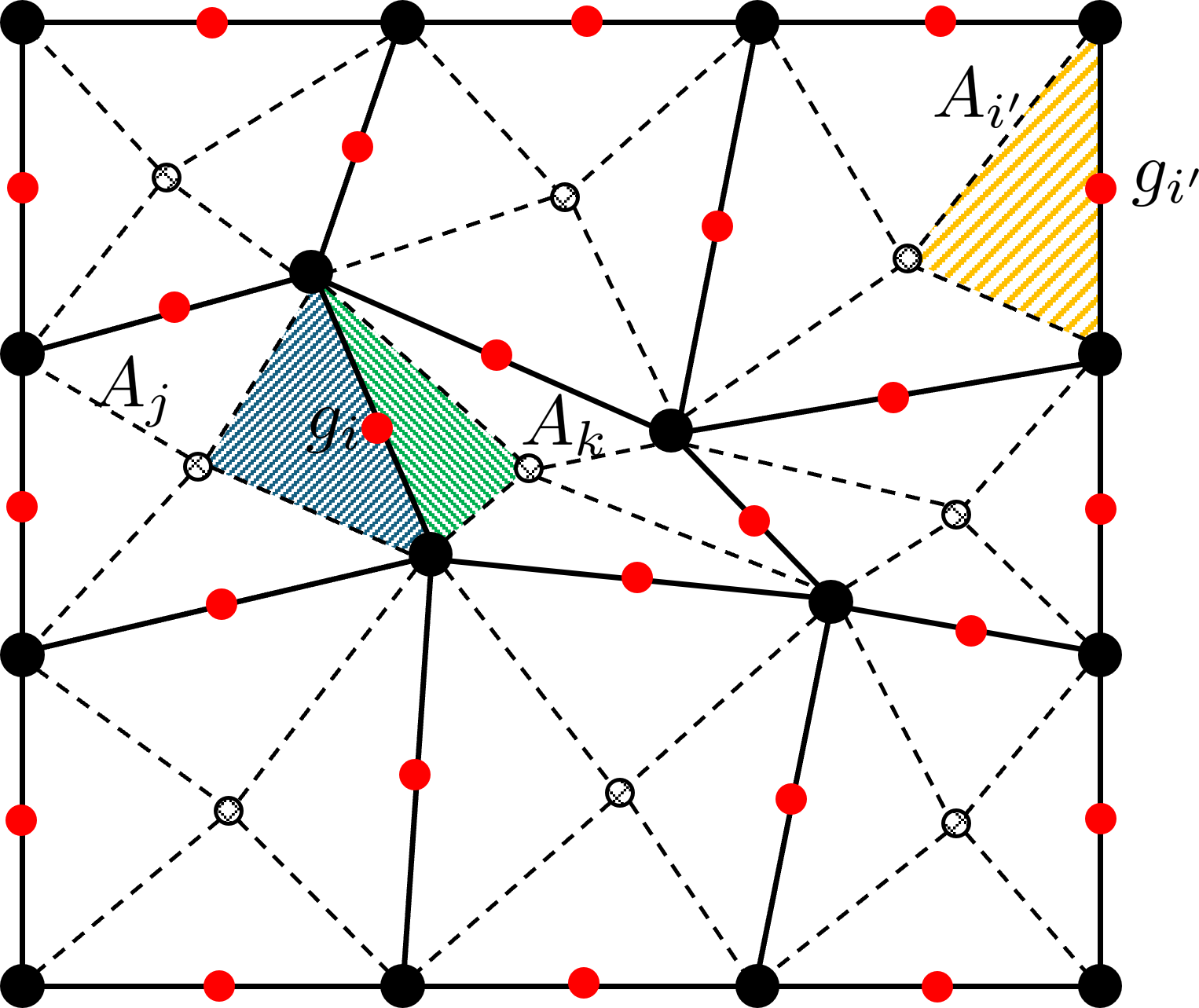}
  \end{center}
   \begin{center}
   (b)
   \end{center}
   \end{minipage}
   \caption{The studied domain is discretized by red points which represent gaussian quadrature points. Notation $A_i$ represents element ID and $g_i$ represents edge quadrature ID: (a). ES-FEM formulation of Constant Strain Triangle element, totally $8$ related elements and $16$ edge quadrature points; (b). ES-FEM formulation of bilinear Quadrilateral element, totally $9$ related elements and $24$ edge quadrature points. The blue and green regions represent shared region of element $A_j$ and $A_k$ by edge quadrature point $g_i$, respectively. The orange region represents exclusive region of element $A_{i^{\prime}}$ contributing to edge quadrature $g_{i^{\prime}}$.}
   \label{fig2-1:ES-FEM_CST_QUAD_formulation}
\end{figure}

Instead of using gaussian quadrature points within elements, Edge-based Smoothed Finite Element Method (ES-FEM, \cite{liu2009edge}) placed quadrature points at elementary edges (see red points in Figure.\ref{fig2-1:ES-FEM_CST_QUAD_formulation}), which possesses a more accurate stiffness than "over-stiff" FEM and "over-soft" NS-FEM (Node-based Smoothed FEM) so that precise capture of crack tip with mesh-independence becomes possible. Besides, except for locations of quadrature points, fundamental computational framework of FEM is preserved therefore computational efficiency is obviously advantageous than meshfree methods. 

The displacement at edge quadrature points are influenced by related nodes in neighboring elements, i.e.,
\begin{eqnarray}
\bm{u} \left(g_i \right) = \sum_{j \in \mathcal{N}^{g_i} } \left( \sum_{k \in \mathcal{M}^{A_j} } N_k^{A_j}(g_i) \bm{u}_k \cdot w_j \right), \qquad w_j = \frac{A_j}{\sum_{m \in \mathcal{L}^{g_i}} A_m} 
\label{eq:disp_ES_FEM}
\end{eqnarray}
where $N_k^{A_j}$ is $k$-th shape function of element $A_j$, $\bm{u}_k$ denotes displacement of node $k$, $A_j$ denotes element ID, $\mathcal{N}^{g_i}$ and $\mathcal{L}^{g_i}$ are the neighboring element sets of edge quadrature $g_i$, $\mathcal{M}^{A_j}$ denotes connectivity nodes of element $A_j$, $w_j$ is the elementary area weight of neighboring elements.  It is important to note that there are $2$ neighboring elements when edge quadrature is within studied domain while there are $1$ neighboring element when edge quadrature is on the boundary of domain, such as quadrature $g_{i^{\prime}}$.

Accordingly, the infinitesimal strain is defined as follows,
\begin{eqnarray}
\bm{\varepsilon} = \left(\varepsilon_{11}, \varepsilon_{22}, \gamma_{12} \right)^T = \A_j \left( \A_k \left[B_k^{A_j} \right]  w_{A_j}  \right) \cdot \left\{\bm{u} \right\}
\label{eq:strain_discretization}
\end{eqnarray}
where $\left[B_k^{A_j} \right]$ is stress-strain matrix with respect to node $k$ of element $A_j$, $\A$ denotes element assembly operator in FEM, $\left(\cdot \right)^T$ is vector transpose operator. In Constant Strain Triangle (CST) formulation, Eq.(\ref{eq:strain_discretization}) is written as,
\begin{eqnarray}
&\bm{\varepsilon} = \begin{Bmatrix}
\varepsilon_{11} \\
\varepsilon_{22} \\
\gamma_{12}
\end{Bmatrix} = \left[ \begin{array}{cc|cc}
\frac{\partial N_i^{e1}}{\partial x} w_{A_1}, & 0, & \frac{\partial N_i^{e_2}}{\partial x}  w_{A_2}, & 0 \\
0, & \frac{\partial N_i^{e_1}}{\partial y} w_{A_1}, & 0, & \frac{\partial N_i^{e_2}}{\partial y} w_{A_2} \\
\frac{\partial N_i^{e_1}}{\partial y} w_{A_1}, & \frac{\partial N_i^{e_1}}{\partial x} w_{A_1}, & \frac{\partial N_i^{e_2}}{\partial y} w_{A_2}, & \frac{\partial N_i^{e_2}}{\partial x} w_{A_2}
\end{array} \right]_{i=1\sim 3} \begin{Bmatrix}
u_i^{e1} \\
v_i^{e1} \\
u_i^{e2} \\
v_i^{e2} \\
\end{Bmatrix}_{i=1 \sim 3} ~.
\label{eq:CST_strain_discretization}
\end{eqnarray}
and in bilinear Quadrilateral (QUAD) formulation, Eq.(\ref{eq:strain_discretization}) is written as,
\begin{eqnarray}
&\bm{\varepsilon} = \begin{Bmatrix}
\varepsilon_{11} \\
\varepsilon_{22} \\
\gamma_{12}
\end{Bmatrix} = \left[ \begin{array}{cc|cc}
\frac{\partial N_i^{e1}}{\partial x} w_{A_1}, & 0, & \frac{\partial N_i^{e_2}}{\partial x}  w_{A_2}, & 0 \\
0, & \frac{\partial N_i^{e_1}}{\partial y} w_{A_1}, & 0, & \frac{\partial N_i^{e_2}}{\partial y} w_{A_2} \\
\frac{\partial N_i^{e_1}}{\partial y} w_{A_1}, & \frac{\partial N_i^{e_1}}{\partial x} w_{A_1}, & \frac{\partial N_i^{e_2}}{\partial y} w_{A_2}, & \frac{\partial N_i^{e_2}}{\partial x} w_{A_2}
\end{array} \right]_{i=1\sim 4} \begin{Bmatrix}
u_i^{e1} \\
v_i^{e1} \\
u_i^{e2} \\
v_i^{e2} \\
\end{Bmatrix}_{i=1\sim 4} ~.
\label{eq:QUAD_strain_discretization}
\end{eqnarray}
in which, $w_{A_j}$ is elementary area ratio of neighboring elements, $u^{e_j}$, $v^{e_j}$ denote nodal $x-$ and $y-$ displacements of element $e_j$, $N_i^{e_j}$ is $i-$the nodal shape function of element $e_j$. It is noted that the dimensions of stress-strain matrix are different with respect to Constant Strain Triangle element and bilinear Quadrilateral element.

The internal force in Constant Strain Triangle ES-FEM is formulated as follows,
\begin{eqnarray}
\bm{f}_{int} = \begin{Bmatrix}
f_{xi}^{e1} \\
f_{yi}^{e1} \\
f_{xi}^{e2} \\
f_{yi}^{e2}
\end{Bmatrix}_{i= 1\sim 3} = \left[ \begin{array}{cc|cc}
\frac{\partial N_i^{e1}}{\partial x} w_{A_1}, & 0, & \frac{\partial N_i^{e_2}}{\partial x}  w_{A_2}, & 0 \\
0, & \frac{\partial N_i^{e_1}}{\partial y} w_{A_1}, & 0, & \frac{\partial N_i^{e_2}}{\partial y} w_{A_2} \\
\frac{\partial N_i^{e_1}}{\partial y} w_{A_1}, & \frac{\partial N_i^{e_1}}{\partial x} w_{A_1}, & \frac{\partial N_i^{e_2}}{\partial y} w_{A_2}, & \frac{\partial N_i^{e_2}}{\partial x} w_{A_2}
\end{array} \right]_{i=1\sim 3}^T \begin{Bmatrix}
\sigma_{11} \\
\sigma_{22} \\
\sigma_{12}
\end{Bmatrix} \cdot \sum_j \frac{A_j}{3}
\label{eq:CST_fint_discretization}
\end{eqnarray}
and the assemblage of internal force in bilinear Quadrilateral ES-FEM is as follows,
\begin{eqnarray}
\bm{f}_{int} = \begin{Bmatrix}
f_{xi}^{e1} \\
f_{yi}^{e1} \\
f_{xi}^{e2} \\
f_{yi}^{e2}
\end{Bmatrix}_{i= 1\sim 4} =
\left[ \begin{array}{cc|cc}
\frac{\partial N_i^{e1}}{\partial x} w_{A_1}, & 0, & \frac{\partial N_i^{e_2}}{\partial x}  w_{A_2}, & 0 \\
0, & \frac{\partial N_i^{e_1}}{\partial y} w_{A_1}, & 0, & \frac{\partial N_i^{e_2}}{\partial y} w_{A_2} \\
\frac{\partial N_i^{e_1}}{\partial y} w_{A_1}, & \frac{\partial N_i^{e_1}}{\partial x} w_{A_1}, & \frac{\partial N_i^{e_2}}{\partial y} w_{A_2}, & \frac{\partial N_i^{e_2}}{\partial x} w_{A_2}
\end{array} \right]_{i=1\sim 4}^T \begin{Bmatrix}
\sigma_{11} \\
\sigma_{22} \\
\sigma_{12}
\end{Bmatrix} \cdot \sum_j \frac{A_j}{4}
\label{eq:QUAD_fint_discretization}
\end{eqnarray}
in which, $f_x^{e_j}$ and $f_y^{e_j}$ indicate nodal force of element $e_j$ with regard to $x$ and $y$ directions, respectively; $\sigma_{11}, \sigma_{22}, \sigma_{12}$ are plane Cauchy stress.

\subsection{Single crack-tip tracking in Crack Element Model}
Understanding how materials fracture requires careful evaluation of the energy released as cracks grow, known as the fracture energy release rate $\mathcal{G}$. This value is central to predicting whether a crack remains static or extends further. In linear elastic systems, the critical threshold $\mathcal{G}_c$ serves as an indicator of material toughness and is mathematically linked to the stress intensity factor. Moreover, in cases where multiple crack-driving forces are present, such as opening, sliding, or tearing, $\mathcal{G}$ helps distinguish between these failure modes.

\begin{figure}[htp]
\begin{minipage}{0.3\linewidth}
   \begin{center}
      \includegraphics[height=1.8in]{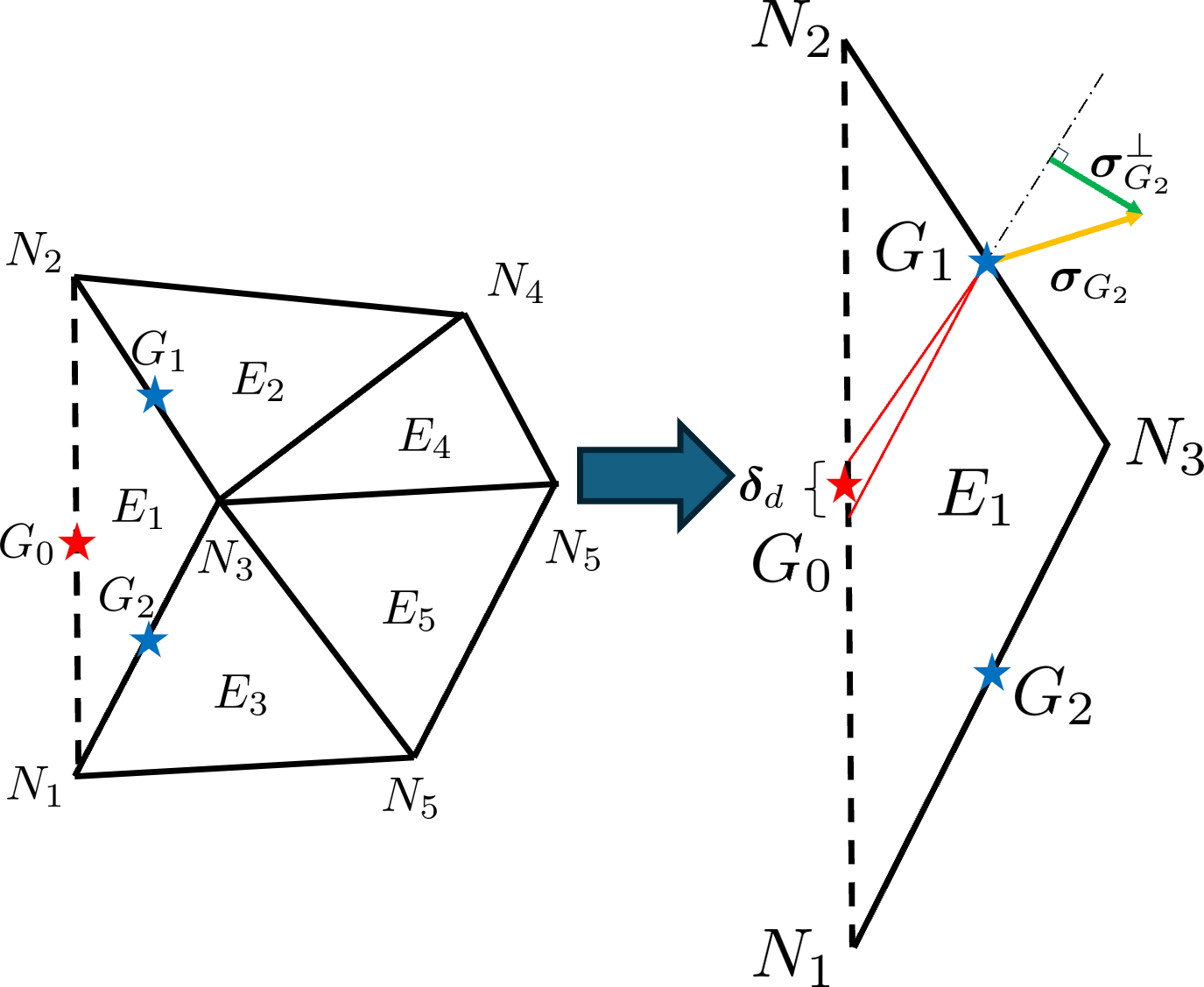}
  \end{center}
   \begin{center}
   (a)
   \end{center}
   \end{minipage}
   \hfill
\begin{minipage}{0.6\linewidth}
   \begin{center}
      \includegraphics[height=1.8in]{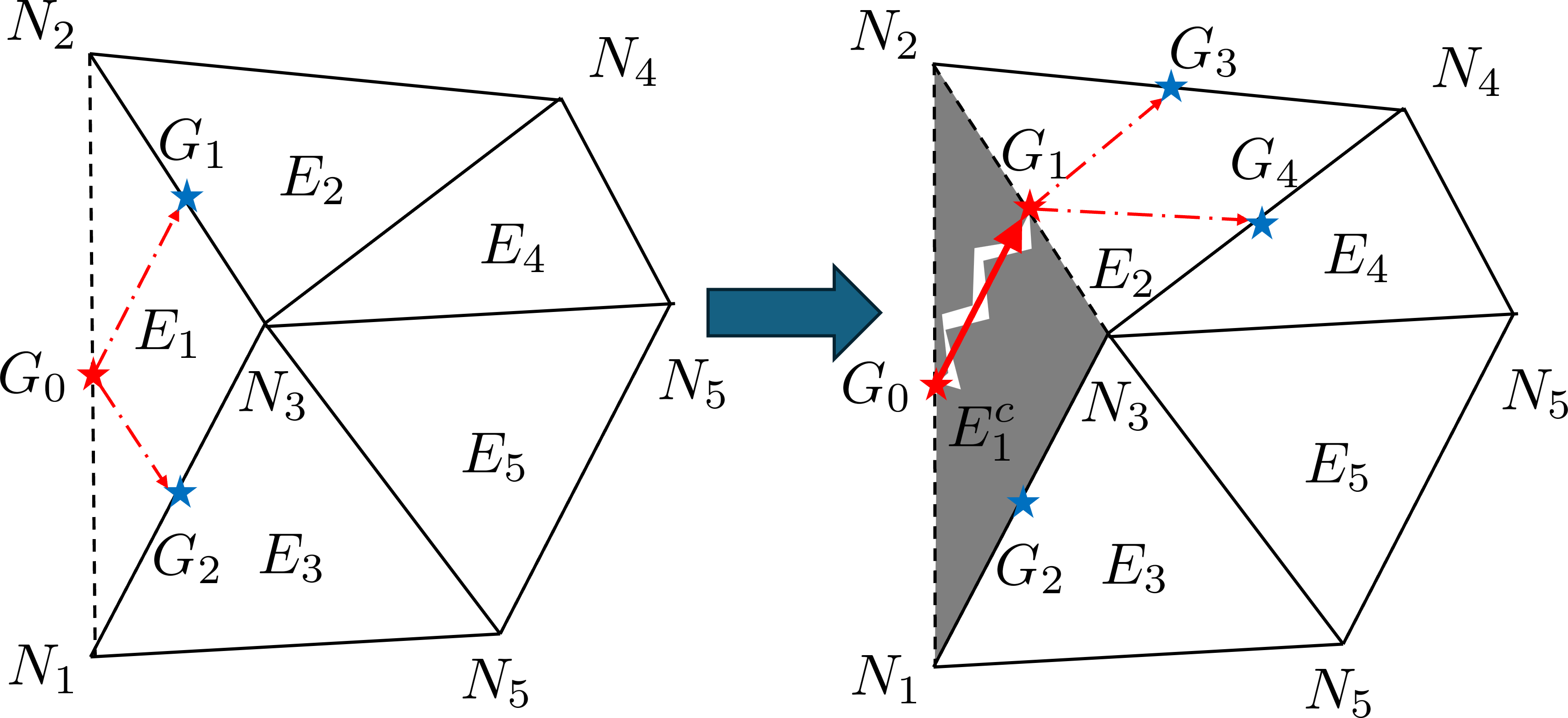}
  \end{center}
   \begin{center}
   (b)
   \end{center}
   \end{minipage}
   \caption{CST element with single crack tip tracking in two-dimensional Crack Element Model: (a). computation of fracture energy release rate $\mathcal{G}$ in CST element, in which $G_i$ is edge quadrature point ID, $E_i$ is element ID, $N_i$ is node ID, $\bm{\sigma}^{\perp}$ is normal projection of Cauchy stress $\bm{\sigma}$ at quadrature location; (b). single crack propagation in CST element, in which red star represents current crack tip, blue stars represent crack tip candidates, grey element denotes the element under completely cracked, dashed red lines represent possible crack path while solid red line represents cracked path within the element. }
   \label{fig2-2:CST_Gc_compute_propagation}
\end{figure}

The single crack-tip tracking algorithm in two-dimensional Crack Element Model was proposed (\cite{xie2025practical}) to resolve difficulties that other numerical methods encountered. In Constant Strain Triangle (CST) element, as shown in Figure.\ref{fig2-2:CST_Gc_compute_propagation}(a), the fracture energy release rate $\mathcal{G}_{G_0}$ at quadrature point $G_0$ is computed as,
\begin{eqnarray}
\mathcal{G}_{G_1-G_0} = \frac{\bm{\delta}_d \cdot \bm{\sigma}_{G_1}^{\perp}}{2}, \qquad \mathcal{G}_{G_2-G_0} = \frac{\bm{\delta}_d \cdot \bm{\sigma}_{G_2}^{\perp}}{2}, \qquad \text{and} \quad \bm{\delta}_d = \left(  \bm{u}_{N_2} -  \bm{u}_{N_1} \right) \cdot \mathcal{H}\left(\frac{\|  \bm{x}_{N_2} -  \bm{x}_{N_1} \|}{\|  \bm{X}_{N_2} -  \bm{X}_{N_1} \|} - 1 \right) 
\nonumber
\label{eq:fracture_energy_release_rate}
\end{eqnarray}
in which, $\bm{\delta}_d$ denotes edge stretch at current crack tip quadrature location, $\bm{u}$ is nodal displacement, $\bm{x}$ and $\bm{X}$ are spatial and reference nodal coordinates, respectively, $\mathcal{H}\left(\cdot \right)$ is Heaviside function and $\left\| \cdot \right\|$ is Euclidean norm. 
Since there are two possible crack paths in Constant Strain Triangle element, all possible conditions based on computed fracture energy release rate above are listed below,
\begin{itemize}
\item If $\mathcal{G}_{G_1 - G_0}$ < $\mathcal{G}_c$ and $\mathcal{G}_{G_2 - G_0}$ < $\mathcal{G}_c$, then No new crack path forms.
\item If any one of the quadrature candidates satisfies the critical fracture energy release rate $\mathcal{G}_c$ while another doesn't, such as $\mathcal{G}_{G_1 - G_0}$ > $\mathcal{G}_c$ > $\mathcal{G}_{G_2 - G_0}$, then a new crack path from $G_0$ to $G_1$ forms (as shown in Figure.\ref{fig2-2:CST_Gc_compute_propagation}(b)).
\item If both $\mathcal{G}_{G_1 - G_0}$ > $\mathcal{G}_c$ and $\mathcal{G}_{G_2 - G_0}$ > $\mathcal{G}_c$ and fracture energy release rate at one of the quadrature candidate is obviously greater than another, such as $\mathcal{G}_{G_1 - G_0}$ > $\gamma \left(\mathcal{G}_{G_1 - G_0} - \mathcal{G}_{G_2 - G_0}\right)$, then a new crack path from $G_0$ to $G_1$ forms.
\end{itemize}
in which, $\gamma$ is termed as split ratio and is to avoid incorrect crack path due to fluctuation in stress fields. 

\begin{figure}[htp]
	\centering
            \begin{center}
            \includegraphics[height=3.5in]{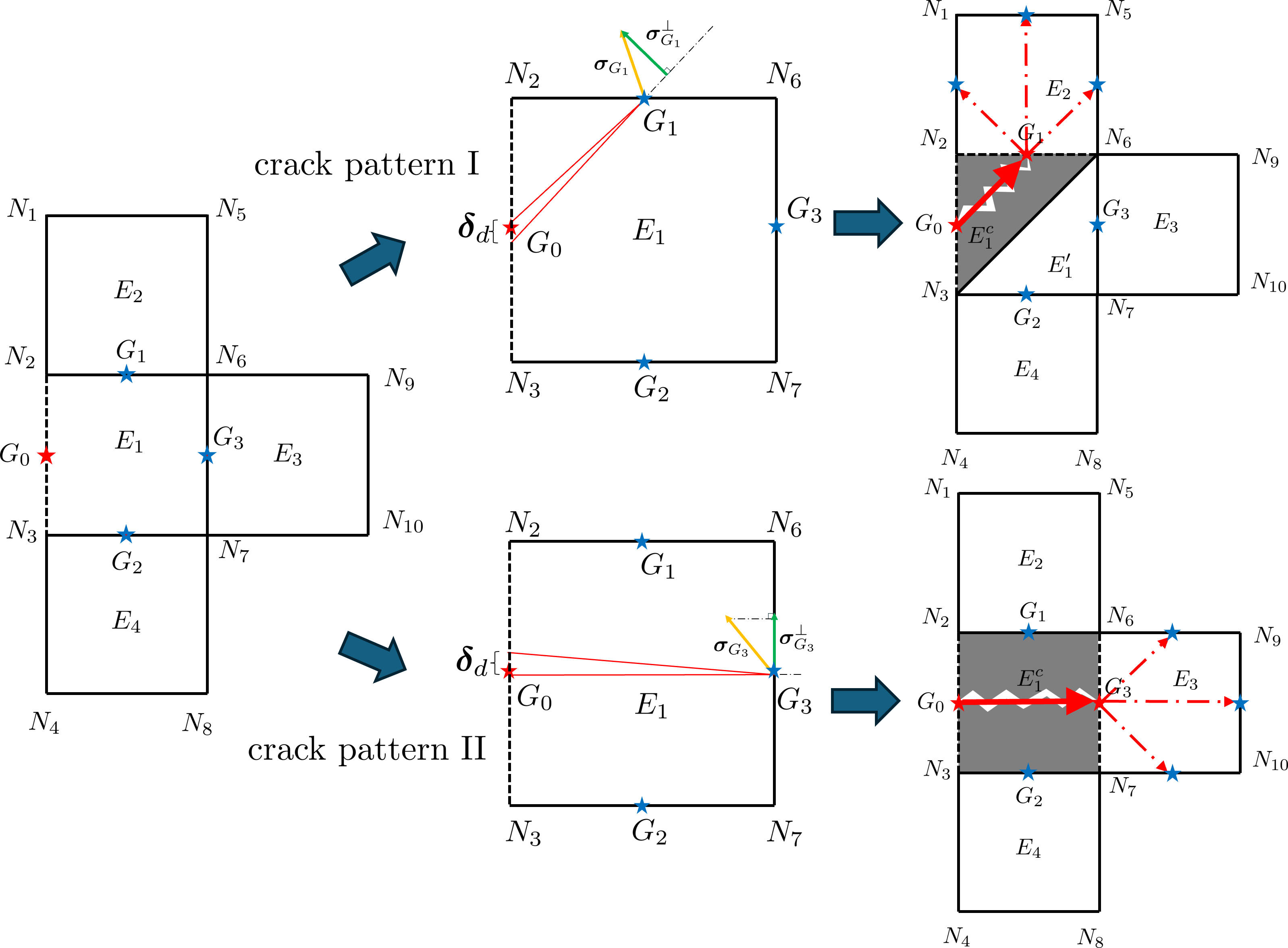}
            \end{center}
        \caption{Quadrilateral element with single crack tip tracking in two-dimensional Crack Element Mode, two possible crack patterns: crack pattern I illustrates a crack path forms from quadrature $G_0$ to $G_1$ or $G_2$ at the adjacent edges; crack pattern II illustrates a crack path forms from quadrature $G_0$ to $G_3$ at opposite edge.}
        \label{fig2-3:QUAD_Gc_compute_propagation}
\end{figure}

In bilinear Quadrilateral (QUAD) element, as shown in Figure.\ref{fig2-3:QUAD_Gc_compute_propagation}, the fracture energy release rate $\mathcal{G}_{G_0}$ for possible crack paths are computed as follows,
\begin{eqnarray}
\mathcal{G}_{G_1-G_0} = \frac{\bm{\delta}_d \cdot \bm{\sigma}_{G_1}^{\perp}}{2}, \qquad \mathcal{G}_{G_2-G_0} = \frac{\bm{\delta}_d \cdot \bm{\sigma}_{G_2}^{\perp}}{2}, \qquad \mathcal{G}_{G_3-G_0} = \frac{\bm{\delta}_d \cdot \bm{\sigma}_{G_3}^{\perp}}{2} ~.
\nonumber
\end{eqnarray}\\
accordingly possible cracking conditions based on computed fracture energy release rate in quadrilateral element is listed below,
\begin{itemize}
\item If $\mathcal{G}_{G_1 - G_0}$ < $\mathcal{G}_c$,  $\mathcal{G}_{G_2 - G_0}$ < $\mathcal{G}_c$ and $\mathcal{G}_{G_3 - G_0}$ < $\mathcal{G}_c$, then No new crack path forms.
\item If any one of the three quadrature candidates satisfies the critical fracture energy release rate $\mathcal{G}_c$ while the other two don't, such as $\mathcal{G}_{G_1 - G_0}$ > $\mathcal{G}_c$ but $\mathcal{G}_{G_2 - G_0}$ < $\mathcal{G}_c$, $\mathcal{G}_{G_3 - G_0}$ < $\mathcal{G}_c$, then a new crack path from $G_0$ to $G_1$ forms.
\item If any two of the three quadrature candidates satisfy the critical fracture energy release rate $\mathcal{G}_c$ while the other one does not, such as $\mathcal{G}_{G_1 - G_0}$ > $\mathcal{G}_c$, $\mathcal{G}_{G_2 - G_0}$ > $\mathcal{G}_c$ but $\mathcal{G}_{G_3 - G_0}$ < $\mathcal{G}_c$, an extra criterion is introduced to distinguish $\mathcal{G}_{G_1 - G_0}$ and $\mathcal{G}_{G_2 - G_0}$, i.e., when one of them is obviously greater than another, such as $\mathcal{G}_{G_1 - G_0}$ > $\gamma \left(\mathcal{G}_{G_1 - G_0} - \mathcal{G}_{G_2 - G_0}\right)$, then a new crack path forms from $G_0$ to $G_1$.
\item If all three quadrature candidates satisfy the critical fracture energy release rate $\mathcal{G}_c$, such as $\mathcal{G}_{G_1 - G_0}$ > $\mathcal{G}_c$, $\mathcal{G}_{G_2 - G_0}$ > $\mathcal{G}_c$, and $\mathcal{G}_{G_3 - G_0}$ > $\mathcal{G}_c$, then extra criteria are introduced to distinguish $\mathcal{G}_{G_1 - G_0}$, $\mathcal{G}_{G_2 - G_0}$ and $\mathcal{G}_{G_3 - G_0}$, i.e., when one of them is obviously greater than the others, such as $\mathcal{G}_{G_1 - G_0}$ > $\gamma \left(\mathcal{G}_{G_1 - G_0} - \mathcal{G}_{G_2 - G_0}\right)$ and $\mathcal{G}_{G_1 - G_0}$ > $\gamma \left(\mathcal{G}_{G_1 - G_0} - \mathcal{G}_{G_3 - G_0}\right)$, then a new crack path forms from $G_0$ to $G_1$.
\end{itemize}
in which, $\gamma$ is termed as split ratio and is to avoid incorrect crack path due to fluctuation in stress fields.

\section{Multiple Crack-tips Tracking}
Crack branching is a key phenomenon in dynamic fracture, where a single propagating crack divides into multiple paths. Experimental studies using materials like PMMA, glass, and concrete have documented that the onset of branching is closely tied to dynamic instabilities near the crack tip. The mechanisms are strongly influenced by factors such as loading rate, material brittleness, and structural geometry. On the theoretical side, branching is generally explained through energy-based criteria, where the energy release rate exceeds the threshold required for creating new crack surfaces. Stress intensity factor-based criteria and local symmetry arguments have also been employed to predict branching. However, these analytical approaches struggle to fully capture the complex, often asymmetric paths seen in experiments.

Numerical methods, such as Finite Element Method, phase-field, and peridynamics, have been validated to simulate crack branching, in which phase-field and peridynamics are capable to model complex crack patterns without crack tracking and predefined criteria. On the contrary, Finite Element Method or element-based methods require strict and intricate criteria, which are usually specific rather than universal (\cite{belytschko2003dynamic}, \cite{linder2009finite}). 

The two-dimensional Crack Element Model with single crack-tip tracking algorithm proposed by Xie et al. (\cite{xie2025practical}) only deals with single crack propagation and the extended three-dimensional Crack Element Model (\cite{xie2025gpu}) demonstrates unexpected capability of capturing crack branching. Inspired by mechanism of three-dimensional crack tracking algorithm, the Multiple Crack-tips Tracking algorithm in two-dimensional CEM (MCT-2D-CEM) is proposed in present study.
\begin{figure}[htp]
        \centering
        \begin{minipage}{0.45\linewidth}
            \begin{center}
            \includegraphics[height=2.4in]{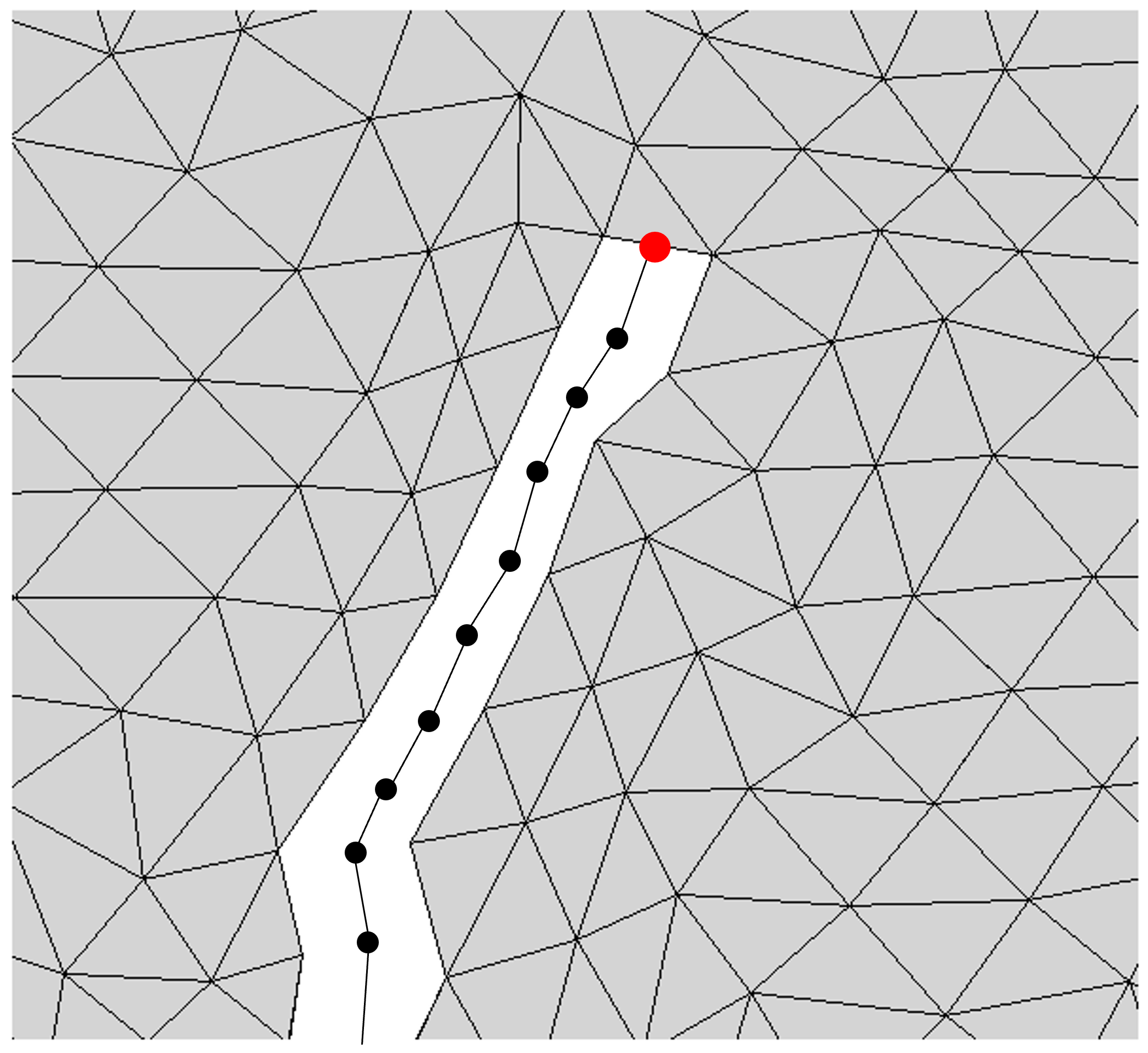}
            \end{center}
            \begin{center}
            (a)
            \end{center}
        \end{minipage}
        \hfill
        \begin{minipage}{0.45\linewidth}
            \begin{center}
            \includegraphics[height=2.4in]{MCT.png}
            \end{center}
            \begin{center}
            (b)
            \end{center}
        \end{minipage}   
        \caption{(a). Single Crack-tip Tracking algorithm in two-dimensional CEM: red solid circle denotes current crack tip while black solid circles denote past crack tips, the solid line connecting solid circles denotes single crack path; (b). Multiple Crack-tips Tracking algorithm in two-dimensional CEM: solid circles with three different color are edge quadrature points on free surfaces after crack propagates, three edge quadrature points with the most fracture energy release rate exceeding critical fracture energy release rate $\mathcal{G}_c$ are marked by red and orange while other candidate quadrature points are marked by green.}
        \label{fig3-1: SCT-MCT-illustration}
\end{figure}

The single crack-tip tracking algorithm in two-dimensional Crack Element Model is so straightforward that only one edge quadrature point is tracked when the crack propagates, as the red solid circle shown in Figure.\ref{fig3-1: SCT-MCT-illustration}(a). Once the computed fracture energy release at the current crack tip (red solid circle) exceeds critical fracture energy release rate, i.e., $\mathcal{G} > \mathcal{G}_c$, the element is fractured and crack tip moves to next edge quadrature. Therefore a single crack path is formed by connecting a sequential of previous crack-tips, as the black solid line shown in Figure.\ref{fig3-1: SCT-MCT-illustration}(a). Contrary to single crack tip tracking algorithm, the newly proposed Multiple Crack-tips Tracking (MCT) algorithm considers all edge quadrature points along the free surfaces (marked by purple line in Figure.\ref{fig3-1: SCT-MCT-illustration}(b)), selects the edge quadrature points satisfying fracture criteria as current crack-tip candidates and picks the edge quadrature point with the largest fracture energy release rate among current crack-tip candidates as the current crack tip.

For example, as shown in Figure.\ref{fig3-1: SCT-MCT-illustration}(b), the MCT algorithm computes and sorts all edge quadrature points (including green points, orange points and the red point) on free surface, then selects the three edge quadrature points (numbered by $1,2,3$) whose computed fracture energy release rate exceed critical fracture energy release rate, i.e., $\mathcal{G}_1 > \mathcal{G}_c, \ \mathcal{G}_2 > \mathcal{G}_c, \ \mathcal{G}_3 > \mathcal{G}_c$, then the No.$2$ edge quadrature point is picked among the three crack-tip candidates as crack tip since the following fracture criteria are satisfied,
\begin{eqnarray}
\begin{aligned}
&\mathcal{G}_2 > \mathcal{G}_1,\quad \mathcal{G}_2 > \mathcal{G}_3 \nonumber \\
&\mathcal{G}_2 > \gamma\left(\mathcal{G}_2-\mathcal{G}_1\right) \nonumber \\
&\mathcal{G}_2 > \gamma\left(\mathcal{G}_2-\mathcal{G}_3\right) ~. \nonumber
\end{aligned}
\end{eqnarray}
in which, the split ratio $\gamma$ is used to smooth stress waves in dynamic simulation.

\section{Representative Numerical Examples}
Four representative two-dimensional transient dynamic fracturing problems are conducted in quasi-brittle material to demonstrate that the various crack patterns, including single crack, crack branching and even fragmentation, are captured by the MCT-2D-CEM algorithm. The Kalthoff-Winkler plate benchmark example is used to show that the MCT-2D-CEM algorithm can accurately trace the single crack with micro-cracks since tracking of multiple crack-tips is enabled in the new algorithm. Two classic crack branching benchmark examples are discussed: one is with Neumann boundary condition and another is with Dirichlet boundary condition. Furthermore, a two-dimensional fragmentation benchmark example is provided to test the capability limit of the MCT-2D-CEM in tracking multiple crack paths, whose results are strongly encouraging and greatly exceed initial expectation. To be consistent, all benchmark examples are implemented using Constant Strain Triangle element.

\subsection{Dynamic Kalthoff-Winkler plate}
The Kalthoff-Winkler experiment involves a rectangular plate with two symmetric notches, typically positioned at a fixed angle and distance from the impact point. When a high-speed projectile strikes the specimen between the notches, it generates a dynamic stress field that is symmetric and predominantly governed by shear forces. This leads to Mode I (in-plane tension) and Mode II (in-plane shear) fractures, which are known to produce crack propagation paths inclined at predictable angles relative to the initial notch direction - commonly around $70^\circ$ from the notch axis.

A key reason for the single, repeatable crack trajectory is the high strain rate induced by the impact loading. Under such conditions, materials - especially brittle ones like hardened steels, ceramics or PMMA (polymethyl methacrylate) - tend to fracture in a more deterministic and less ductile manner. Brittle fracture under dynamic loading is governed more by the stress state and less by micro-structural variability, allowing the crack to follow the same trajectory in repeated tests. 

The homogeneity and isotropy of the test material also play a crucial role. When the material's internal structure is uniform and free from defects or grain boundaries that could otherwise divert the crack, stress field dominates the fracture behavior. In such cases, crack initiation and propagation are highly reproducible. The notches themselves further guide this process by acting as stress concentrations, ensuring the crack initiation occurs at known, controlled locations on each specimen. 

\begin{figure}[htp]
        \centering
        \begin{minipage}{0.45\linewidth}
            \begin{center}
            \includegraphics[height=2.4in]{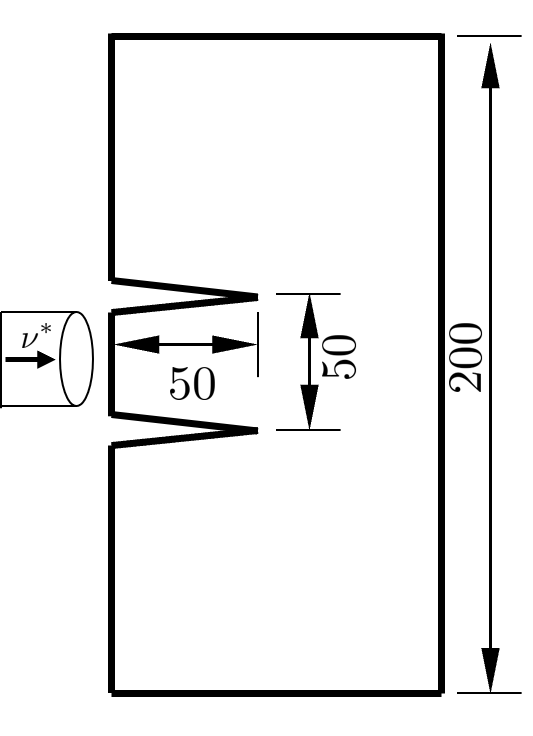}
            \end{center}
            \begin{center}
            (a)
            \end{center}
        \end{minipage}
        \begin{minipage}{0.45\linewidth}
            \begin{center}
            \includegraphics[height=2.4in]{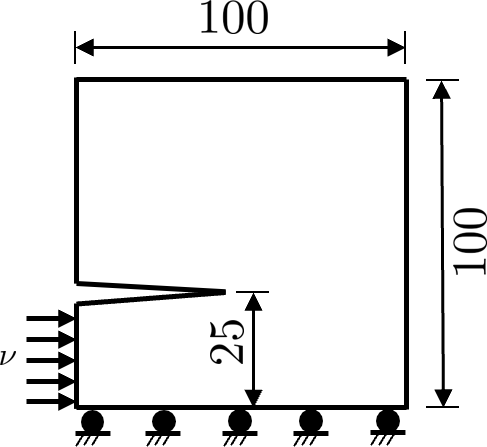}
            \end{center}
            \begin{center}
            (b)
            \end{center}
        \end{minipage}   
        \caption{(a). The whole geometry, boundary condition of Kalthoff-Winkler plate experiment; (b). Upper half of the Kalthoff-Winkler plate.}
        \label{fig5-1: kalthoff-plate}
\end{figure}

The material properties used in simulation are as below: Young's modulus $E=190\ GPa$, poisson ratio $v=0.3$, critical fracture energy release rate $\mathcal{G}_c = 2.213 \times 10^4 J/m^2$, density $\rho=8000 kg/m^3$. The boundary conditions are shown in Figure.\ref{fig5-1: kalthoff-plate}(b), i.e., external impact is horizontally applied at left center part with specific velocity and bottom boundary of the upper half plate are constrained vertically. The applied velocity follows following equation,
\begin{equation}
v =
\begin{cases} 
\frac{t}{t_0}v_0,  & \text{if } \ t \le t_0, \\
v_0, & \text{if } \ t > t_0.
\end{cases}
\end{equation}
in which, $v_0 = 16.5\ m/s$ and $t_0 = 1\ \mu s$.

\begin{figure}[htp]
        \centering
        \begin{minipage}{0.3\linewidth}
            \begin{center}
            \includegraphics[height=2.0in]{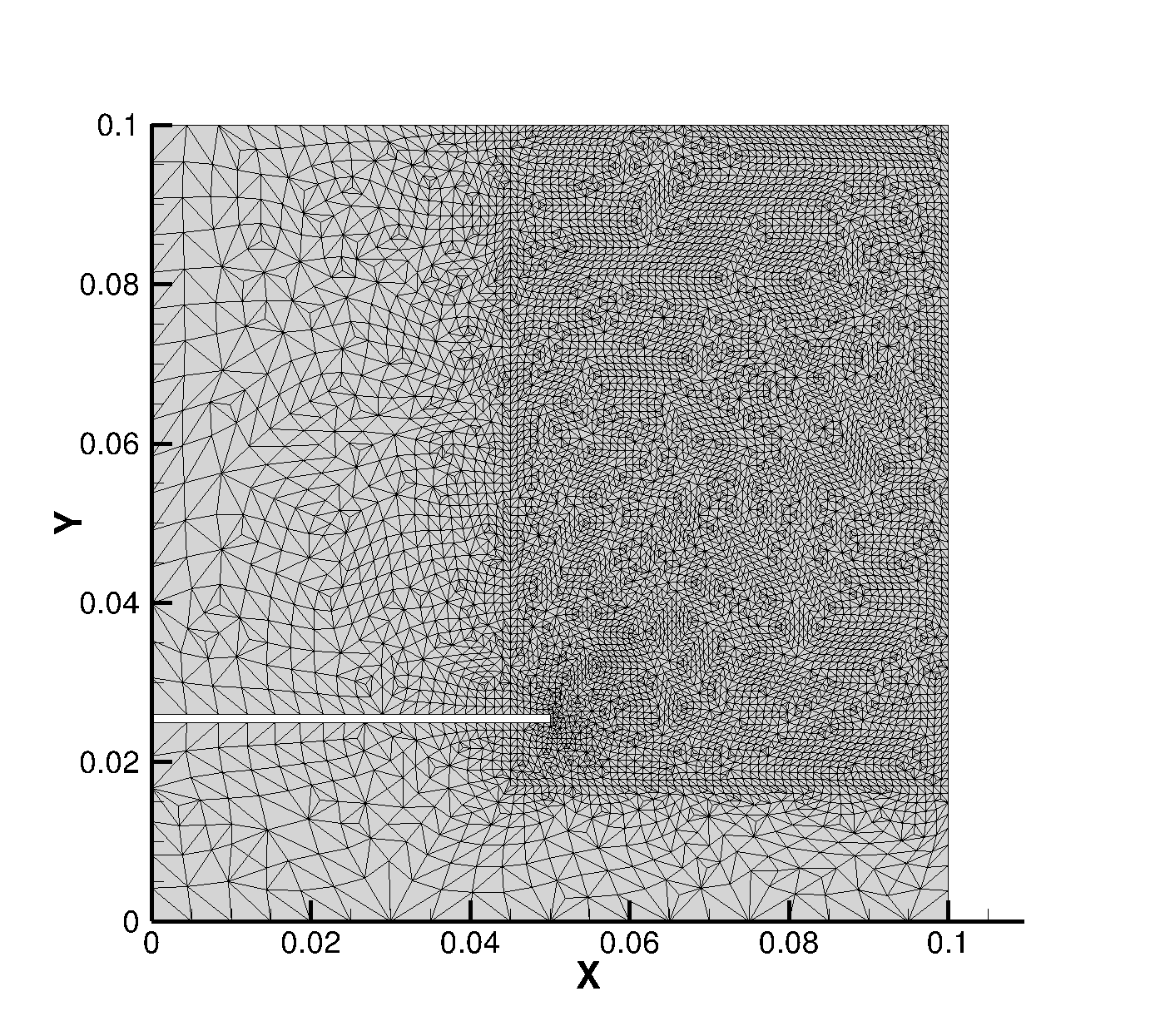}
            \end{center}
            \begin{center}
            (a)
            \end{center}
        \end{minipage}
        \begin{minipage}{0.3\linewidth}
            \begin{center}
            \includegraphics[height=2.0in]{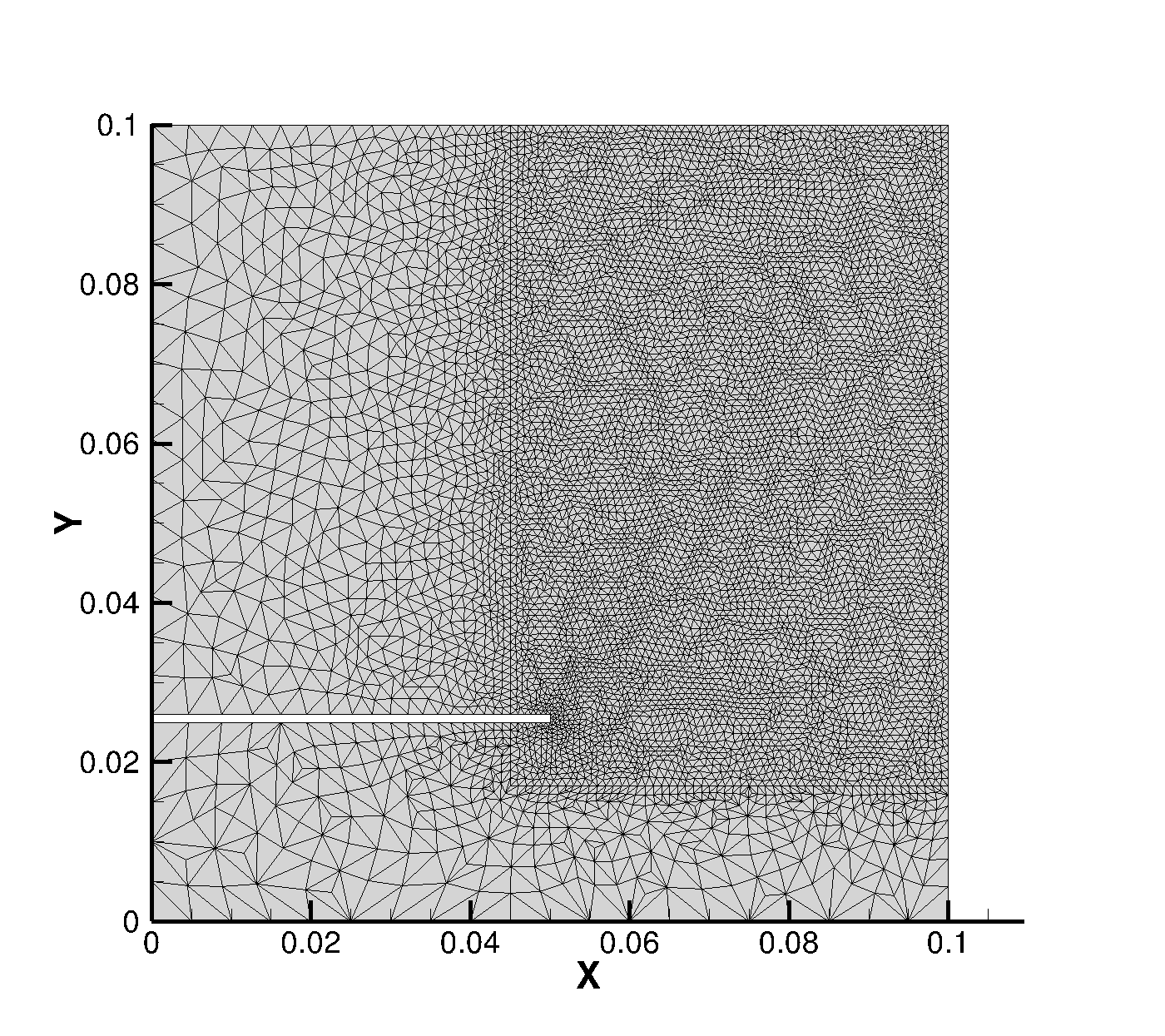}
            \end{center}
            \begin{center}
            (b)
            \end{center}
        \end{minipage}   
        \begin{minipage}{0.3\linewidth}
            \begin{center}
            \includegraphics[height=2.0in]{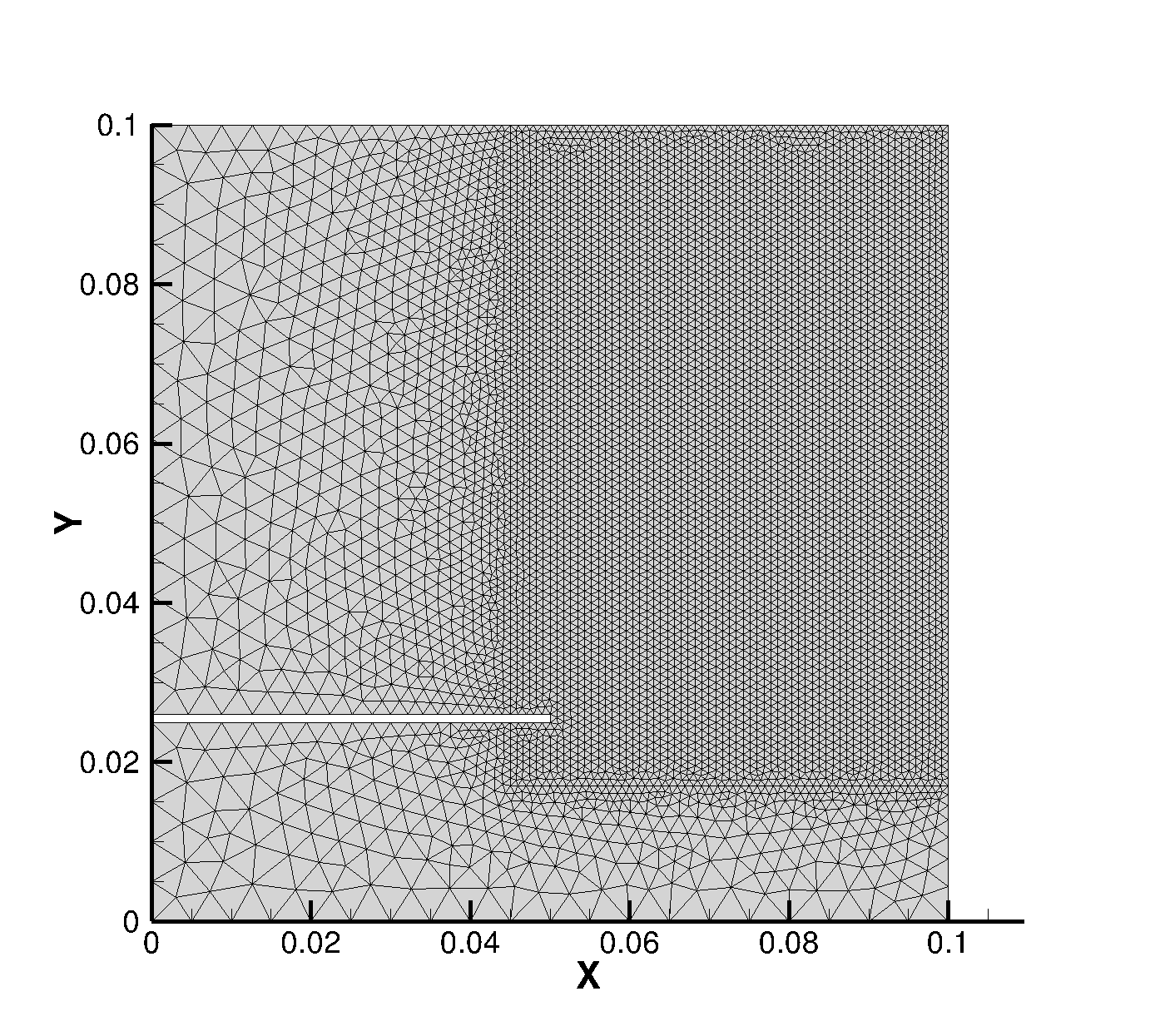}
            \end{center}
            \begin{center}
            (c)
            \end{center}
        \end{minipage}
        \caption{Three representative triangle meshes are illustrated: (a). irregular mesh with $7212$ nodes and $14158$ elements; (b). irregular mesh with $7173$ nodes and $14086$ elements; (c). regular mesh with $6727$ nodes and $13199$ elements.}
        \label{fig5-2: kalthoff-meshes}
\end{figure}

To comprehensively test the MCT-2D-CEM algorithm in capturing single crack pattern, three different meshes are selected to implement simulation: the first two are irregular meshes including one with $7212$ nodes and $14158$ elements and the other with $7173$ nodes and $14086$ elements. Besides, another regular mesh with $6727$ nodes and $13199$ elements is also investigated. The details of three meshes are shown in Figure.\ref{fig5-2: kalthoff-meshes}.

\begin{figure}[htp]
	\centering
        \begin{minipage}{0.3\linewidth}
            \begin{center}
            \includegraphics[height=2.0in]{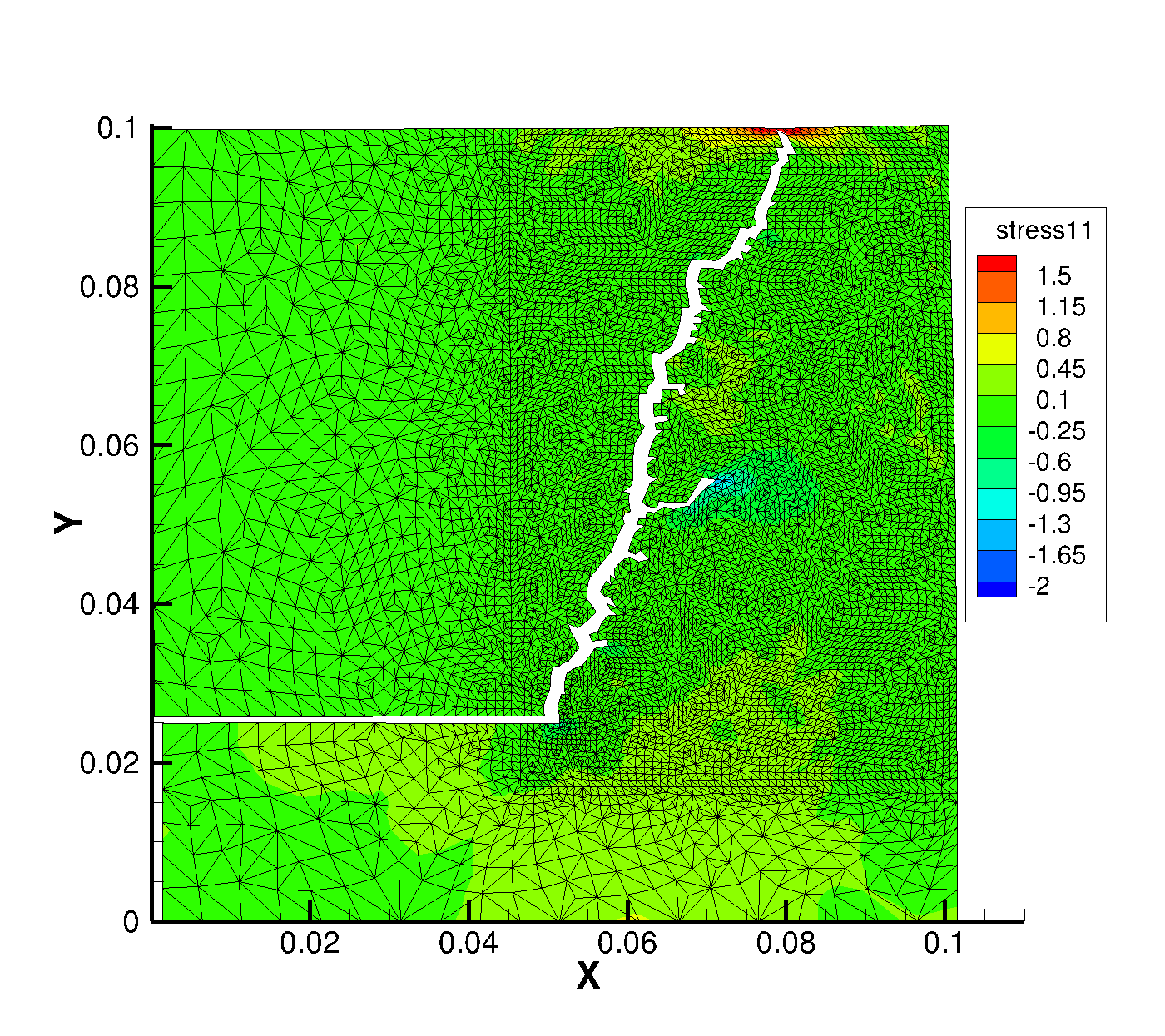}
            \end{center}
            \begin{center}
            (a)
            \end{center}
        \end{minipage}
        \hfill
        \begin{minipage}{0.3\linewidth}
            \begin{center}
            \includegraphics[height=2.0in]{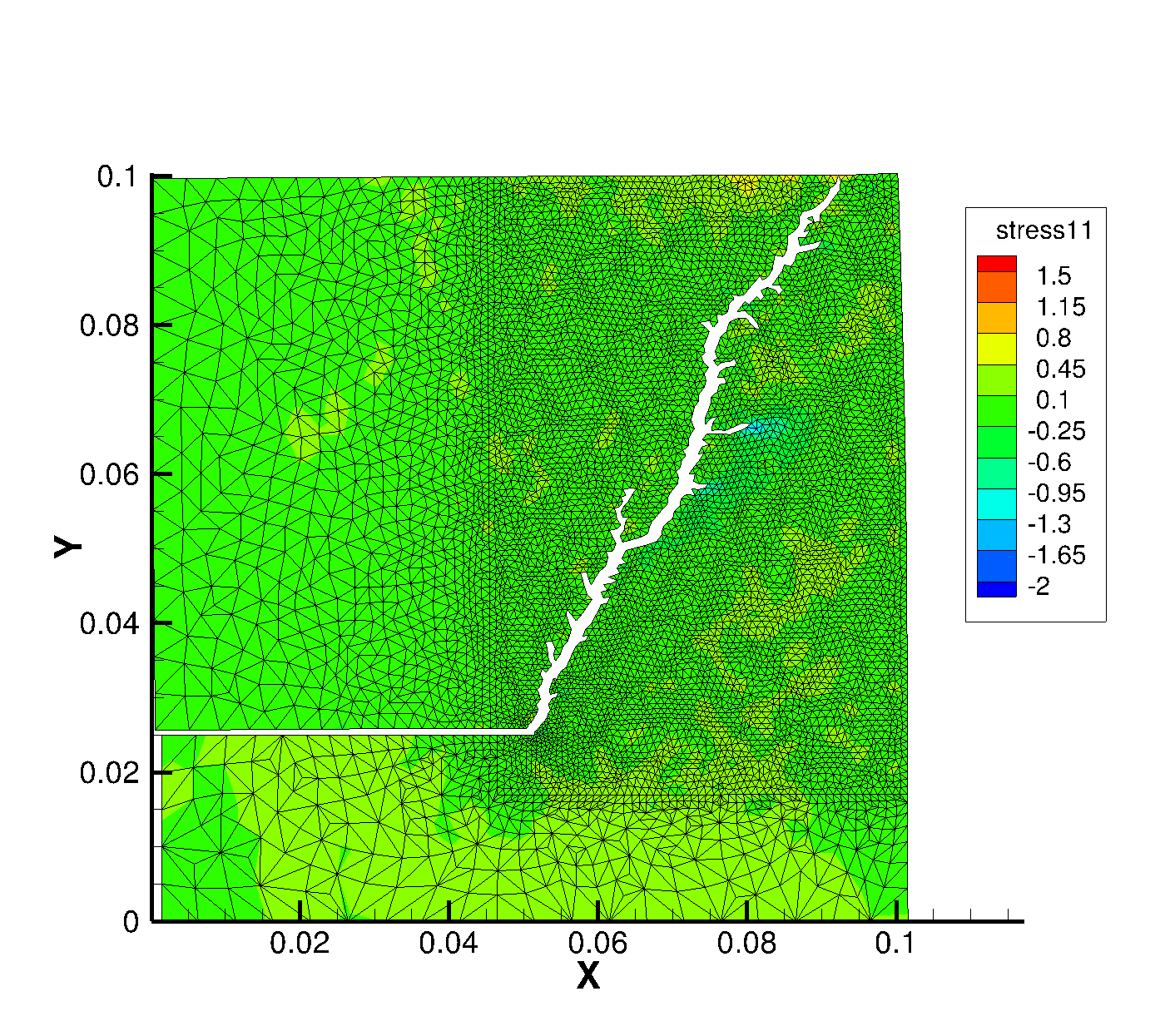}
            \end{center}
            \begin{center}
            (b)
            \end{center}
        \end{minipage}   
        \hfill
        \begin{minipage}{0.3\linewidth}
            \begin{center}
            \includegraphics[height=2.0in]{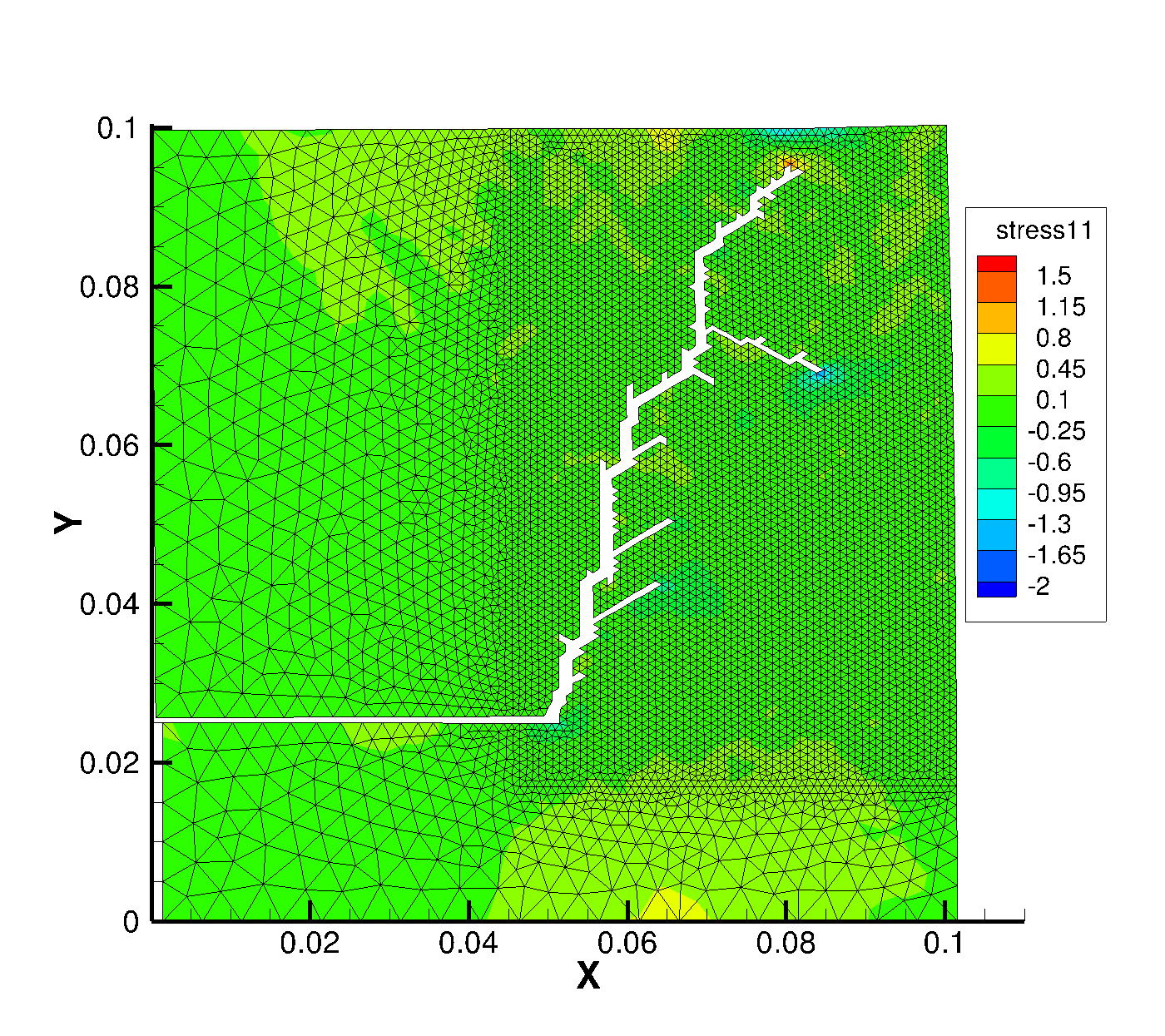}
            \end{center}
            \begin{center}
            (c)
            \end{center}
        \end{minipage}
        \caption{The final crack patterns of the three representative meshes are illustrated: (a). irregular mesh with $7212$ nodes and $14158$ elements; (b). irregular mesh with $7173$ nodes and $14086$ elements; (c). regular mesh with $6727$ nodes and $13199$ elements.}
        \label{fig5-3: Kalthoff-crack-patterns}
\end{figure}

\begin{figure}[htp]
	\centering
        \begin{minipage}{0.3\linewidth}
            \begin{center}
            \includegraphics[height=2.0in]{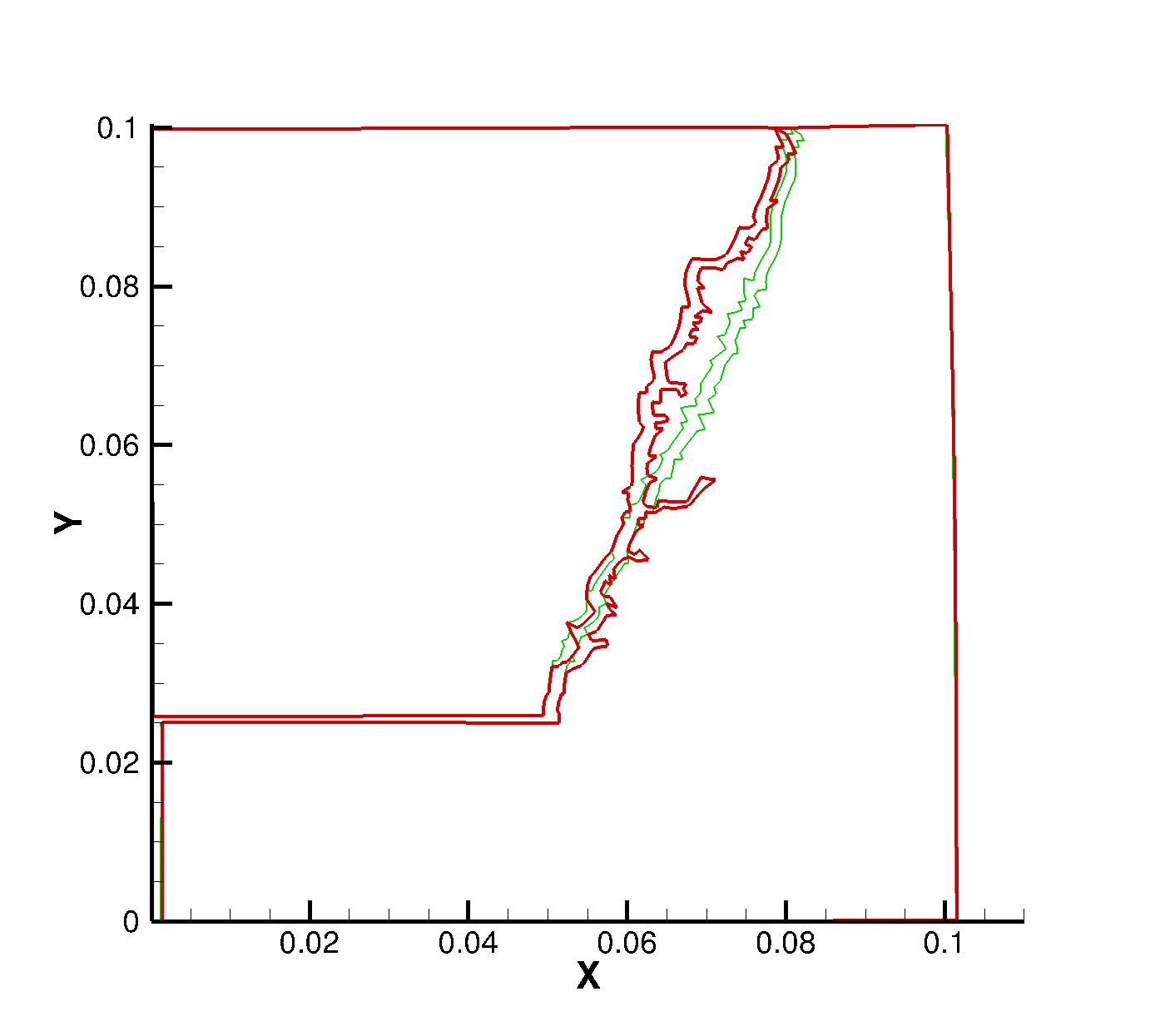}
            \end{center}
            \begin{center}
            (a)
            \end{center}
        \end{minipage}
        \hfill
        \begin{minipage}{0.3\linewidth}
            \begin{center}
            \includegraphics[height=2.0in]{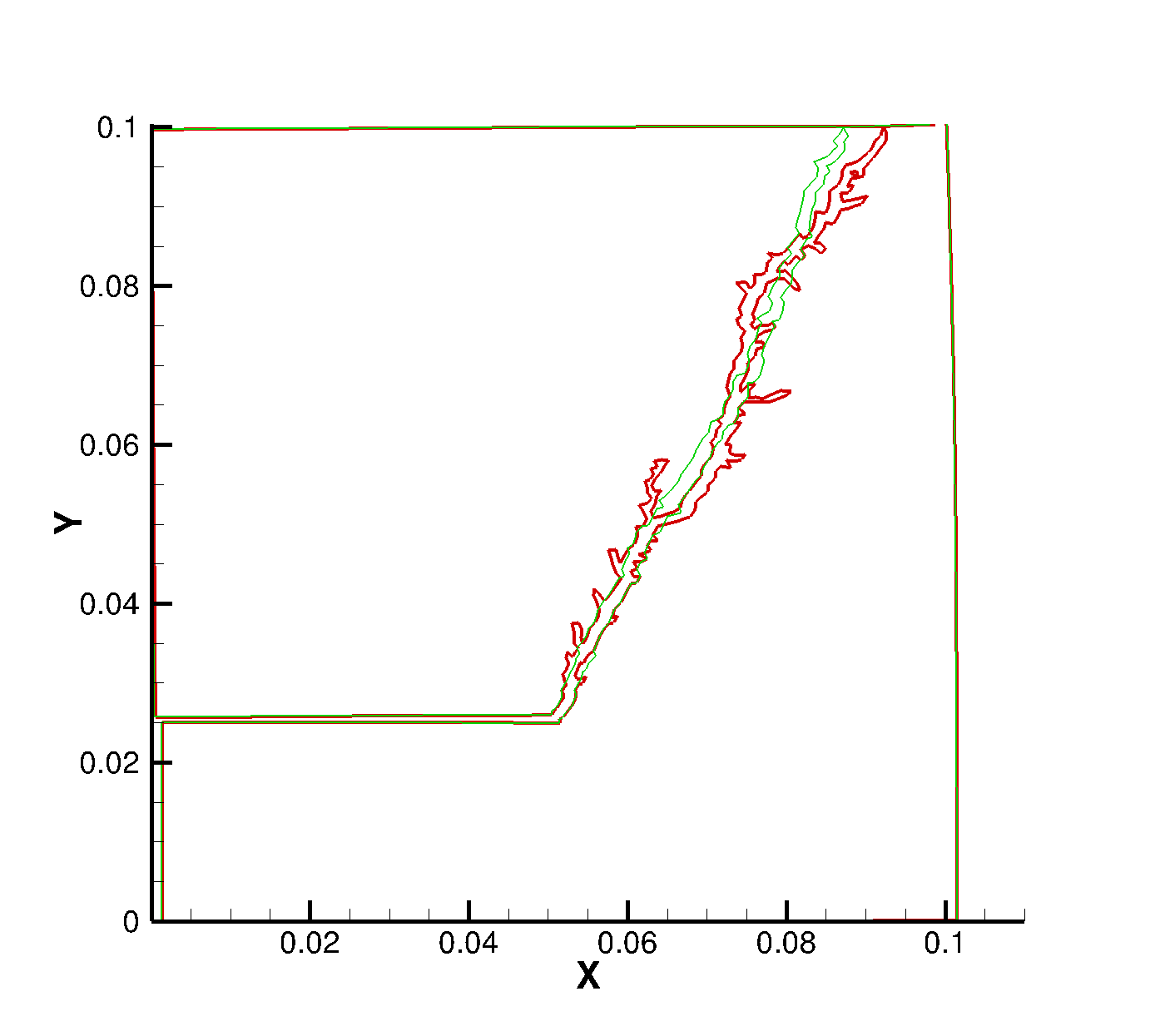}
            \end{center}
            \begin{center}
            (b)
            \end{center}
        \end{minipage}   
        \hfill
        \begin{minipage}{0.3\linewidth}
            \begin{center}
            \includegraphics[height=2.0in]{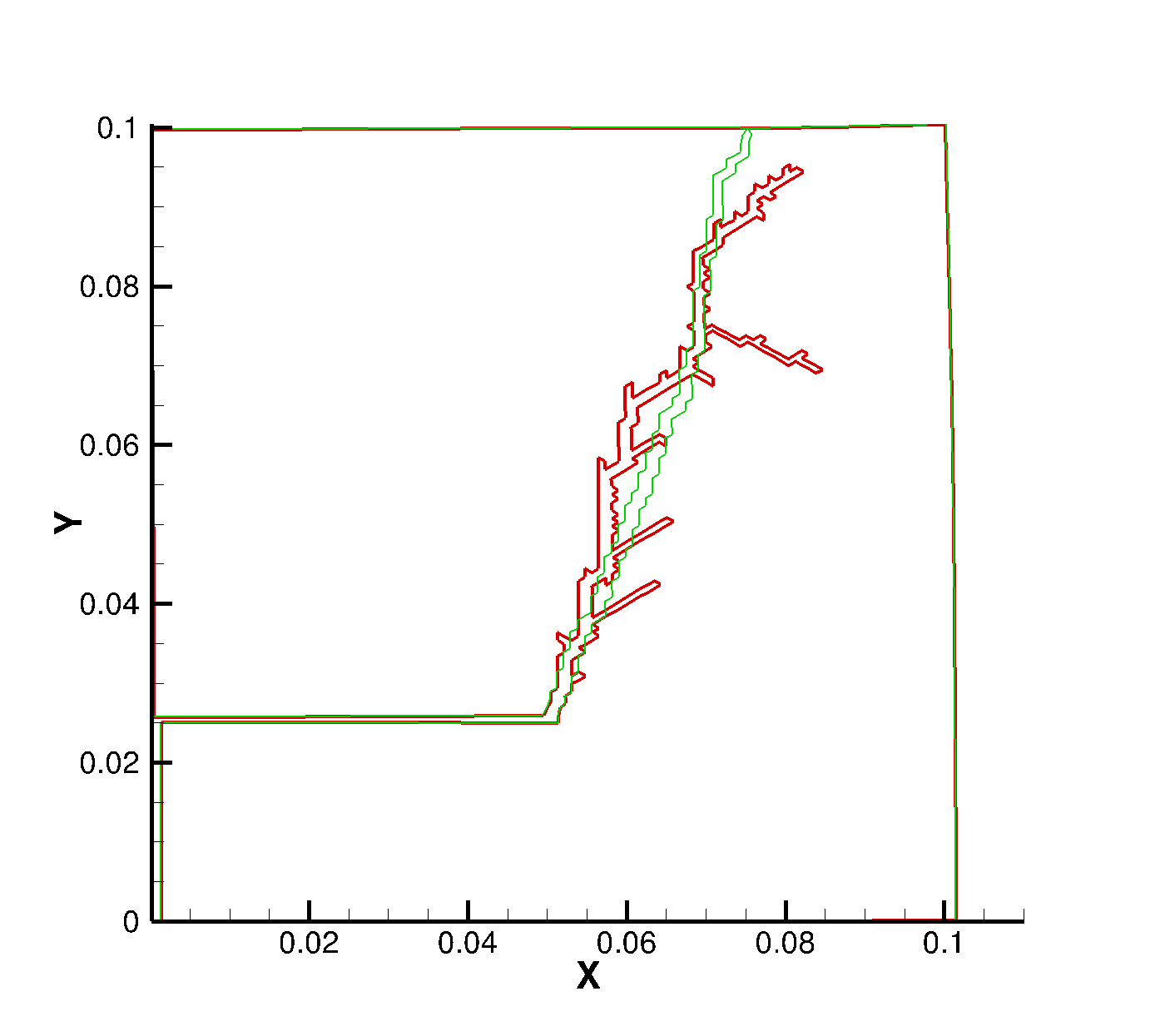}
            \end{center}
            \begin{center}
            (c)
            \end{center}
        \end{minipage}
        \caption{The final crack patterns comparison between mutiple crack tips tracking algorithm and single crack tip tracking algorithm in the three representative meshes are illustrated, in which the red solid line is from multiple crack tips tracking algorithm and the green solid line is from single crack tip tracking algorithm: (a). irregular mesh with $7212$ nodes and $14158$ elements; (b). irregular mesh with $7173$ nodes and $14086$ elements; (c). regular mesh with $6727$ nodes and $13199$ elements.}
        \label{fig5-4: crack-patterns-comparison}
\end{figure}

The final crack patterns and maximum principal stress contour are shown in Figure.\ref{fig5-3: Kalthoff-crack-patterns}. Besides, the crack path comparison between multiple crack tips tracking algorithm and single crack tip tracking algorithm (\cite{xie2025practical}) are provided in Figure.\ref{fig5-4: crack-patterns-comparison}. Compared to the results from the single crack tip tracking algorithm whose crack angle is strongly confined around $65^\circ \sim 70^\circ$, the multiple crack tips tracking algorithm can reproduce an approximated crack pattern as experimental and numerical results (\cite{borden2012phase}, \cite{rabczuk2008discontinuous}, \cite{kalthoff1988failure}, \cite{kalthoff2000modes}, \cite{song2009cracking}), but more micro-cracks are preserved, which cannot be captured by single crack tip tracking algorithms. 

However, due to extra introduced micro-cracks along the main crack path, it leads to stronger fluctuation in stress fields and extra dissipated energy in micro-branch so that more difficulties and confusion in tracking main crack pattern. For example, in Figure.\ref{fig5-4: crack-patterns-comparison}(c), additional micro-cracks dissipated extra energy so that the final main crack tip stops within the plate rather than penetrating the whole upper half plate. 

\begin{figure}[htp]
	\centering
        \begin{minipage}{0.24\linewidth}
            \begin{center}
            \includegraphics[height=1.5in]{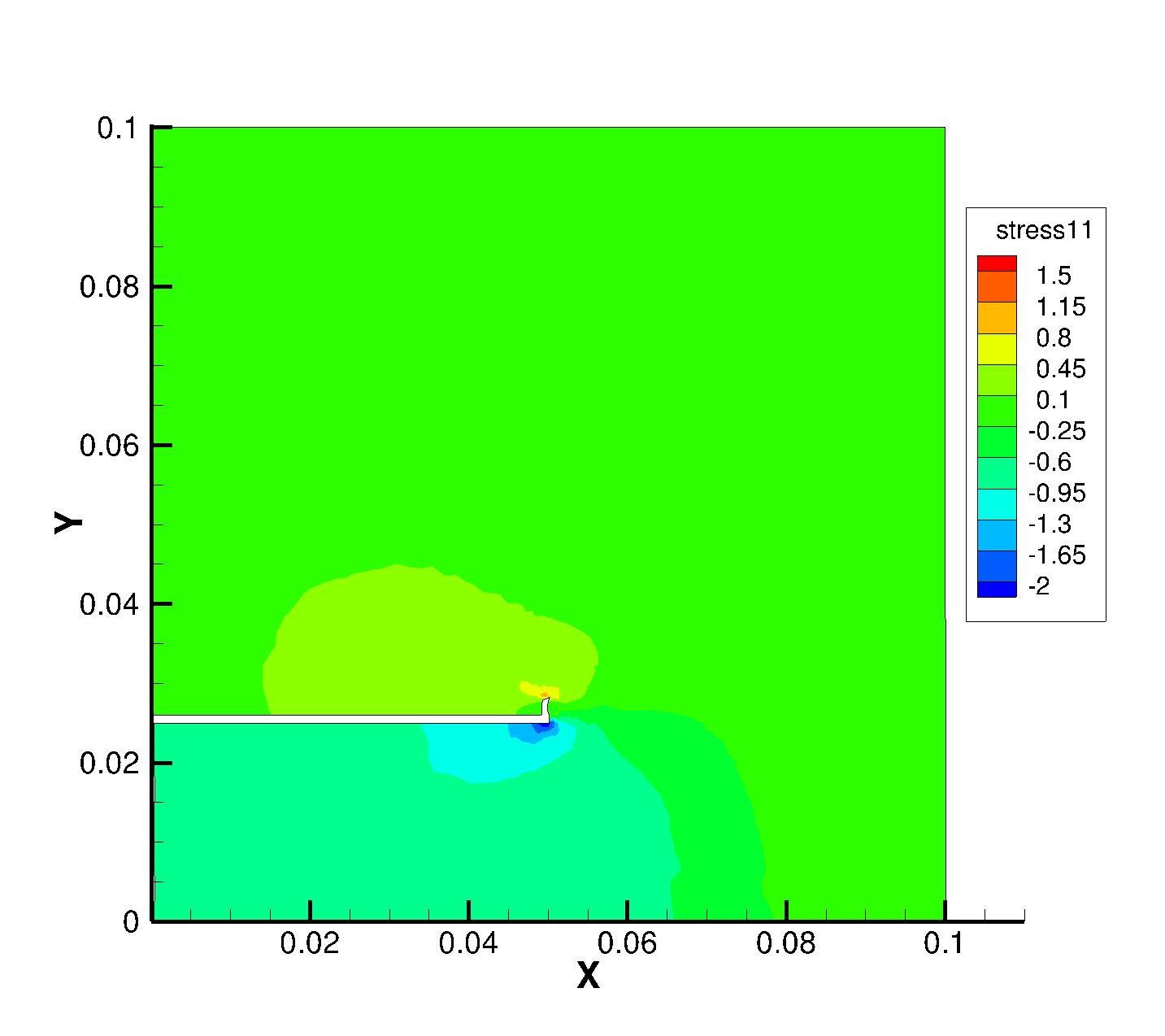}
            (a)
            \end{center}
        \end{minipage}
        \hfill
        \begin{minipage}{0.24\linewidth}
            \begin{center}
            \includegraphics[height=1.5in]{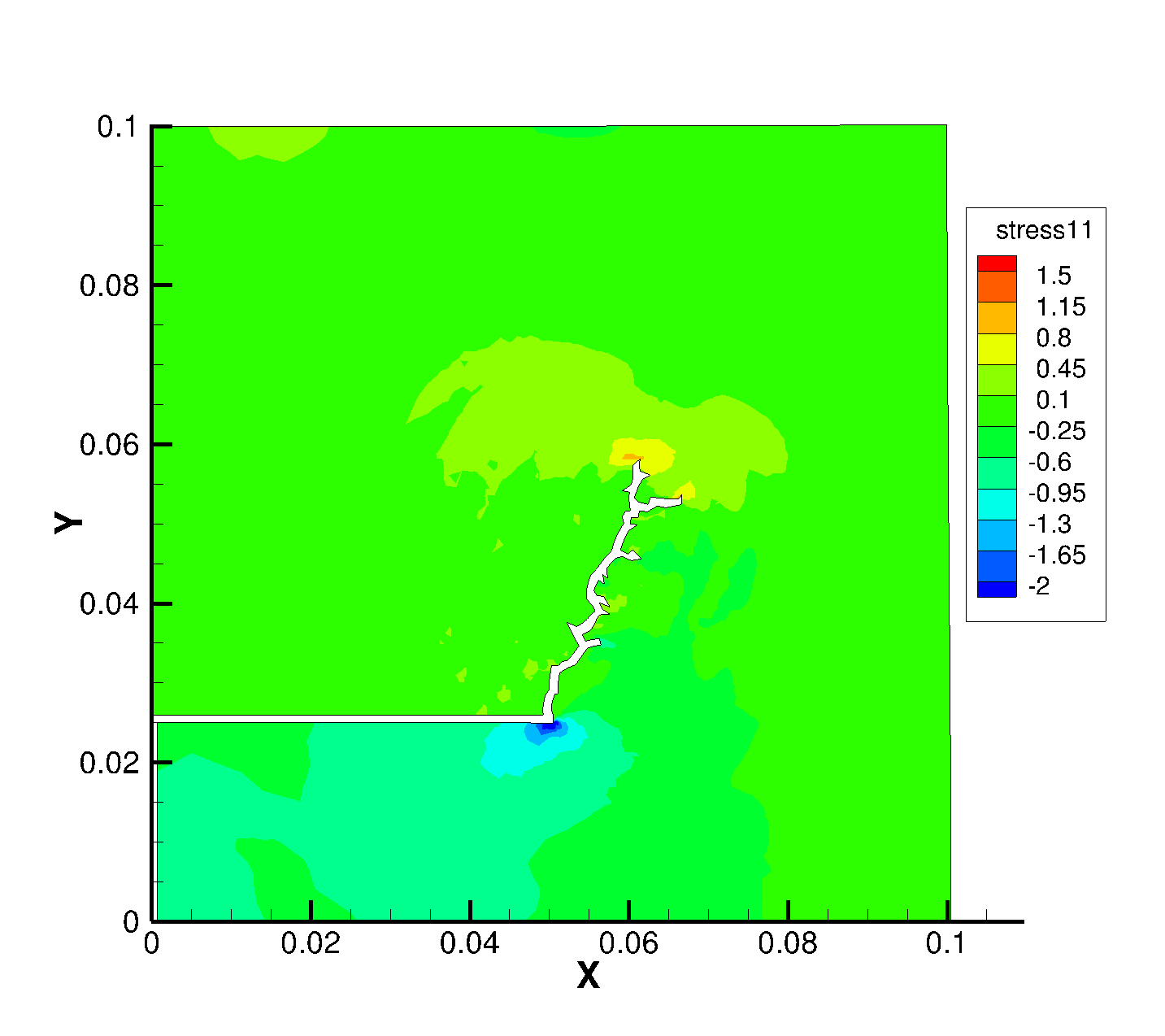}
            (b)
            \end{center}
        \end{minipage}   
        \hfill
        \begin{minipage}{0.24\linewidth}
            \begin{center}
            \includegraphics[height=1.5in]{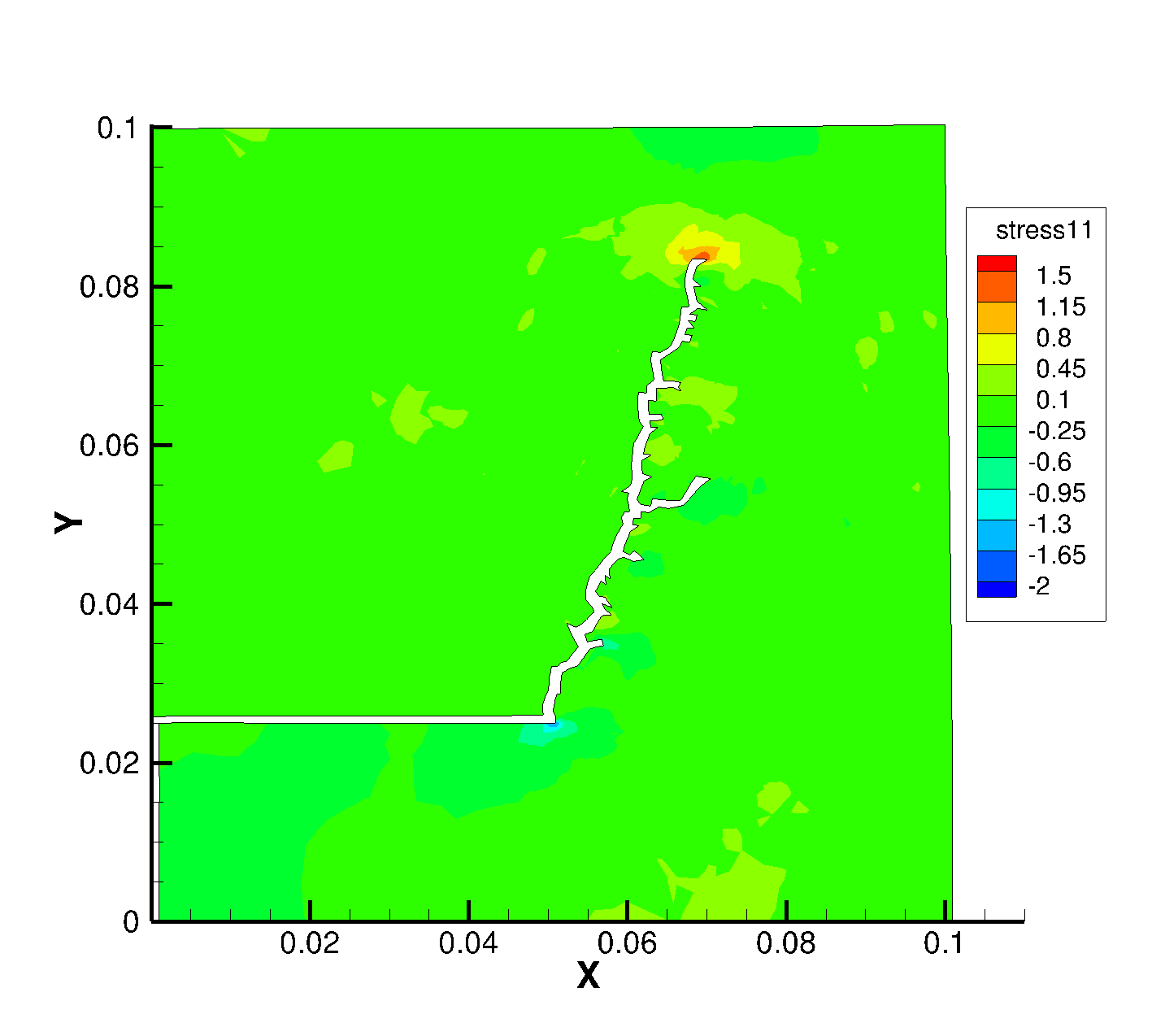}
            (c)
            \end{center}
        \end{minipage}
        \hfill
        \begin{minipage}{0.24\linewidth}
            \begin{center}
            \includegraphics[height=1.5in]{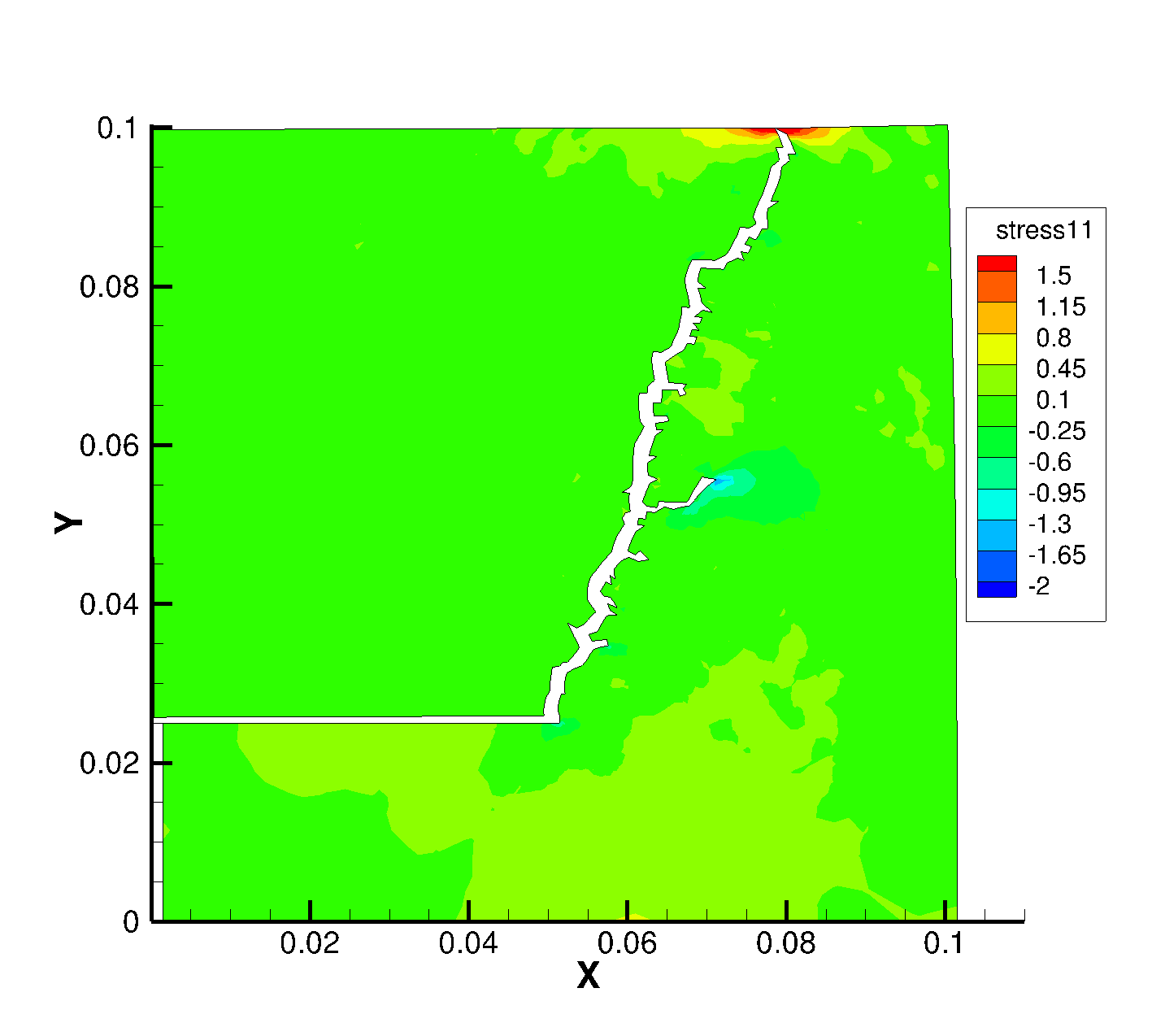}
            (d)
            \end{center}
        \end{minipage}
        \caption{Stress fields and crack propagation patterns for the grid of $7212$ nodes and $14158$ elements at times at (a). $t=25.5\ \mu s$, (b). $t=45\ \mu s$, (c). $t=64\ \mu s$, (d). $t=90\ \mu s$.}
        \label{fig5-5: Kalthoff-crack-evolution-1}
\end{figure}

\begin{figure}[htp]
	\centering
        \begin{minipage}{0.24\linewidth}
            \begin{center}
            \includegraphics[height=1.5in]{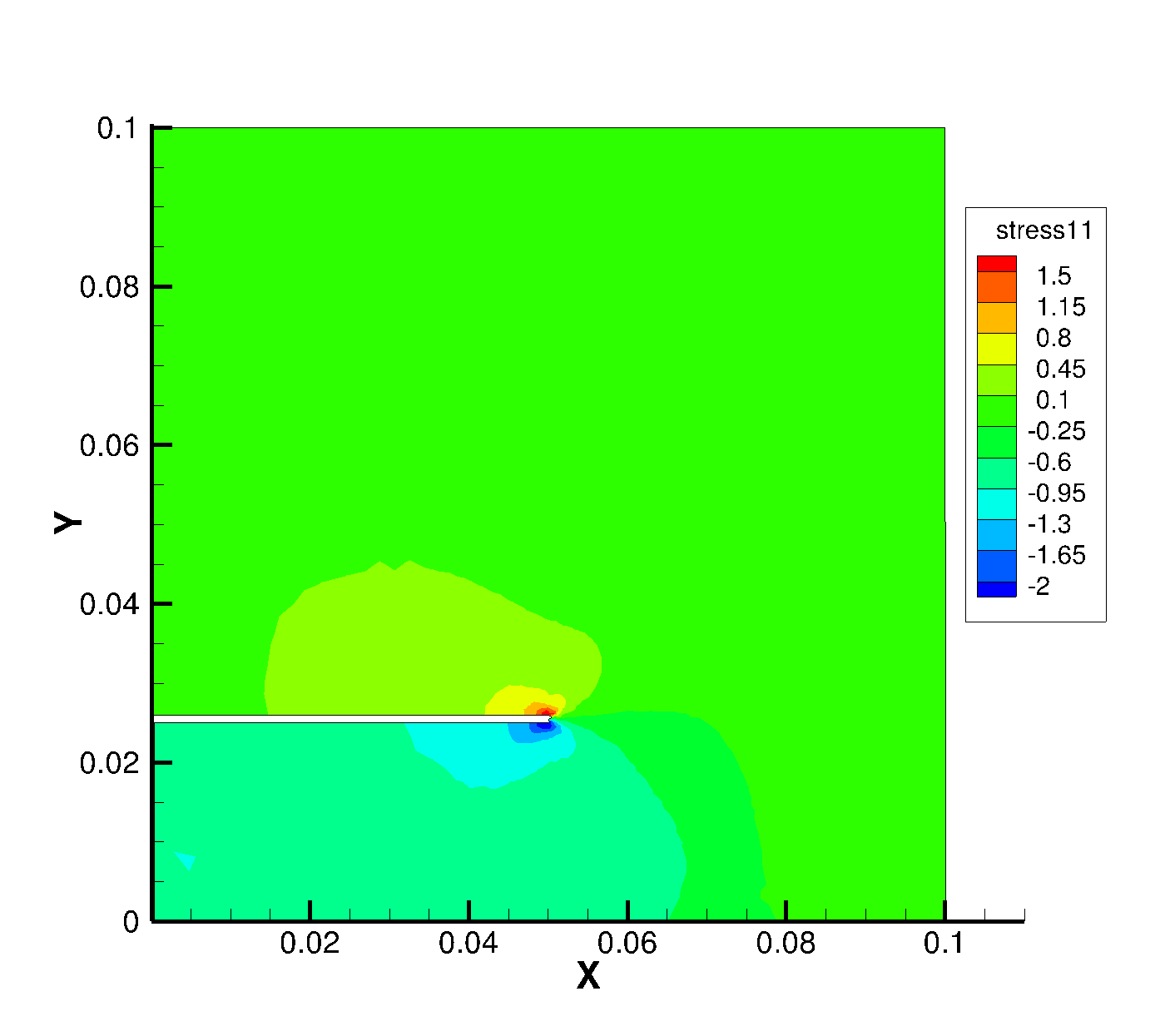}
            (a)
            \end{center}
        \end{minipage}
        \hfill
        \begin{minipage}{0.24\linewidth}
            \begin{center}
            \includegraphics[height=1.5in]{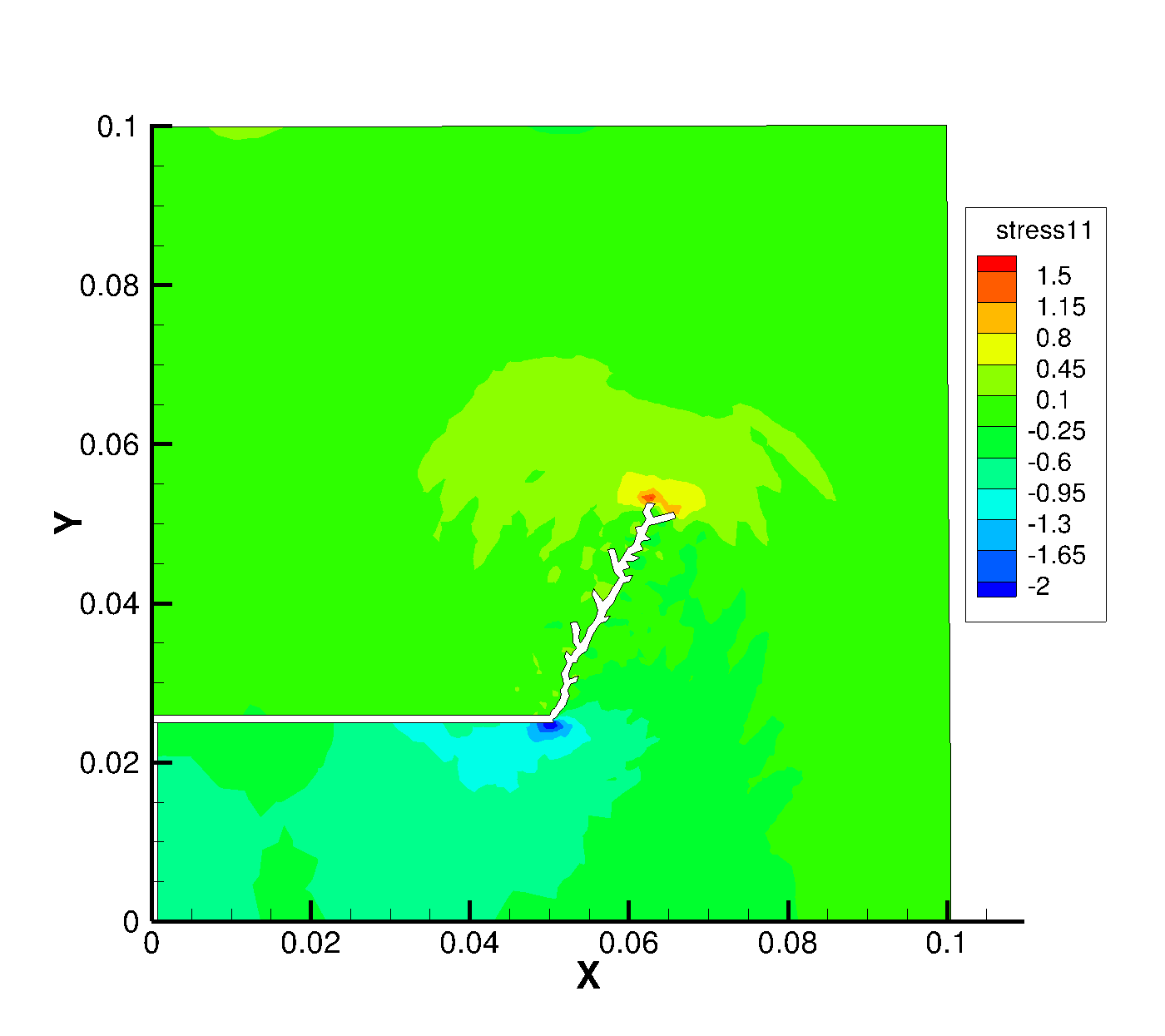}
            (b)
            \end{center}
        \end{minipage}   
        \hfill
        \begin{minipage}{0.24\linewidth}
            \begin{center}
            \includegraphics[height=1.5in]{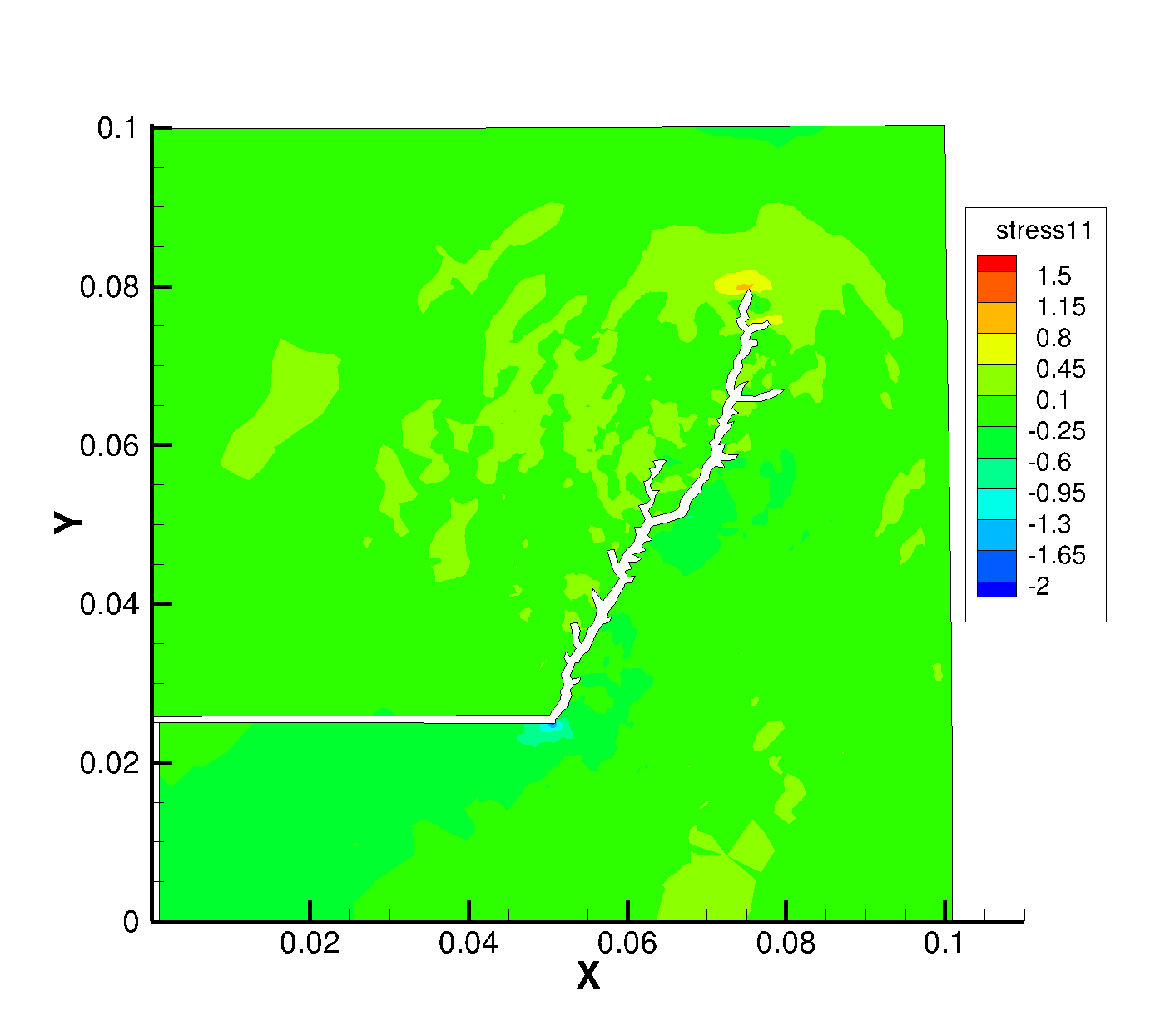}
            (c)
            \end{center}
        \end{minipage}
        \hfill
        \begin{minipage}{0.24\linewidth}
            \begin{center}
            \includegraphics[height=1.5in]{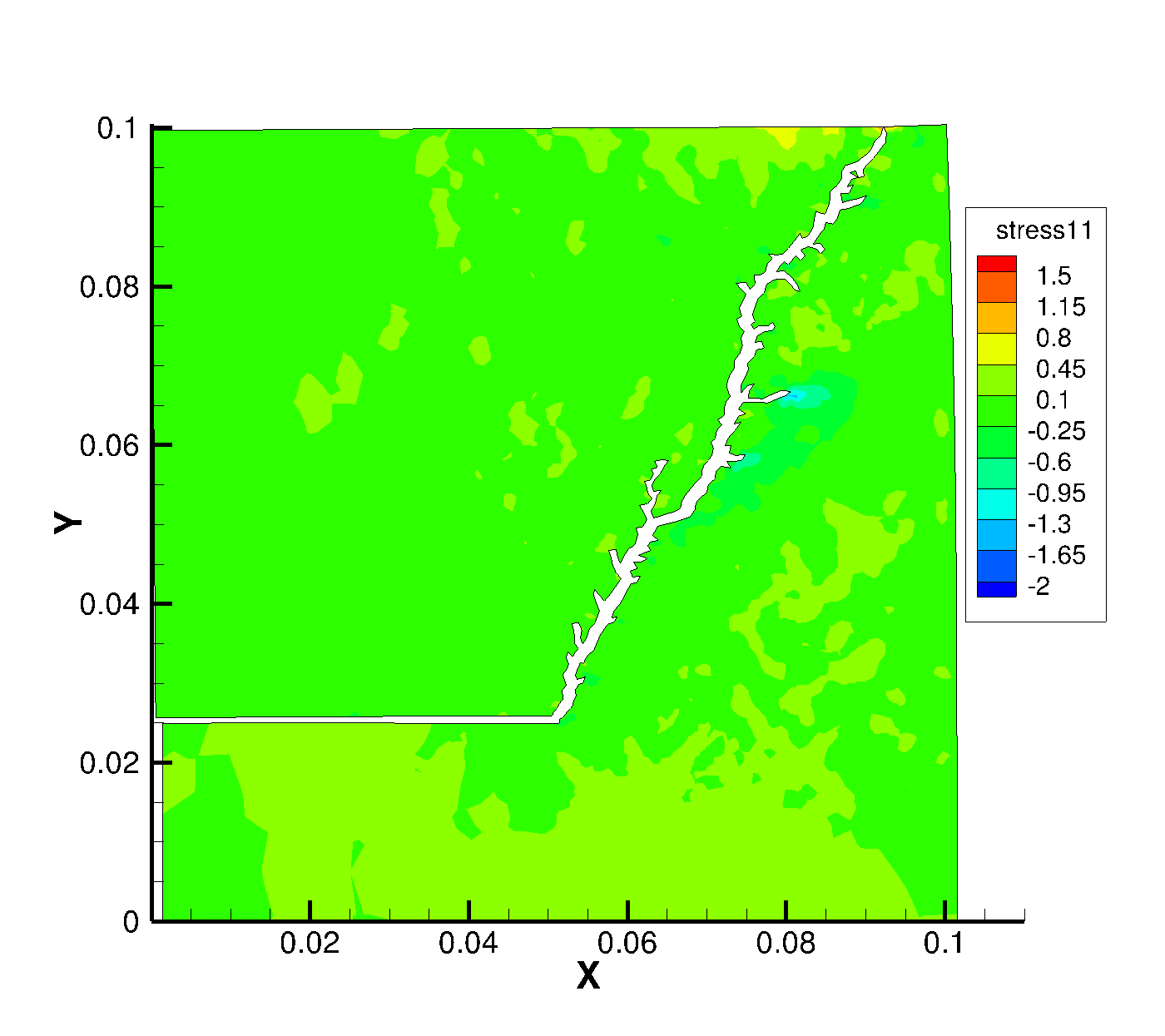}
            (d)
            \end{center}
        \end{minipage}
        \caption{Stress fields and crack propagation patterns for the grid of $7173$ nodes and $14086$ elements at times at (a). $t=25.5\ \mu s$, (b). $t=45\ \mu s$, (c). $t=64\ \mu s$, (d). $t=90\ \mu s$.}
        \label{fig5-6: Kalthoff-crack-evolution-2}
\end{figure}

\begin{figure}[htp]
	\centering
        \begin{minipage}{0.24\linewidth}
            \begin{center}
            \includegraphics[height=1.5in]{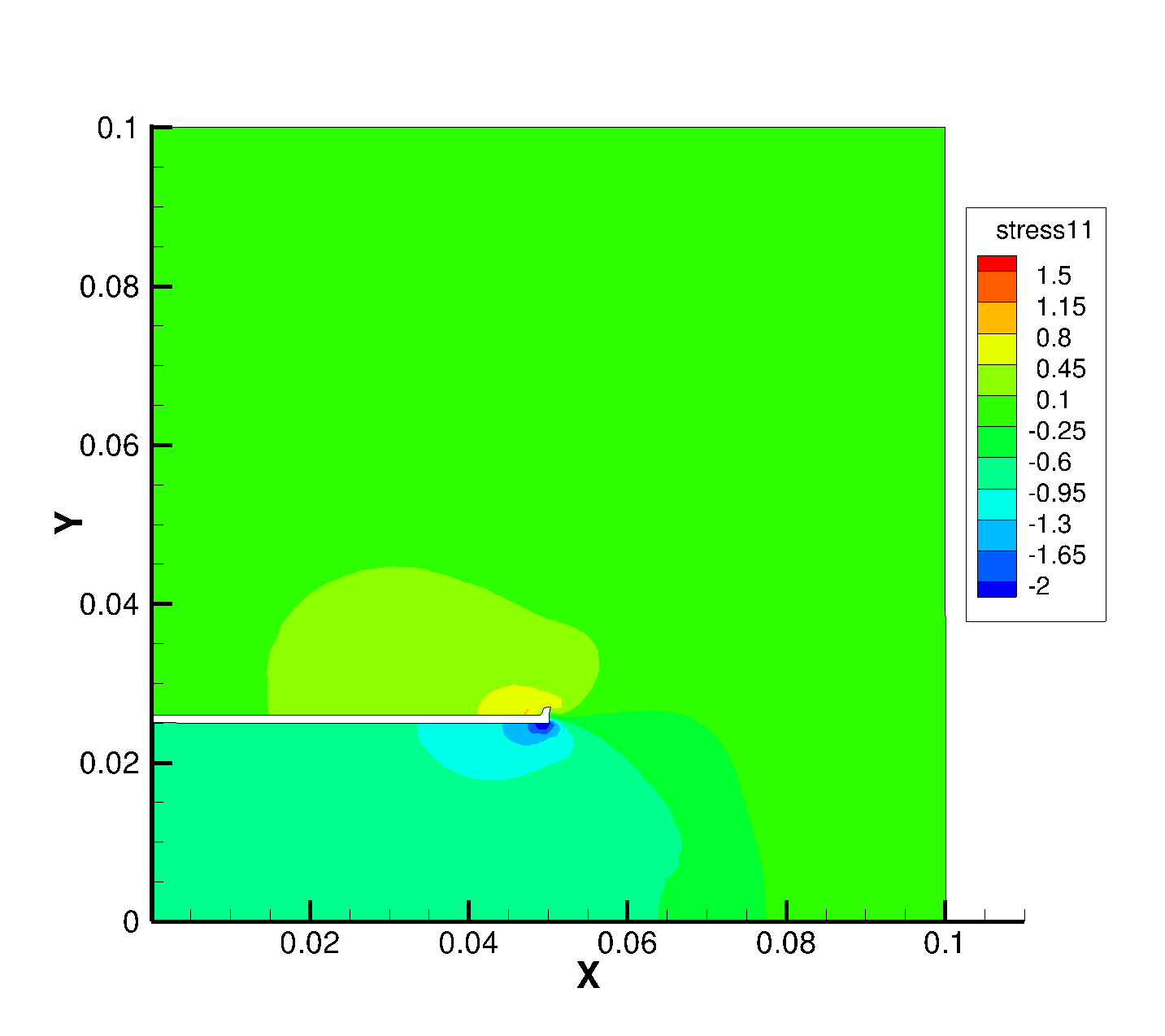}
            (a)
            \end{center}
        \end{minipage}
        \hfill
        \begin{minipage}{0.24\linewidth}
            \begin{center}
            \includegraphics[height=1.5in]{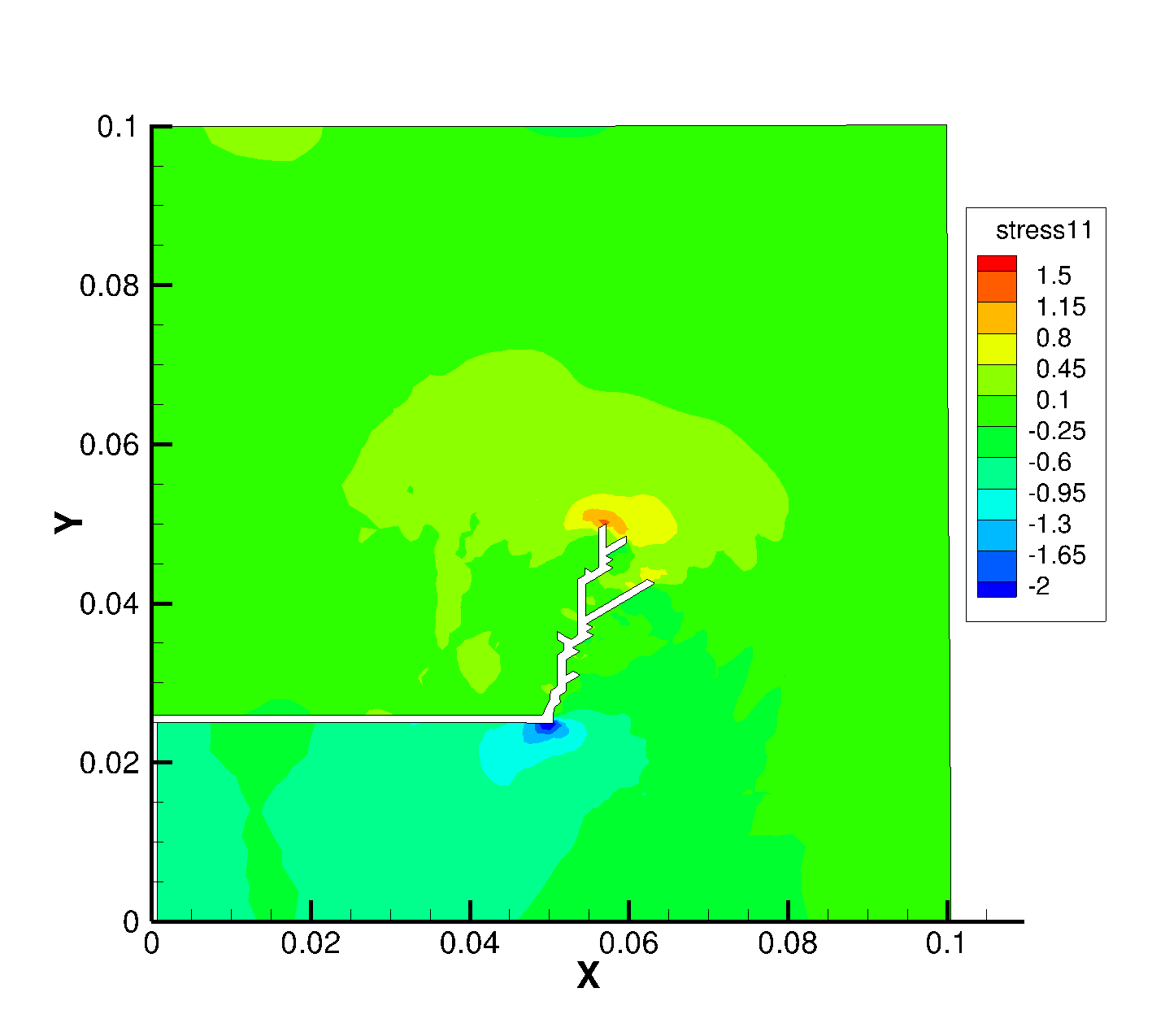}
            (b)
            \end{center}
        \end{minipage}   
        \hfill
        \begin{minipage}{0.24\linewidth}
            \begin{center}
            \includegraphics[height=1.5in]{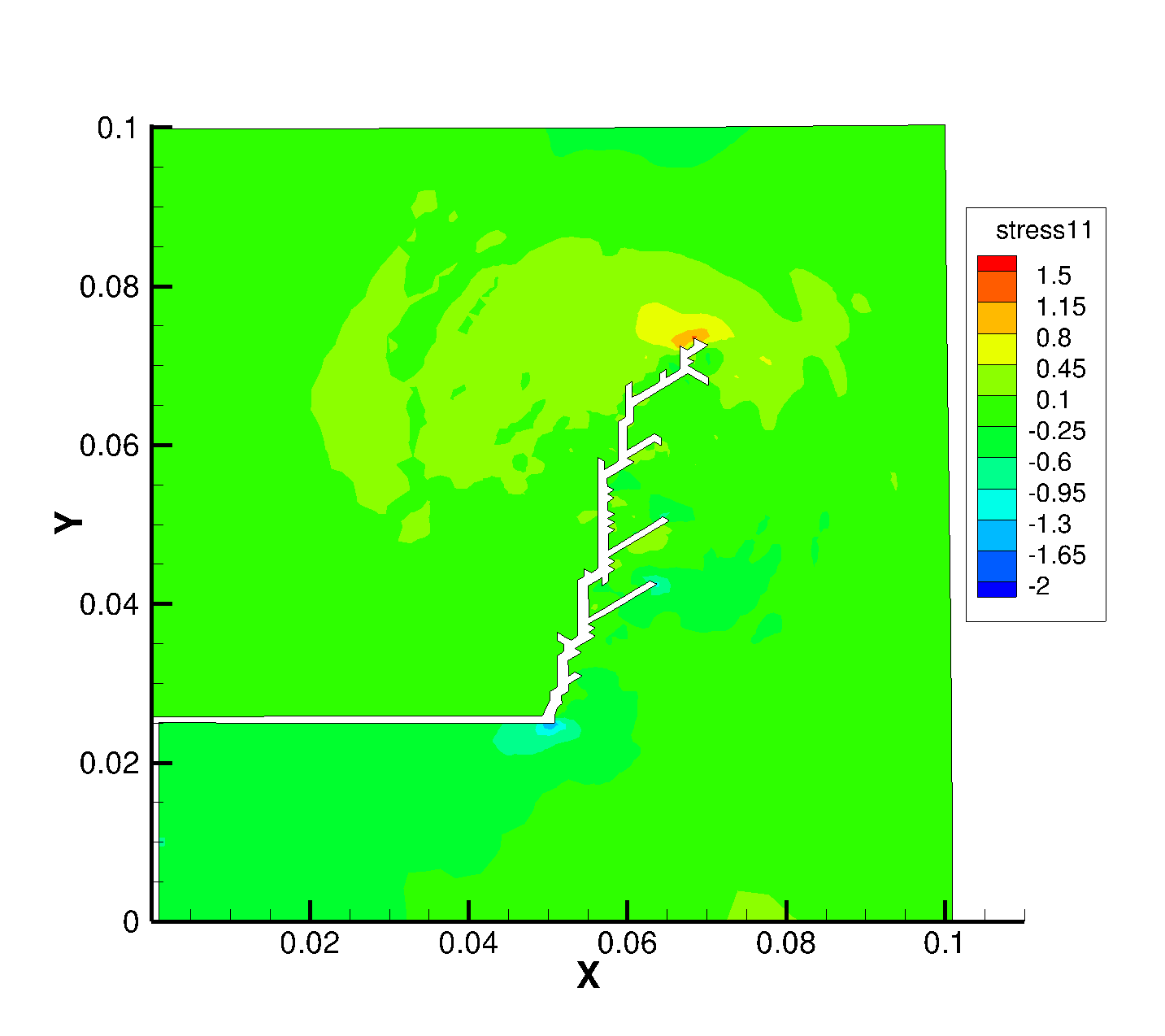}
            (c)
            \end{center}
        \end{minipage}
        \hfill
        \begin{minipage}{0.24\linewidth}
            \begin{center}
            \includegraphics[height=1.5in]{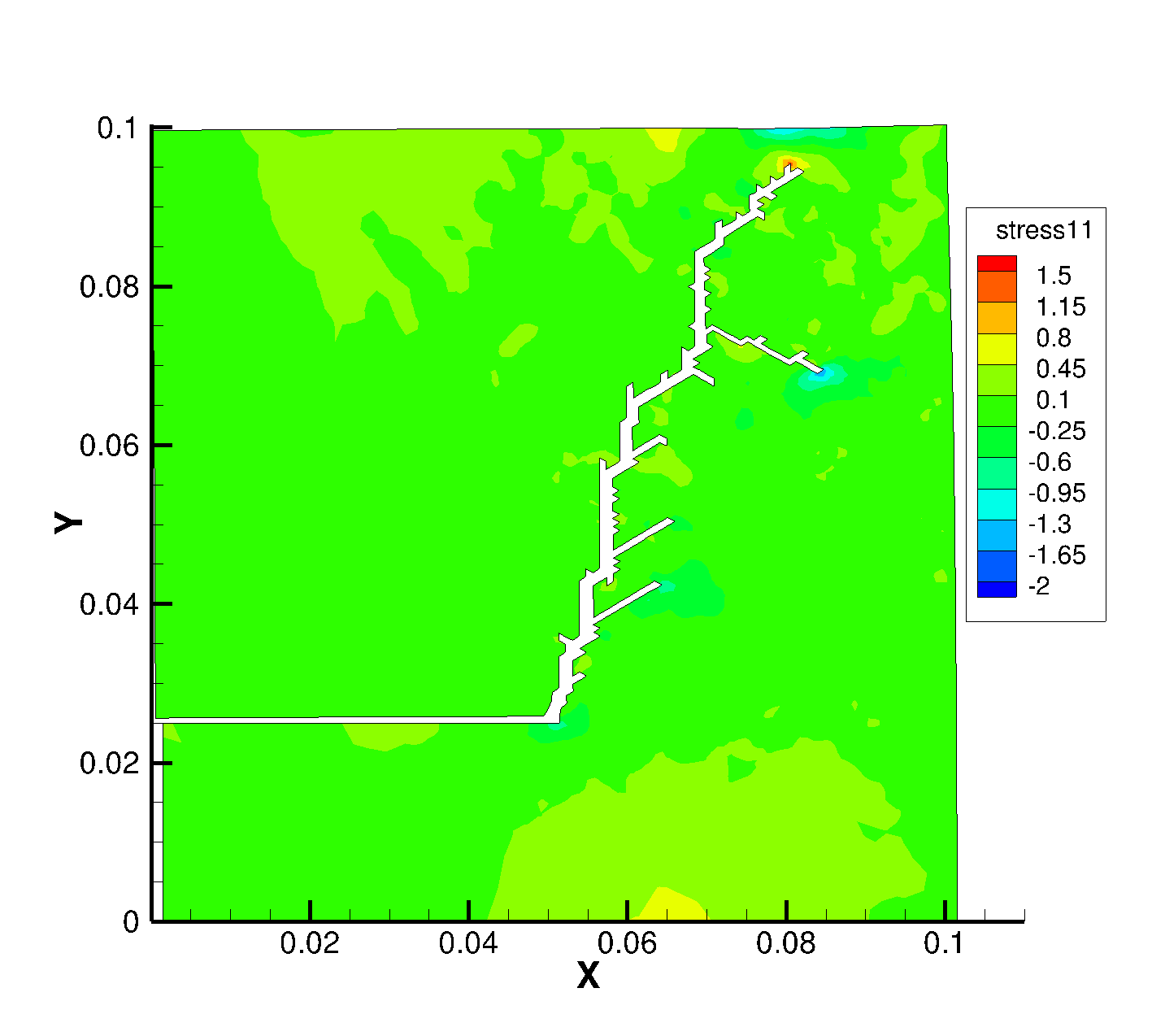}
            (d)
            \end{center}
        \end{minipage}
        \caption{Stress fields and crack propagation patterns for the grid of $6727$ nodes and $13199$ elements at times at (a). $t=25.5\ \mu s$, (b). $t=45\ \mu s$, (c). $t=64\ \mu s$, (d). $t=90\ \mu s$.}
        \label{fig5-7: Kalthoff-crack-evolution-3}
\end{figure}

Furthermore, the development of crack pattern in the grid of $7212$ nodes and $14158$ elements is pictured in Figure.\ref{fig5-5: Kalthoff-crack-evolution-1},
the development of crack pattern in grid of $7173$ nodes and $14086$ elements is pictured in Figure.\ref{fig5-6: Kalthoff-crack-evolution-2},
the development of crack pattern in grid of $6727$ nodes and $13199$ elements is pictured in Figure.\ref{fig5-7: Kalthoff-crack-evolution-3}.
%


\subsection{Crack branching with Neumann boundary condition}
\begin{figure}[htp]
	\centering
            \begin{center}
            \includegraphics[height=2.4in]{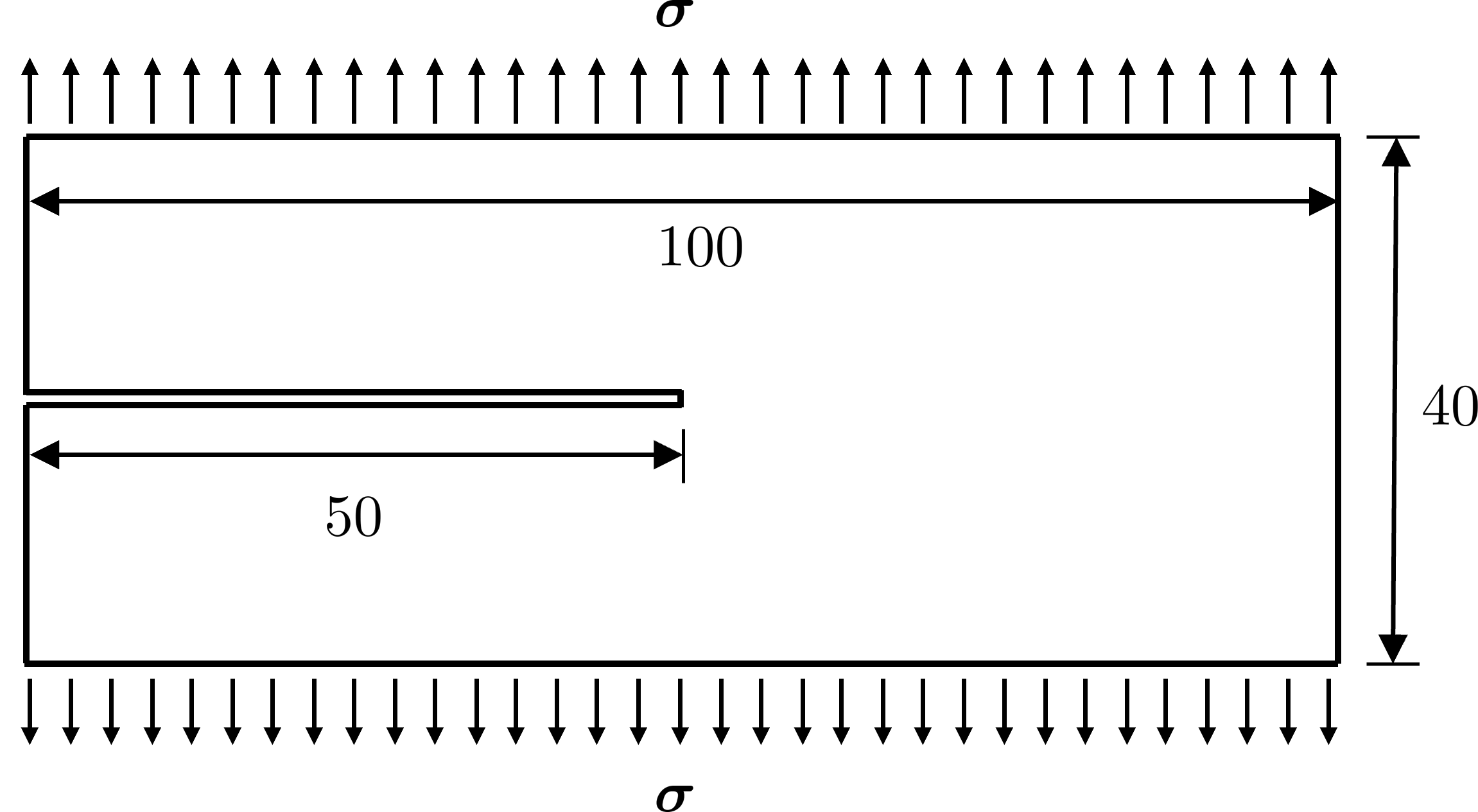}
            \end{center}
        \caption{Geometric dimensions of the two-dimensional pre-notched plate under traction-applied boundary condition. (unit: mm)}
        \label{fig5-8: cbranching-1-geometry}
\end{figure}

Starting with this example, we dive into investigation and discussion of two-dimensional crack branching. The most classic and famous crack branching benchmark example is a long plate under opposite traction on upper and bottom boundaries, and the geometric sizes are shown in Figure.\ref{fig5-8: cbranching-1-geometry}. The two-dimensional plate has dimensions of $0.1 m \times 0.04 m$ and the Neumann/Essential boundary conditions on upper and bottom boundaries are formed by applied opposite traction $\bm{\sigma} = 1 MPa$. During the whole simulation process, the applied traction $\bm{\sigma}$ keeps constant. 

When a plate is subjected to opposite and equal tractions on its top and bottom surfaces therefore the resulting stress field is indeed symmetric. If a crack initiates at the center of the plate or at a notch, it is initially subjected to a Mode I (pure tensile opening) loading under symmetric conditions. However, as the crack propagates dynamically under high loading rates, it may reach a critical speed at which dynamic instabilities emerge, leading to crack branching. Even though the external loading and stress distribution are symmetric, the crack tip becomes unstable as the energy release rate exceeds a critical threshold. The material, especially if brittle, cannot sustain a single crack path efficiently at such high energy input level, and the excessive energy leads to spontaneous branching. This branching is not caused by asymmetry in the stress field but by intrinsic instabilities in the fracture process, amplified by small perturbations such as microstructural inhomogeneities or minor imperfections that break symmetry at the microscale. Once the crack reaches a high enough velocity, typically a significant fraction of the Rayleigh wave speed, micro-branching and then macro-branching occurs as natural mechanisms for dissipating energy. This branching breaks the symmetry of the solution, even though the loading and geometry are symmetric.

\begin{figure}[htp]
	\centering
        \begin{minipage}{0.9\linewidth}
            \begin{center}
            \includegraphics[height=2.0in]{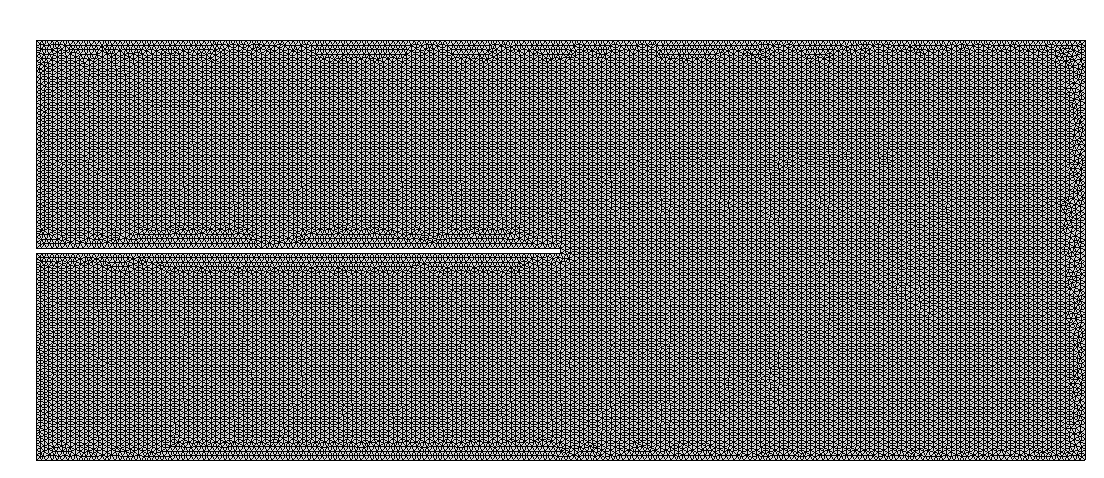}
            \end{center}
            \begin{center}
            (a)
            \end{center}
        \end{minipage}
        \hfill
        \begin{minipage}{0.9\linewidth}
            \begin{center}
            \includegraphics[height=2.0in]{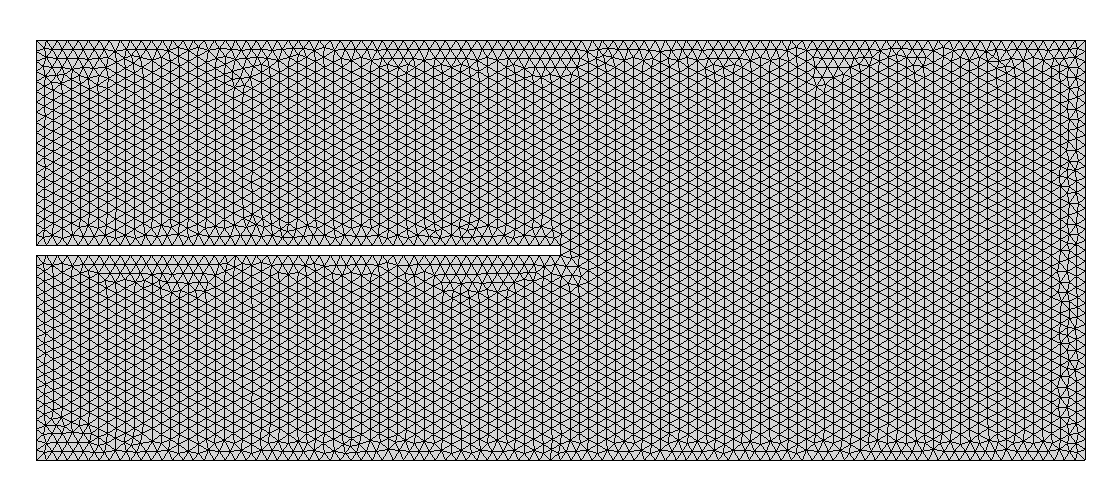}
            \end{center}
            \begin{center}
            (b)
            \end{center}
        \end{minipage}   
        \hfill
        \begin{minipage}{0.9\linewidth}
            \begin{center}
            \includegraphics[height=2.0in]{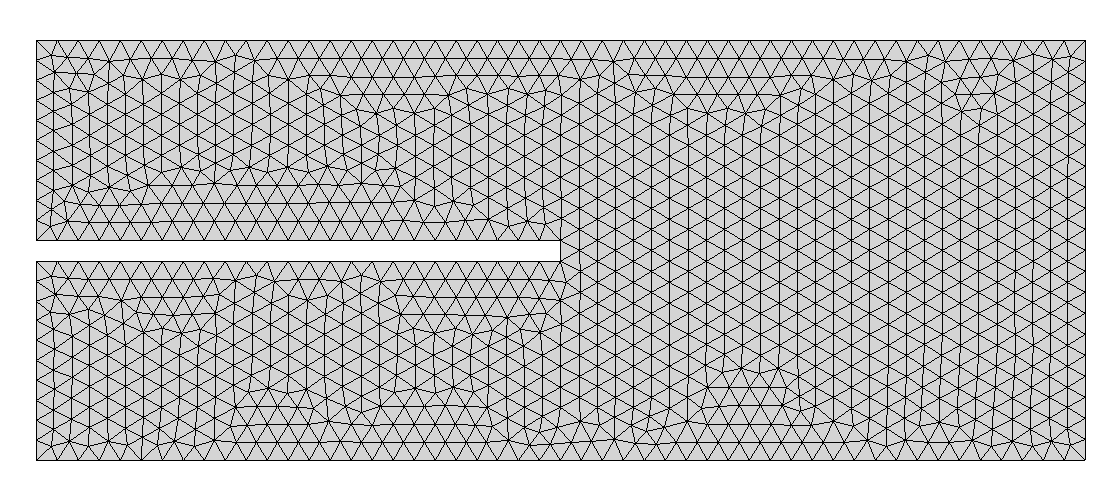}
            \end{center}
            \begin{center}
            (c)
            \end{center}
        \end{minipage}
        \caption{Three different two-dimensional meshes of the plate with Constant Strain Triangle element are illustrated: (a). $37199$ elements and $18981$ nodes; (b). $9287$ elements and $4835$ nodes; (c). $2343$ elements and $1268$ nodes.}
        \label{fig5-9: cb-neumann-meshes}
\end{figure}

Different from three-dimensional benchmark example that very few studies on (\cite{xie2025gpu}, \cite{rabczuk2007three}, \cite{bordas2008three}), there exist a plethora of researches on two-dimensional benchmark example of plate crack branching (\cite{rabczuk2004cracking}, \cite{xu1993void}, \cite{falk2001critical}, \cite{borden2012phase}, \cite{bui2022simulation}). Following these two-dimensional studies, the same material parameters are used in our benchmark example: Young's Modulus $E=32\ GPa$, Poisson ratio $v=0.2$, critical fracture energy release rate $\mathcal{G}_c = 3\ J/m^2$ and density $\rho=2450\ kg/m^3$. Three meshes with from coarse to fine mesh densities are utilized to investigate mesh-dependence of the proposed multiple crack tips tracking algorithm, as shown in Figure.\ref{fig5-9: cb-neumann-meshes}.

\begin{figure}[htp]
	\centering
        \begin{minipage}{0.9\linewidth}
            \begin{center}
            \includegraphics[height=2.0in]{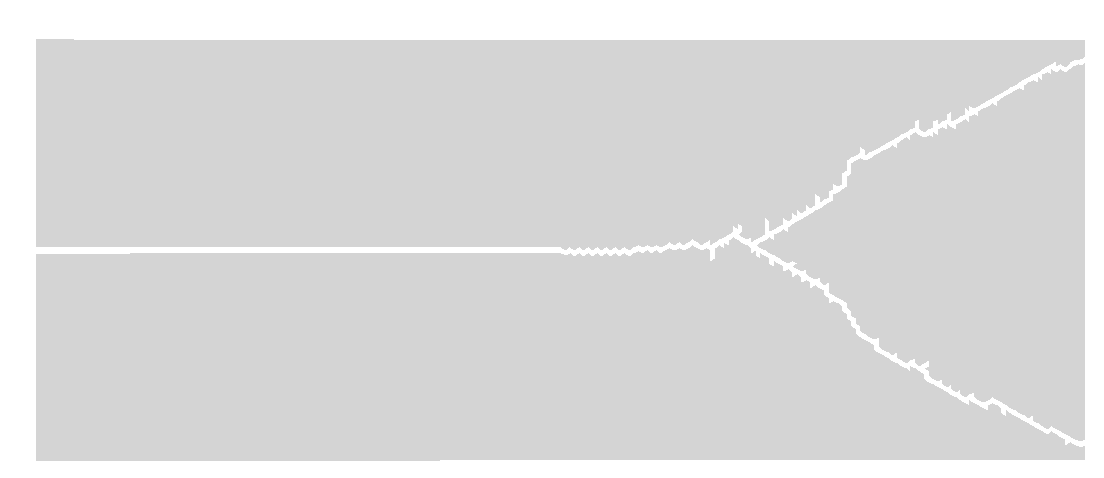}
            \end{center}
            \begin{center}
            (a)
            \end{center}
        \end{minipage}
        \hfill
        \begin{minipage}{0.9\linewidth}
            \begin{center}
            \includegraphics[height=2.0in]{cb-neumann-N4835_E9287-N4835_E9287_crackpattern_crop.png}
            \end{center}
            \begin{center}
            (b)
            \end{center}
        \end{minipage}   
        \hfill
        \begin{minipage}{0.9\linewidth}
            \begin{center}
            \includegraphics[height=2.0in]{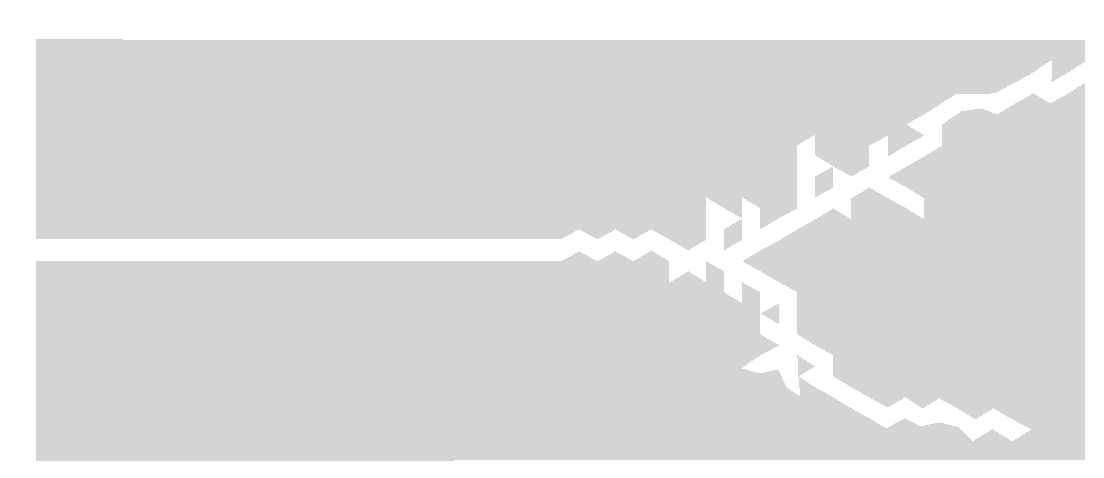}
            \end{center}
            \begin{center}
            (c)
            \end{center}
        \end{minipage}
        \caption{The final crack patterns of three representative Constant Strain Triangle grids are illustrated: (a). $37199$ elements and $18981$ nodes; (b). $9287$ elements and $4835$ nodes; (c). $2343$ elements and $1268$ nodes.}
        \label{fig5-10: cb-neumann-crack-pattern}
\end{figure}

The crack patterns of three meshes at final stage are illustrated in Figure.\ref{fig5-10: cb-neumann-crack-pattern}. It is found that all models with varied mesh densities capture crack branching patterns. Furthermore, in addition to two main branching crack paths, more micro-cracks are captured, which demonstrates the proposed multiple crack tips tracking algorithm is tracking all possible crack tips rather than single one.

\begin{figure}[htp]
	\centering
        \begin{minipage}{0.45\linewidth}
            \begin{center}
            \includegraphics[height=2.8in]{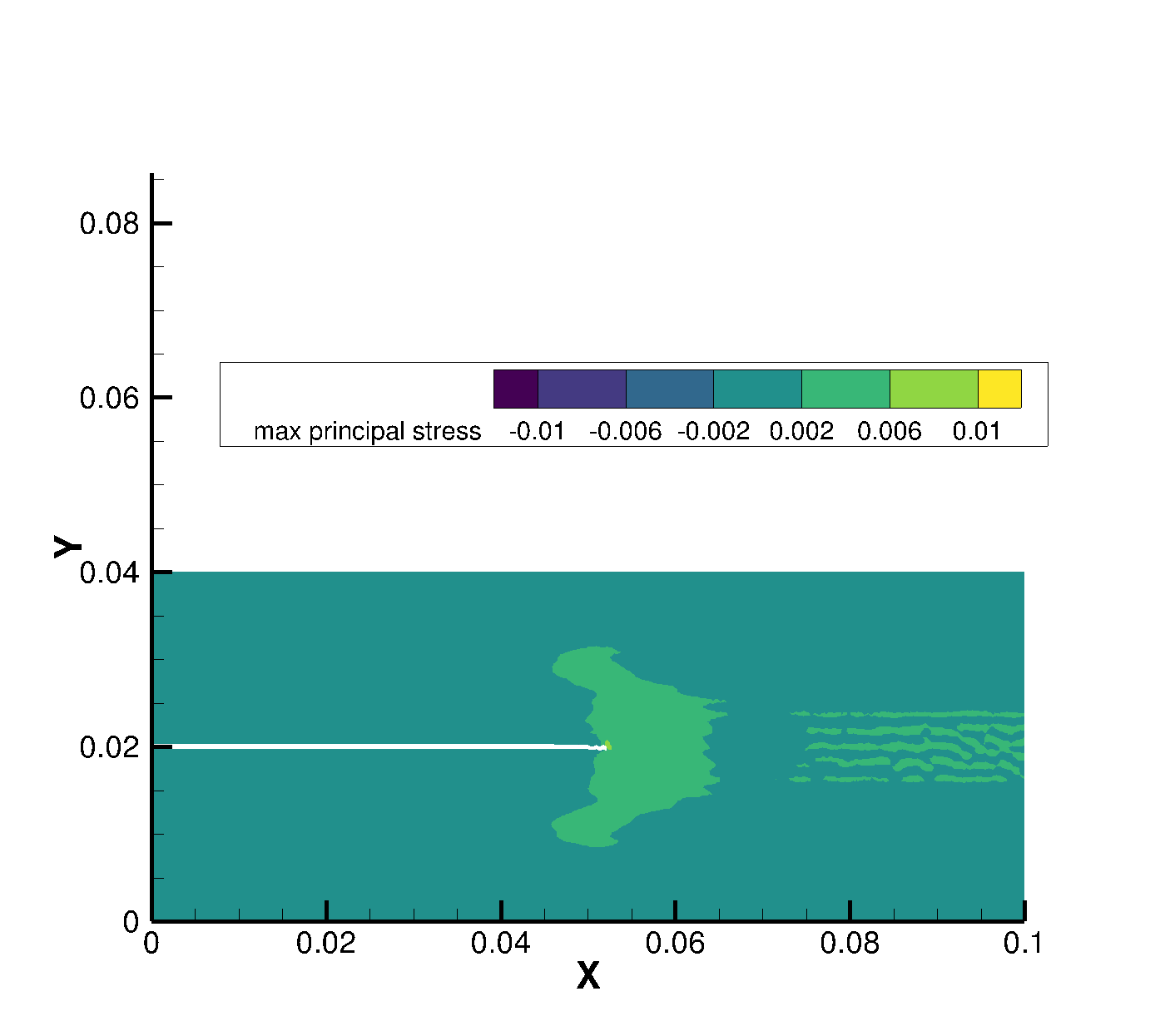}
            \end{center}
            \begin{center}
            (a)
            \end{center}
        \end{minipage}
        \hfill
        \begin{minipage}{0.45\linewidth}
            \begin{center}
            \includegraphics[height=2.8in]{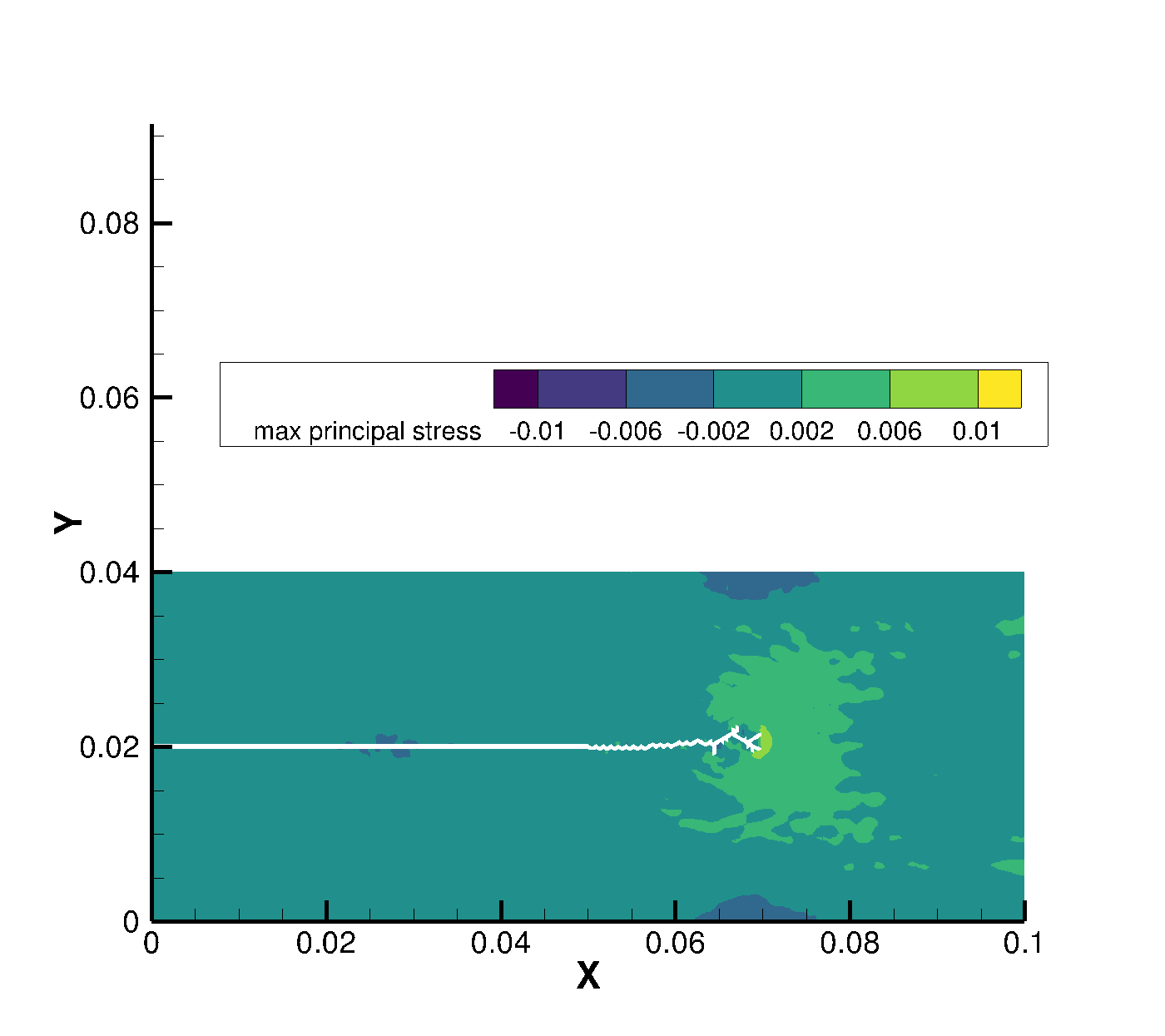}
            \end{center}
            \begin{center}
            (b)
            \end{center}
        \end{minipage}   
        \hfill
        \begin{minipage}{0.45\linewidth}
            \begin{center}
            \includegraphics[height=2.8in]{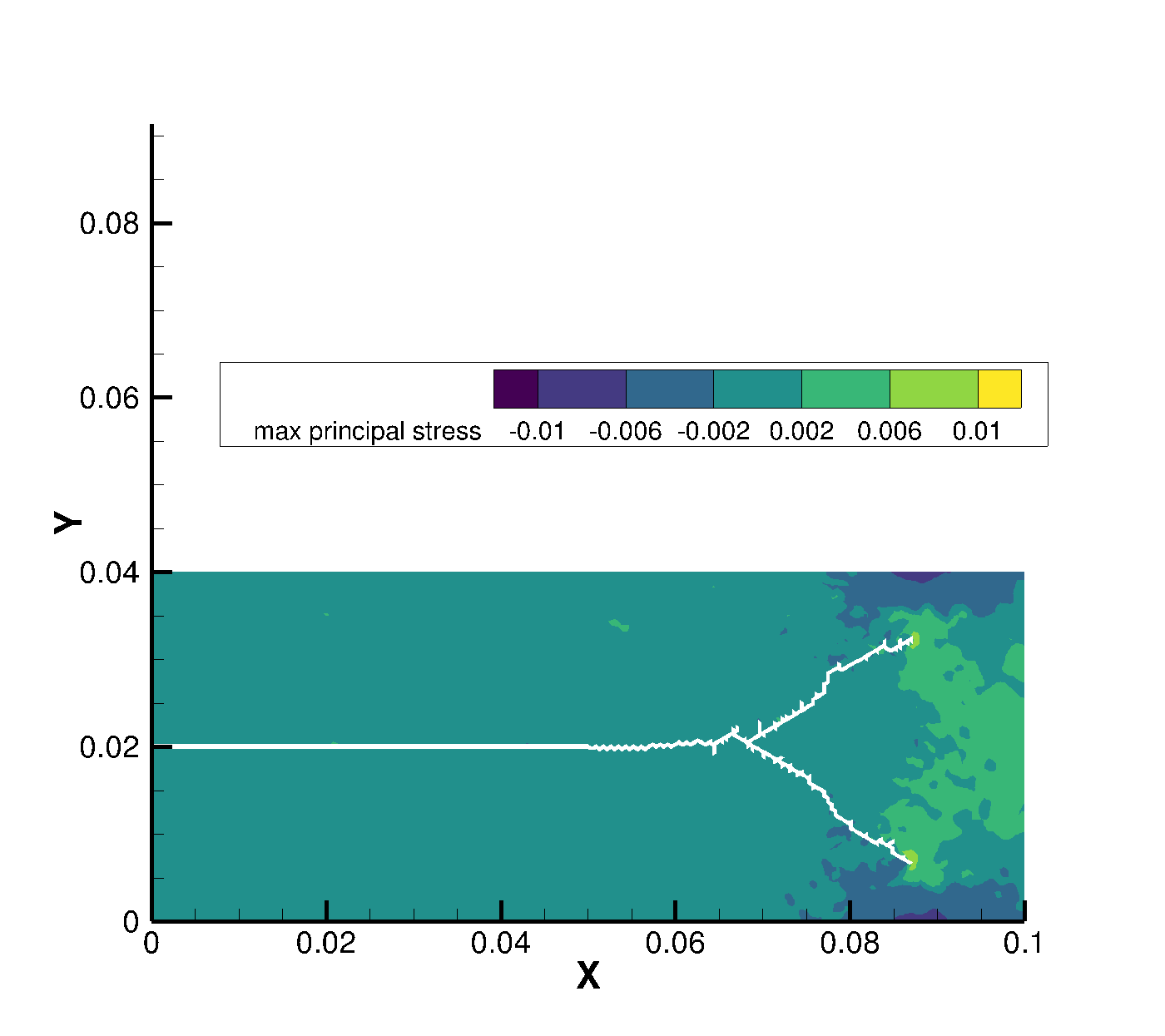}
            \end{center}
            \begin{center}
            (c)
            \end{center}
        \end{minipage}
        \hfill
        \begin{minipage}{0.45\linewidth}
            \begin{center}
            \includegraphics[height=2.8in]{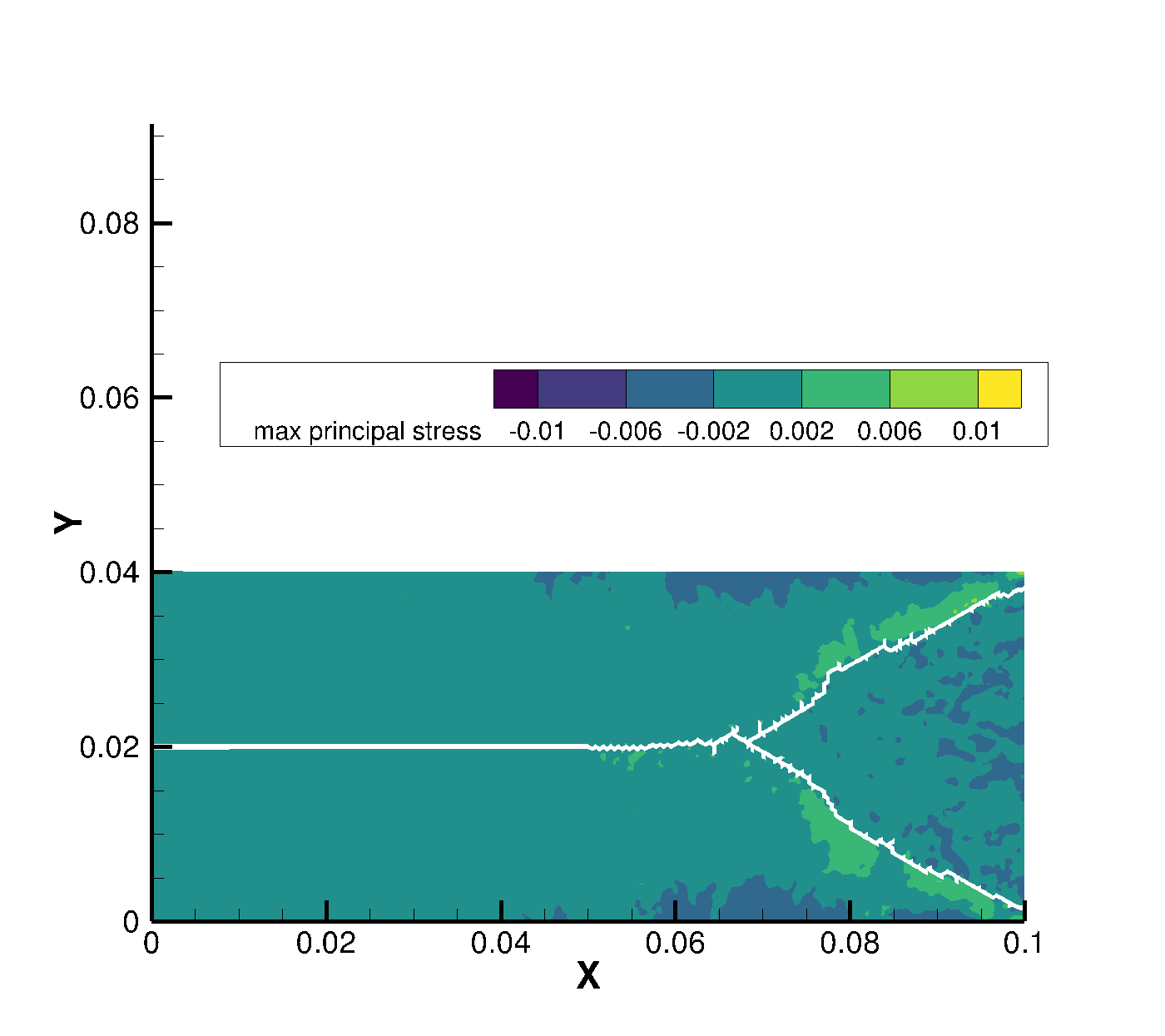}
            \end{center}
            \begin{center}
            (d)
            \end{center}
        \end{minipage}
        \caption{The crack pattern evolution of the mesh with $36922$ elements and $188425$ nodes are provided: (a). time $t=14.4\ \mu s$; (b). time $t=33.6\ \mu s$; (c). time $t=56\ \mu s$; (d). time $t=80\ \mu s$.}
        \label{fig5-11: cb-neumann-mesh1-crack-evolution}
\end{figure}

\begin{figure}[htp]
	\centering
        \begin{minipage}{0.45\linewidth}
            \begin{center}
            \includegraphics[height=2.8in]{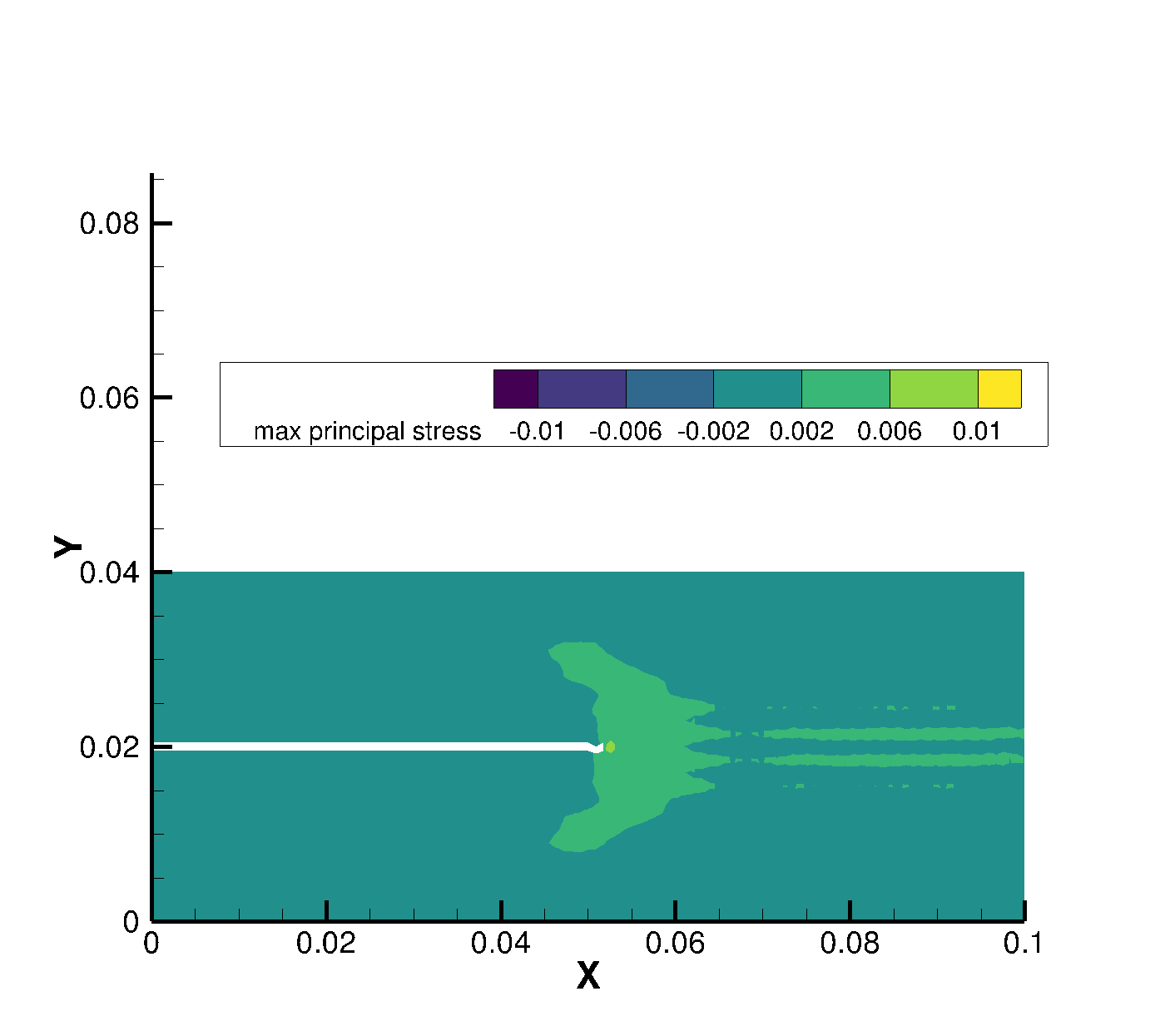}
            \end{center}
            \begin{center}
            (a)
            \end{center}
        \end{minipage}
        \hfill
        \begin{minipage}{0.45\linewidth}
            \begin{center}
            \includegraphics[height=2.8in]{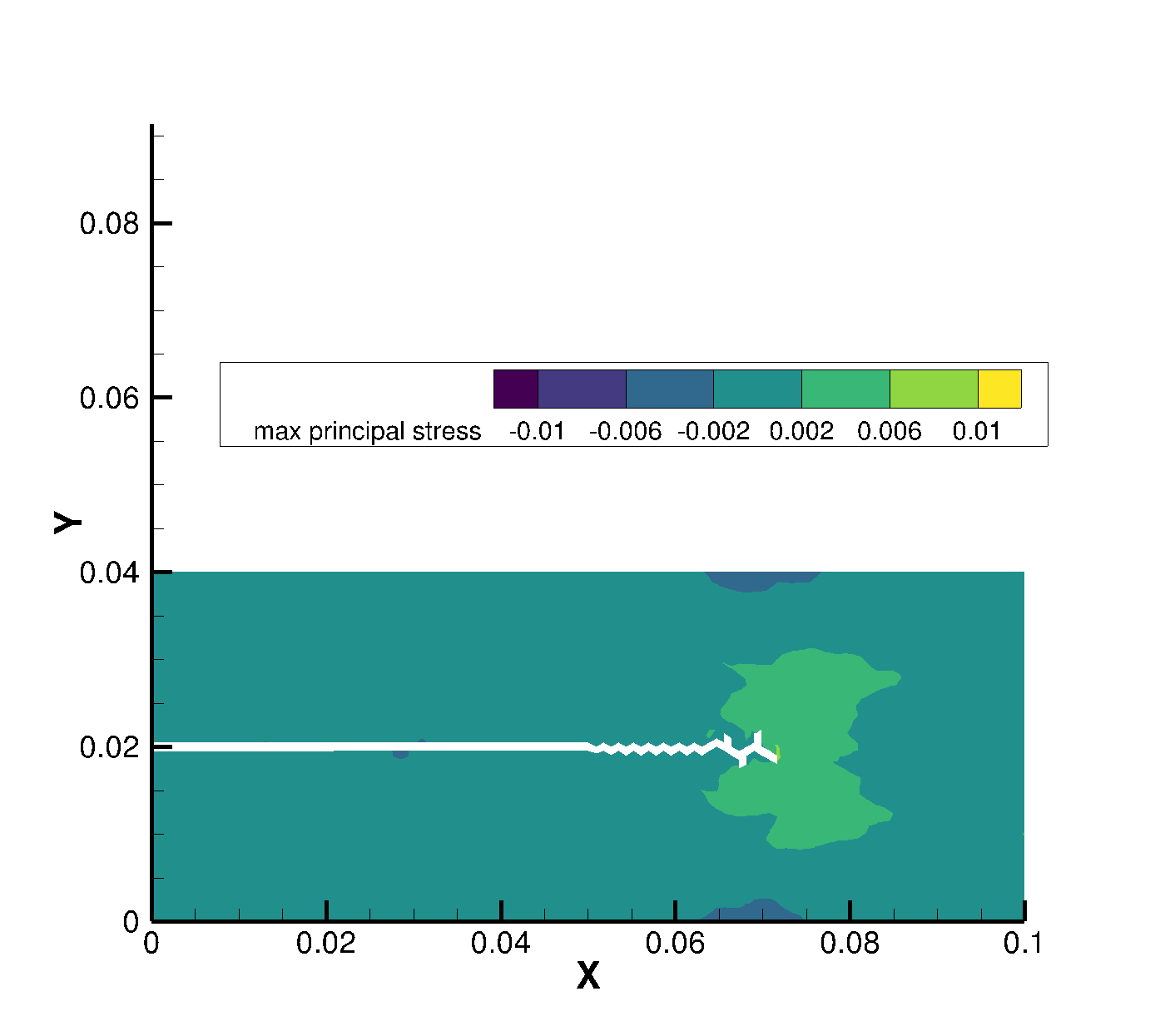}
            \end{center}
            \begin{center}
            (b)
            \end{center}
        \end{minipage}   
        \hfill
        \begin{minipage}{0.45\linewidth}
            \begin{center}
            \includegraphics[height=2.8in]{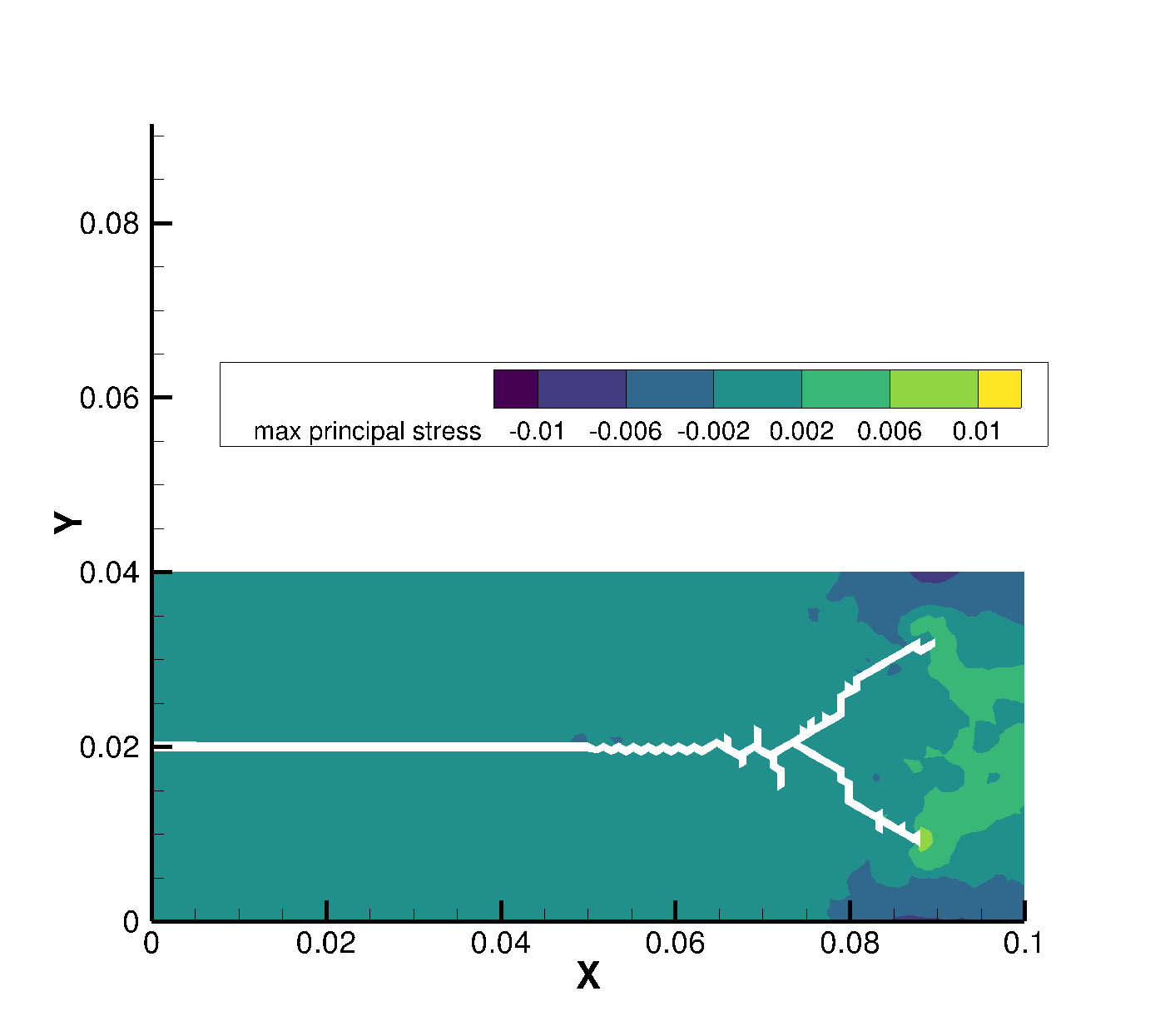}
            \end{center}
            \begin{center}
            (c)
            \end{center}
        \end{minipage}
        \hfill
        \begin{minipage}{0.45\linewidth}
            \begin{center}
            \includegraphics[height=2.8in]{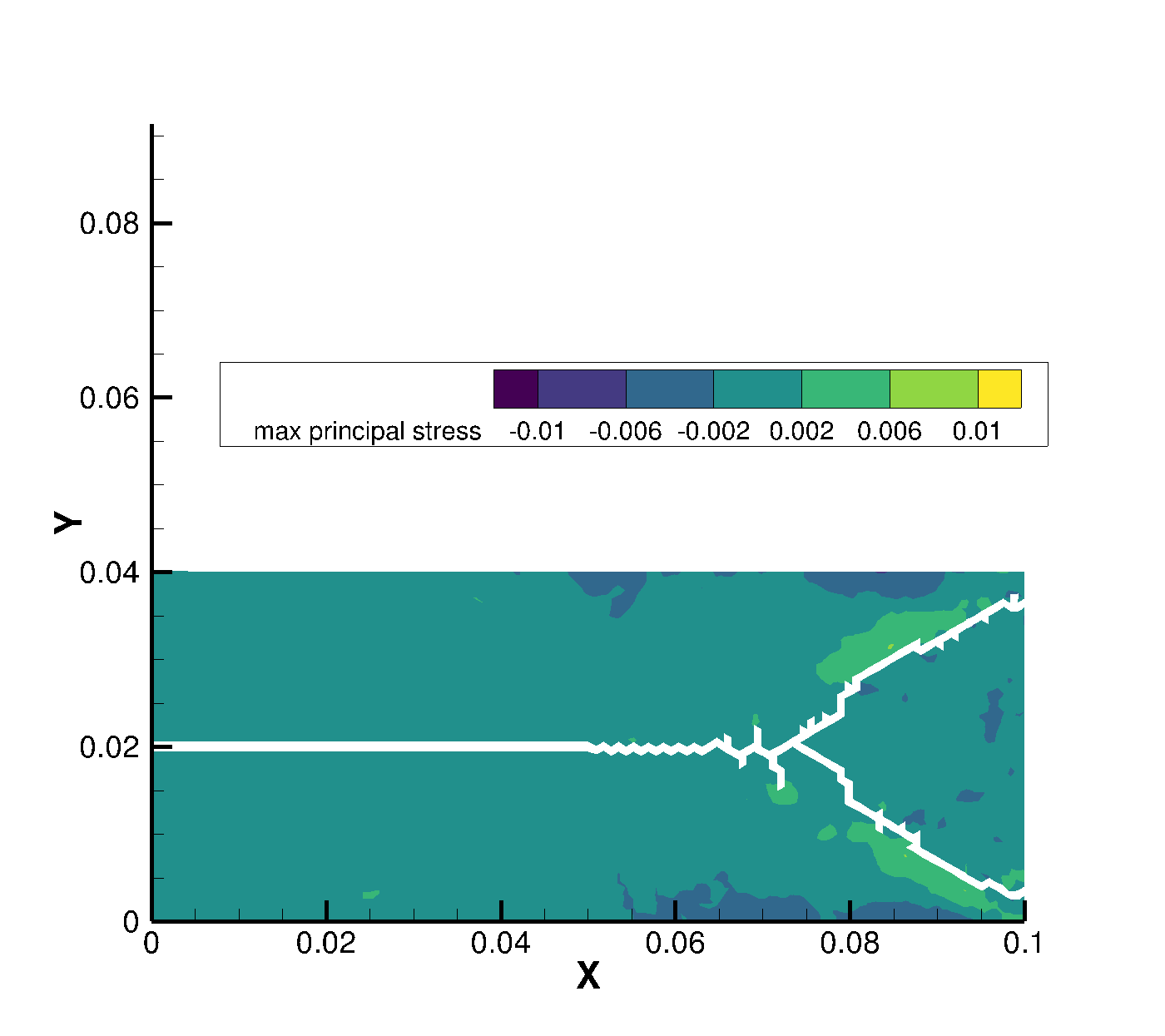}
            \end{center}
            \begin{center}
            (d)
            \end{center}
        \end{minipage}
        \caption{The crack pattern evolution of medium mesh with $9287$ elements and $4835$ nodes are provided: (a). time $t=14.4\ \mu s$; (b). time $t=33.6\ \mu s$; (c). time $t=56\ \mu s$; (d). time $t=80\ \mu s$.}
        \label{fig5-12: cb-neumann-mesh2-crack-evolution}
\end{figure}

\begin{figure}[htp]
	\centering
        \begin{minipage}{0.45\linewidth}
            \begin{center}
            \includegraphics[height=2.8in]{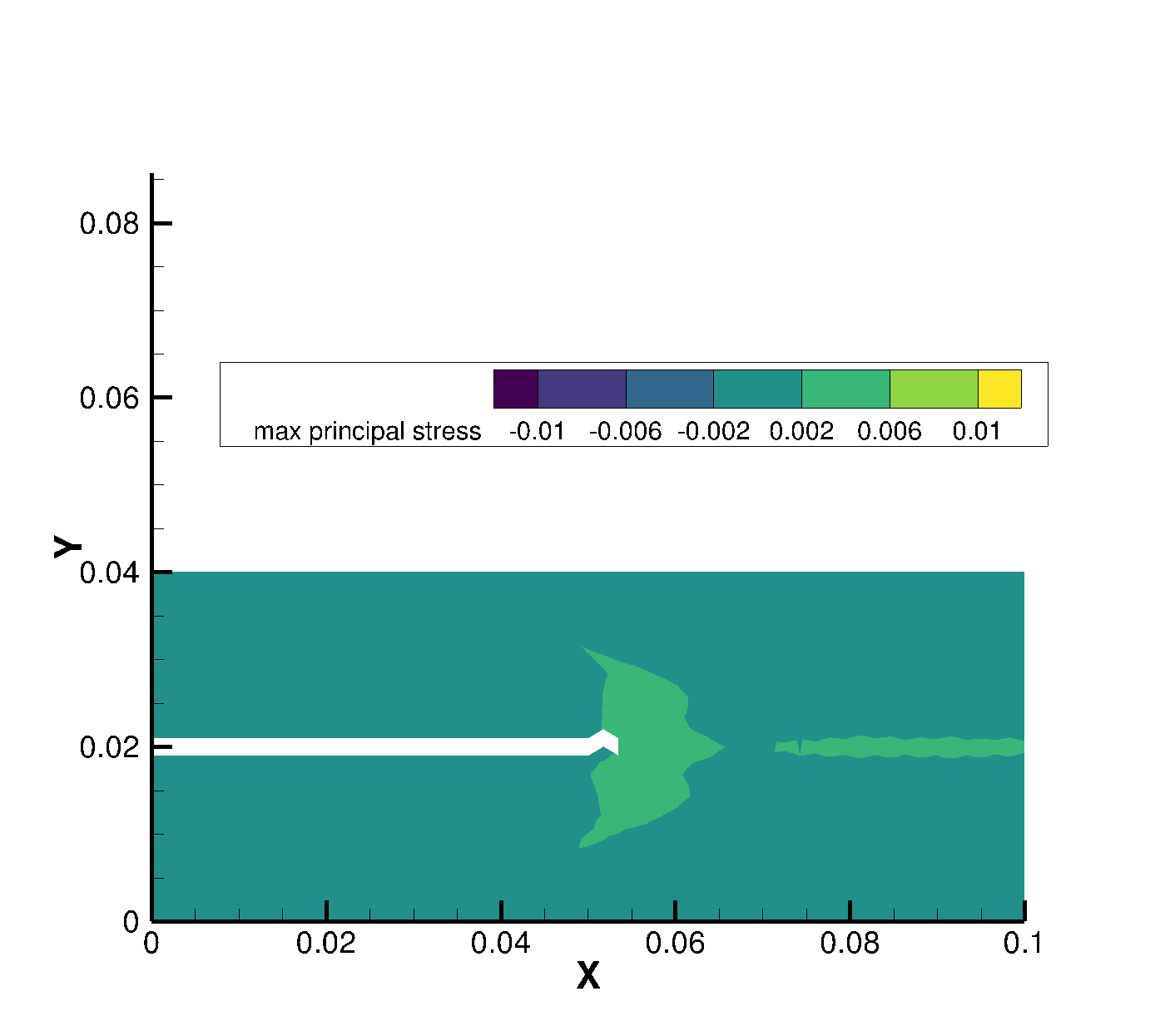}
            \end{center}
            \begin{center}
            (a)
            \end{center}
        \end{minipage}
        \hfill
        \begin{minipage}{0.45\linewidth}
            \begin{center}
            \includegraphics[height=2.8in]{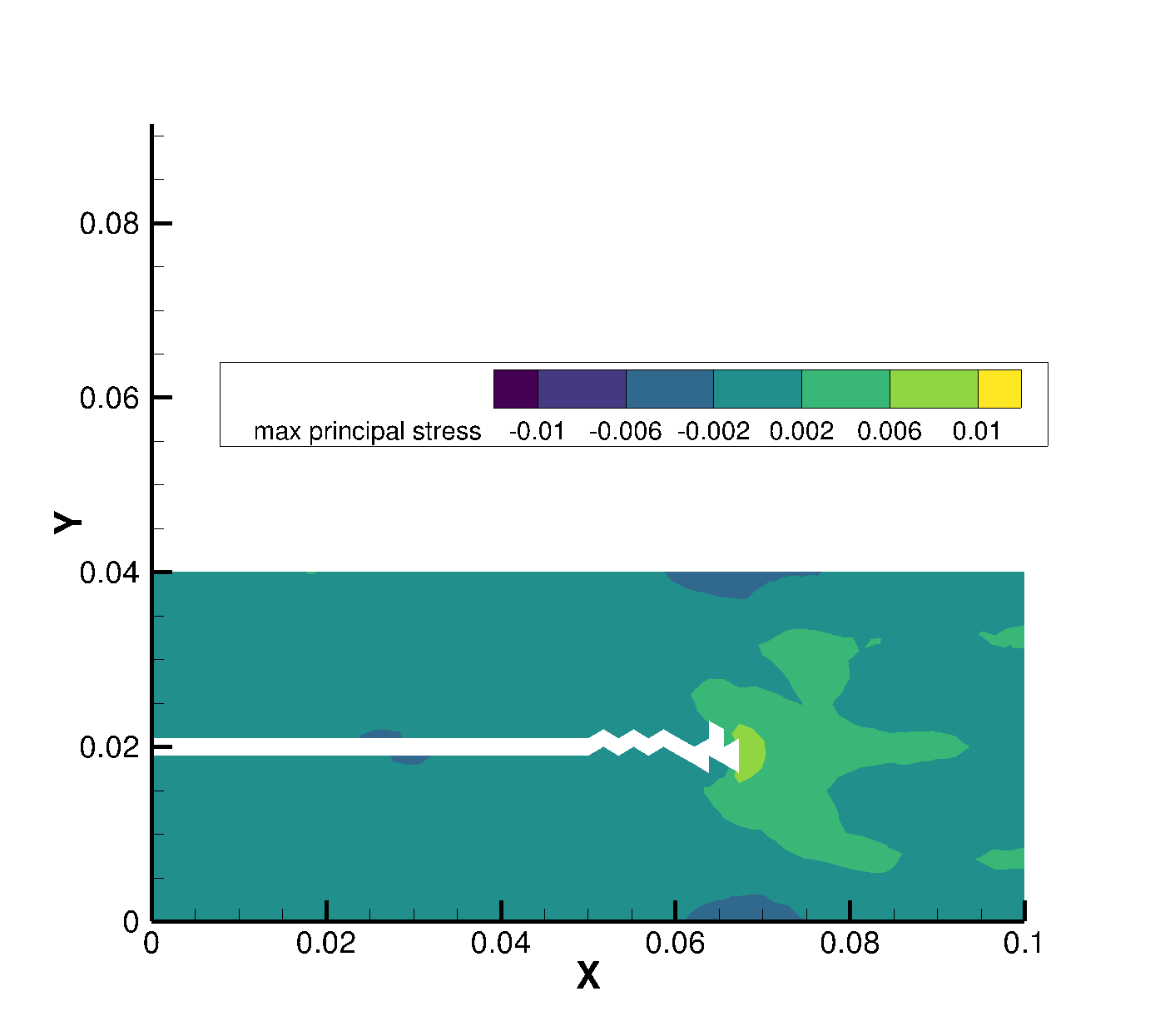}
            \end{center}
            \begin{center}
            (b)
            \end{center}
        \end{minipage}   
        \hfill
        \begin{minipage}{0.45\linewidth}
            \begin{center}
            \includegraphics[height=2.8in]{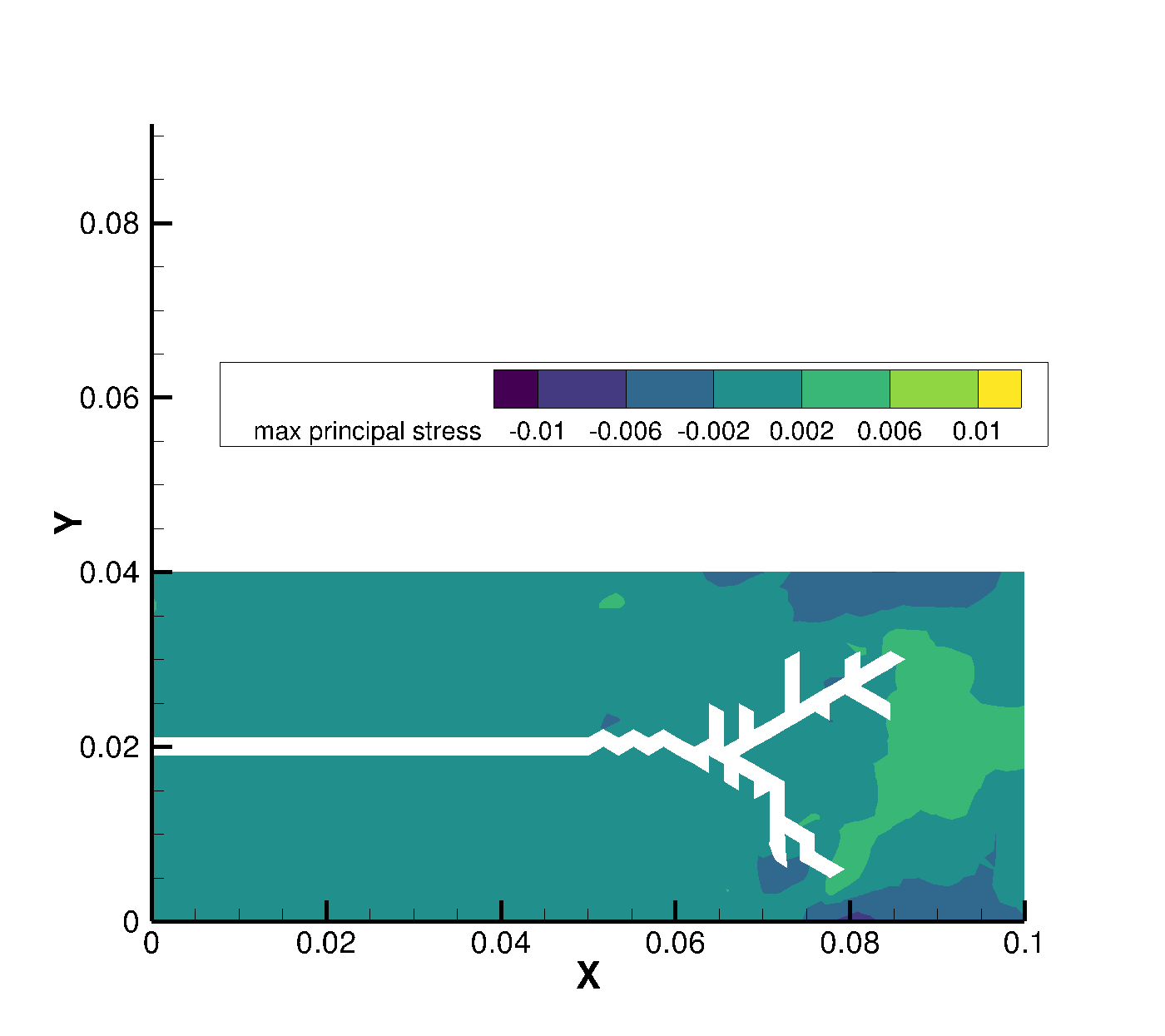}
            \end{center}
            \begin{center}
            (c)
            \end{center}
        \end{minipage}
        \hfill
        \begin{minipage}{0.45\linewidth}
            \begin{center}
            \includegraphics[height=2.8in]{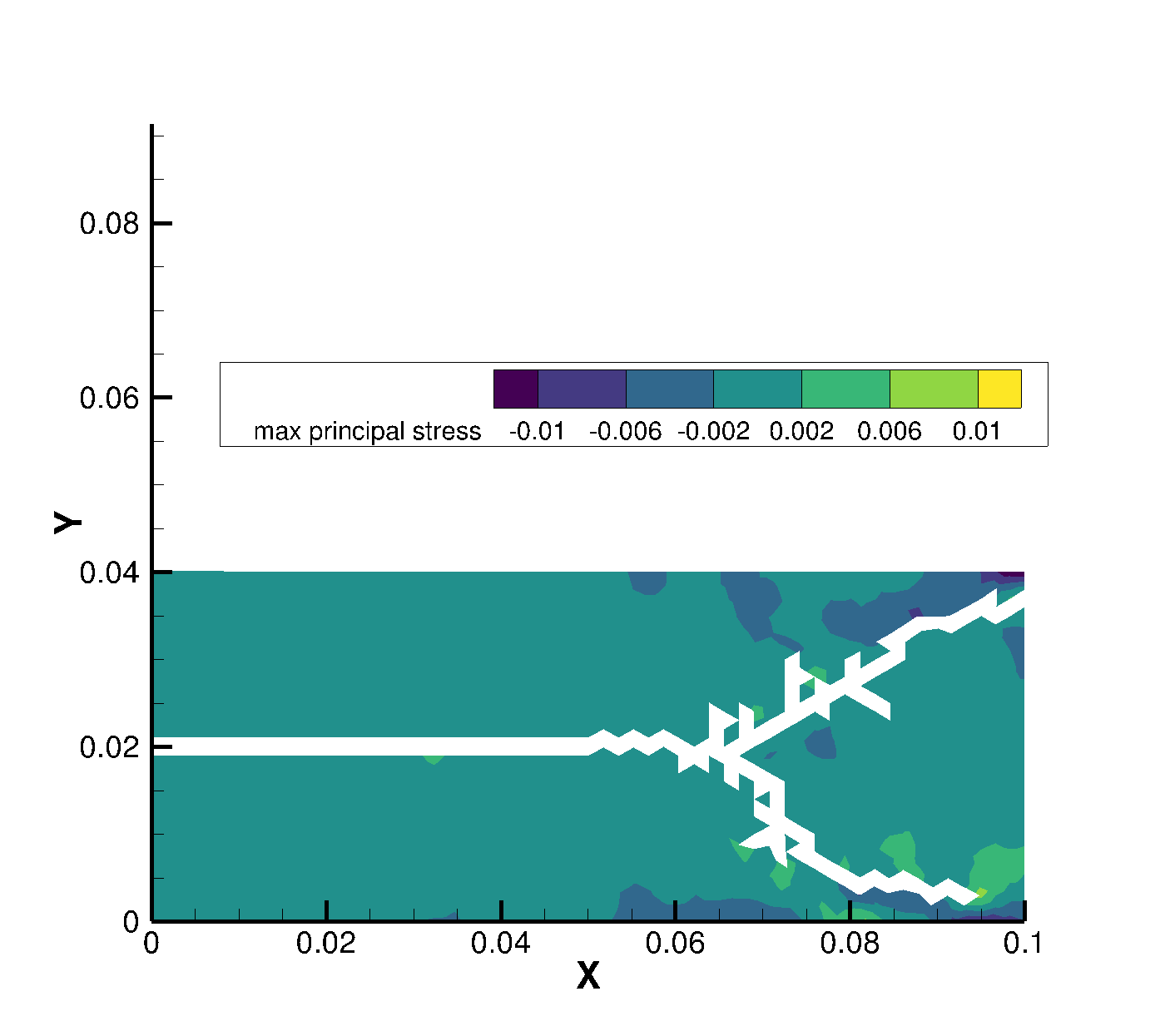}
            \end{center}
            \begin{center}
            (d)
            \end{center}
        \end{minipage}
        \caption{The crack pattern evolution of coarse mesh with $2343$ elements and $1268$ nodes are provided: (a). time $t=14.4\ \mu s$; (b). time $t=33.6\ \mu s$; (c). time $t=56\ \mu s$; (d). time $t=80\ \mu s$.}
        \label{fig5-13: cb-neumann-mesh3-crack-evolution}
\end{figure}

The crack pattern evolution of the mesh with $37199$ elements and $18981$ nodes at different time snaps are shown in Figure.\ref{fig5-11: cb-neumann-mesh1-crack-evolution},
the crack pattern evolution of the mesh with $9287$ elements and $4835$ nodes at different time snaps are shown in Figure.\ref{fig5-12: cb-neumann-mesh2-crack-evolution},
and the crack pattern evolution of the mesh with $2343$ elements and $1268$ nodes at different time snaps are shown in Figure.\ref{fig5-13: cb-neumann-mesh3-crack-evolution}. By comparing crack pattern evolution of three grids, it is found that at time $t=33.6\ \mu s$, the crack branching starts to initiate. However, contrary to clear crack patterns in $37199$ elements model and $9287$ elements model, the crack pattern in the $2343$ elements model is NOT completely symmetric due to relatively coarse mesh.(see Figure.\ref{fig5-13: cb-neumann-mesh3-crack-evolution}(c-d)). 

\begin{figure}[htp]
	\centering
            \begin{center}
            \includegraphics[height=2.4in]{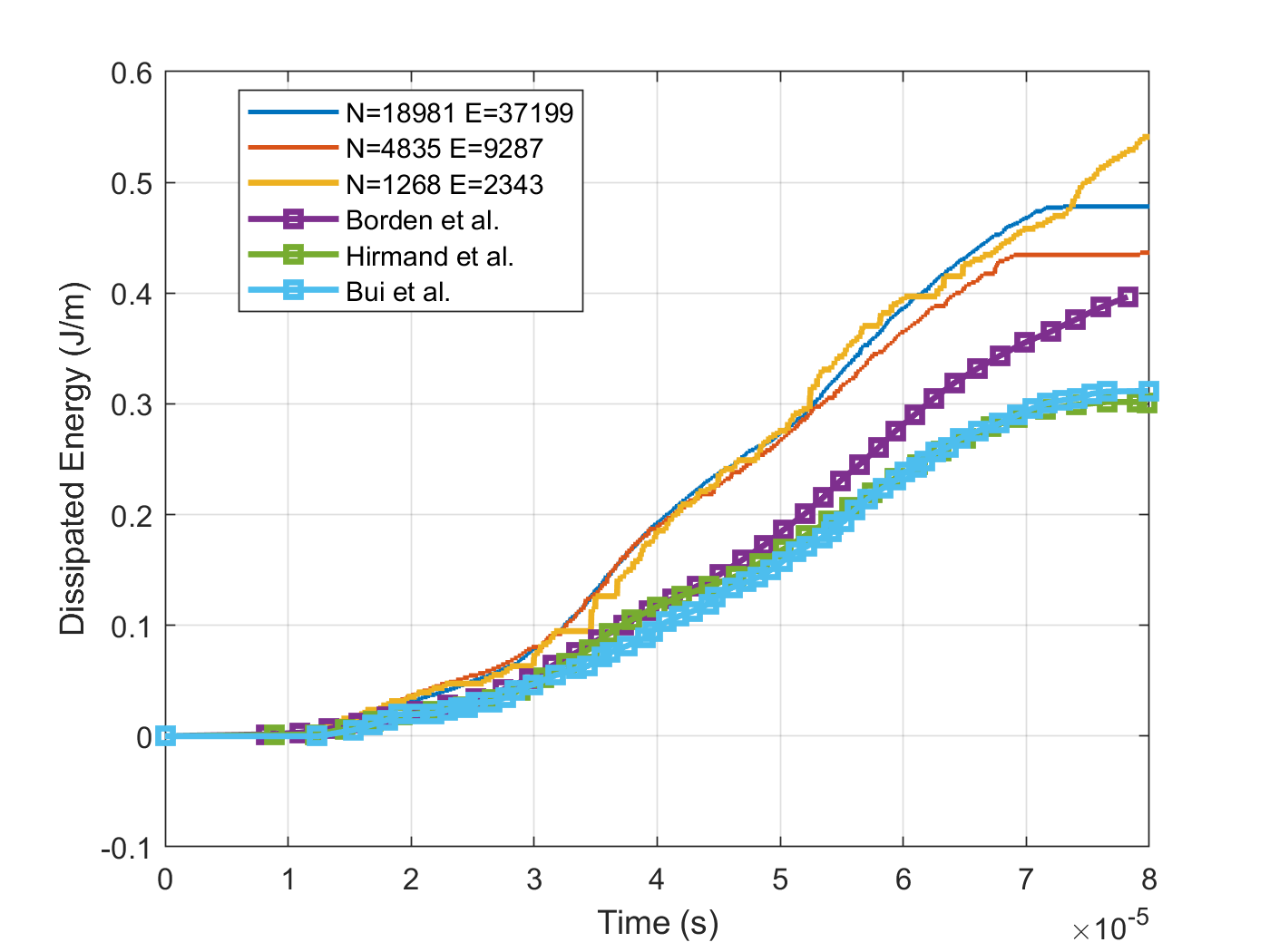}
            \end{center}
        \caption{Comparison of dissipated energy among three representative grids and three reference results(\cite{bui2022simulation}, \cite{borden2012phase}, \cite{hirmand2019block}).}
        \label{fig5-14: cb-neumann-Ud}
\end{figure}

On the other hand, a quantitative dissipated energy comparison among the proposed multiple crack tips tracking algorithm and other numerical methods is made. The reference data from the phase-field model by Borden et al. (\cite{borden2012phase}), local damage model by Bui et al. (\cite{bui2022simulation}) and coehsvie fracture approach by Hirmand et al. (\cite{hirmand2019block}) are included. From the Figure.\ref{fig5-14: cb-neumann-Ud}, it is found that the dissipated energy from the meshes of $37199$ elements, $9287$ elements, and $1268$ elements are obviously higher than other numerical results. It is because that introduction of micro-cracks costs extra energy in the proposed multiple crack tips tracking algorithm.

Generally speaking, the proposed multiple crack tips tracking algorithm is capable to capture two-dimensional crack branching under transient-dynamic loads even with relatively coarse mesh (such as $2343$ elements). It is important to note that different from three-dimensional CEM that has no any explicit criteria of crack branching applied, the Multiple Crack-tips Tracking in two-dimensional CEM (MCT-2D-CEM) introduces extra branch criteria to determine when to branch out.

\subsection{Crack branching with Dirichlet boundary condition}
In addition to crack branching with Neumann boundary condition, we also investigated the benchmark example of crack branching with Dirichlet boundary condition. The onsite experiments were conducted by Ozbolt et al. (\cite{ovzbolt2013dynamic}), which is a transient-dynamic compact tension problem in concrete specimen. Ozbolt et al. studied the influence of various loading rates on crack patterns in plain concrete samples. Six different loading rates from $0.304\ m/s$, $0.491\ m/s$, $1.375\ m/s$ to $3.318\ m/s$, $3.993\ m/s$, $3.967\ m/s$ are investigated by Ozbolt et al. and the corresponding crack patterns vary from mode-I fracture to mixed-mode fracture (see Figure.3 in \cite{ovzbolt2013dynamic}). 

In the dynamic fracture study of concrete compact tension specimens, the transition from single crack propagation at lower loading rates to crack branching at higher rates reflects the fundamental interplay between crack tip dynamics, stress wave interactions, and the rate-dependent nature of concrete’s fracture process.

At a lower loading rate such as $1.375\ m/s$, the applied stress is introduced slowly enough that the crack propagation remains stable and primarily mode-I (opening mode) dominates. In this regime, the strain energy released from the loading is dissipated smoothly at the crack tip, and the surrounding stress field remains nearly symmetric. The crack tip increases steadily but stays below a critical threshold beyond which dynamic instabilities begin to dominate. Since the material response remains quasi-static or mildly dynamic, the fracture process zone around the crack tip evolves in an orderly fashion, and micro-cracking is limited. As a result, the crack propagates along a single, straight path, following the direction of the maximum tensile stress. This behavior is consistent with classical linear elastic fracture mechanics where a stable stress intensity factor governs the growth direction and rate of the crack.

In contrast, when the loading rate is increased significantly, such as $3.318\ m/s$ or $3.993\ m/s$, the applied kinetic energy per unit time rises sharply, leading to rapid stress concentration at the crack tip. The crack tip accelerates quickly and may approach or exceed the critical velocity associated with branching instabilities. As the crack moves faster, it cannot dissipate the incoming energy efficiently through a single propagation front. Instead, the excessive energy causes a redistribution of the stress field, often amplified by inertial effects and dynamic stress wave reflections within the specimen. These inertial forces, which are negligible at low loading rates, begin to resist the deformation locally and cause asymmetry in the crack tip stress field, breaking the symmetry necessitating stable single-crack growth.

Moreover, the stress waves generated at the crack tip during rapid propagation interact with the specimen boundaries and the evolving crack itself. These interactions introduce local stress concentrations off the main crack path, potentially triggering secondary cracks. The competition between the primary crack front and these localized instabilities give rise to crack branching, a phenomenon well-documented in brittle materials under dynamic loading. Branching allows the material to dissipate the surplus strain energy more broadly, reducing the energy density at any single crack tip and thus stabilizing the overall fracture process.

\begin{figure}[htp]
	\centering
        \begin{minipage}{0.45\linewidth}
            \centering
            \includegraphics[height=2.8in]{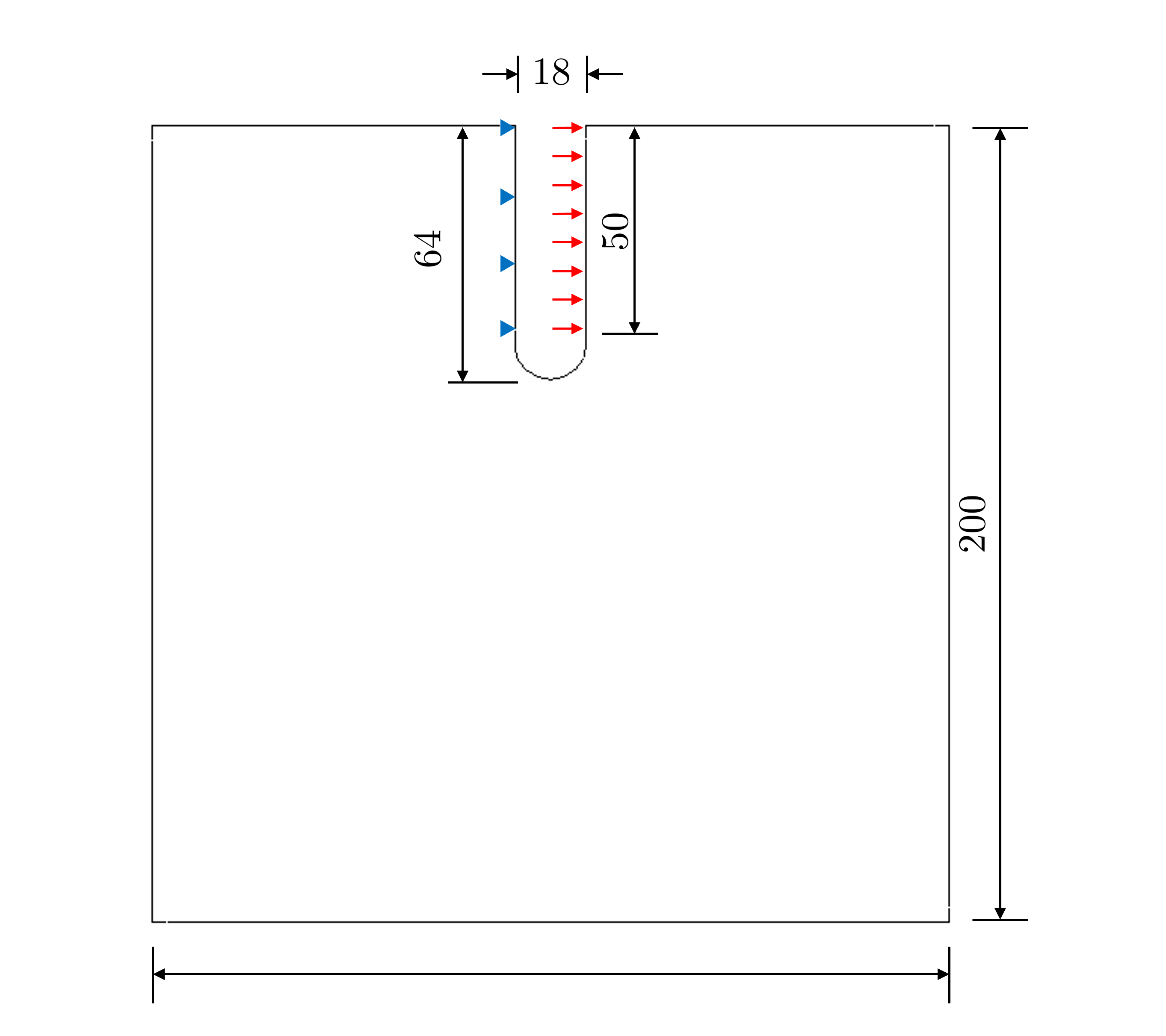}
            \begin{center}
            (a)
            \end{center}
        \end{minipage}
        \hfill
        \begin{minipage}{0.45\linewidth}
            \centering
            \includegraphics[height=2.8in]{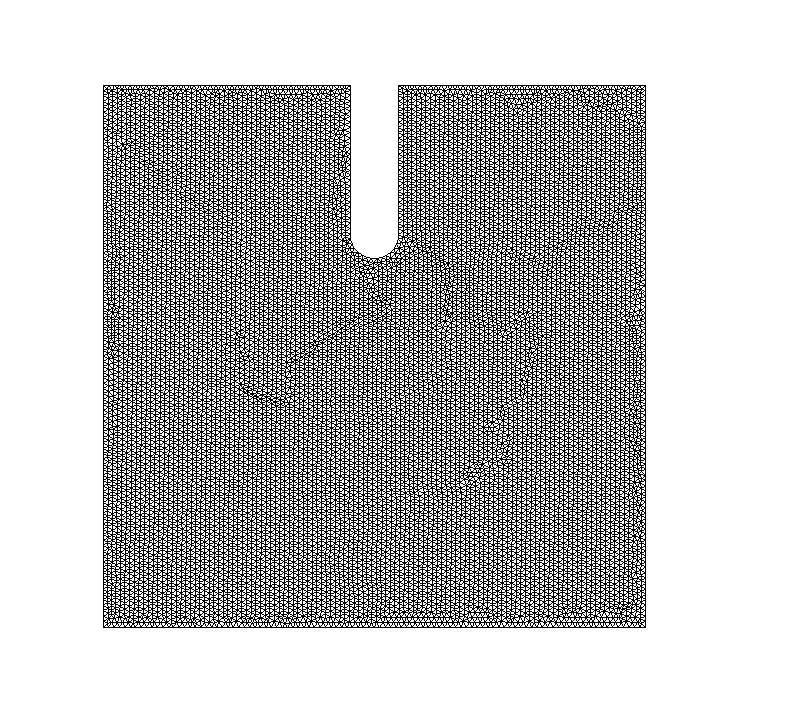}
            \begin{center}
            (b)
            \end{center}
        \end{minipage}   
        \hfill
        \begin{minipage}{0.45\linewidth}
            \centering
            \includegraphics[height=2.8in]{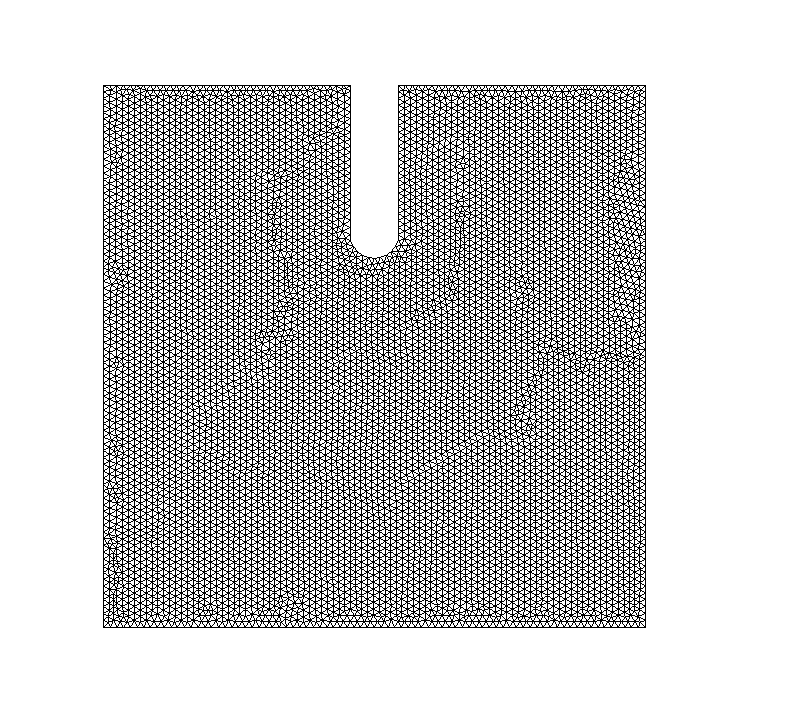}
            \begin{center}
            (c)
            \end{center}
        \end{minipage}  
        \hfill
        \begin{minipage}{0.45\linewidth}
            \centering
            \includegraphics[height=2.8in]{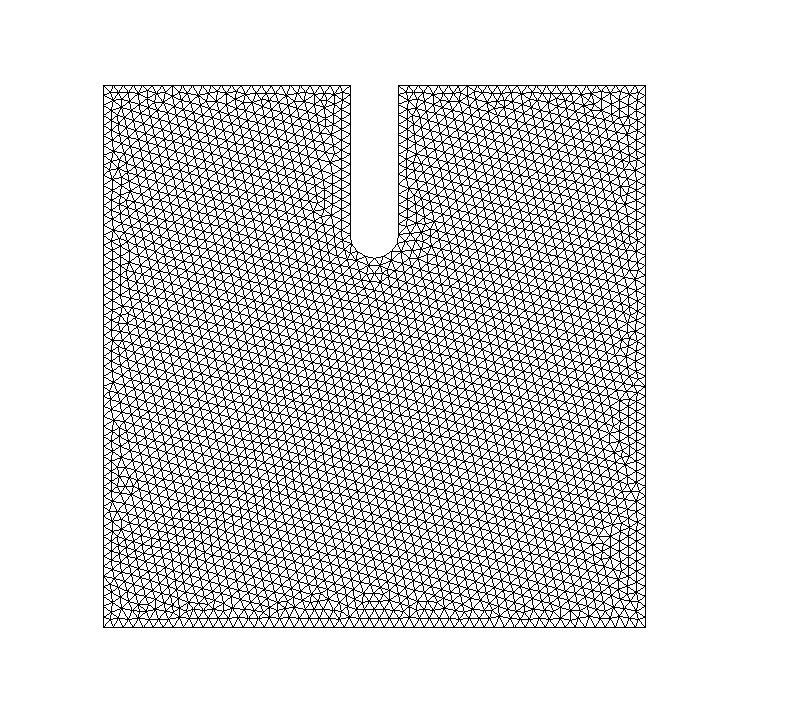}
            \begin{center}
            (d)
            \end{center}
        \end{minipage}  
        \caption{(a). The model geometric dimensions (unit: mm) and boundary conditions of compact tension experiment. (b). The mesh with $11939$ Nodes and $23413$ Elements; (c). The mesh with $7643$ Nodes and $14914$ Elements; (d). The mesh with $3563$ Nodes and $6874$ Elements.}
        \label{fig5-15: cb-dirichlet-model-grids}
\end{figure}

To further examine capability of the proposed Multiple Crack-tips Tracking algorithm in two-dimensional CEM (MCT-2D-CEM), three representative applied velocities are selected: Case 1 with a velocity of $1.375\ m/s$, Case 2 with a velocity of $3.318\ m/s$ and Case 3 with a velocity of $3.993\ m/s$. The determined velocity is applied on right surface of pre-notch in the sample, marked by red arrows, and the left surface of pre-notch is constrained horizontally, marked by blue triangles. The applied velcity is ramp-up firstly, then keep constant, as formulated as follows,
\begin{equation}
v =
\begin{cases} 
\frac{t}{t_0}v_0,  & \text{if }\ t \le t_0, \\
v_0, & \text{if } \ t > t_0.
\end{cases}
\end{equation}
in which, $t_0 = \ 100 \mu s$ and $v_0 = 1.375\ m/s, \  3.318\ m/s, \  3.993 \ m/s$ for three cases, respectively. 
Other geometric dimensions and boundary conditions are illustrated in Figure.\ref{fig5-15: cb-dirichlet-model-grids}(a).

Three different meshes of Constant Strain Triangle element, i.e., fine mesh of $11939$ nodes and $23413$ elements, medium mesh of $7643$ nodes and $14914$ elements and coarse mesh of $3563$ nodes and $6874$ elements, are utilized to discretize the studied domain, as shown in Figure.\ref{fig5-15: cb-dirichlet-model-grids}(b-d). 

\begin{figure}[htp]
	\centering
        \begin{minipage}{0.3\linewidth}
            \begin{center}
            \includegraphics[height=1.3in]{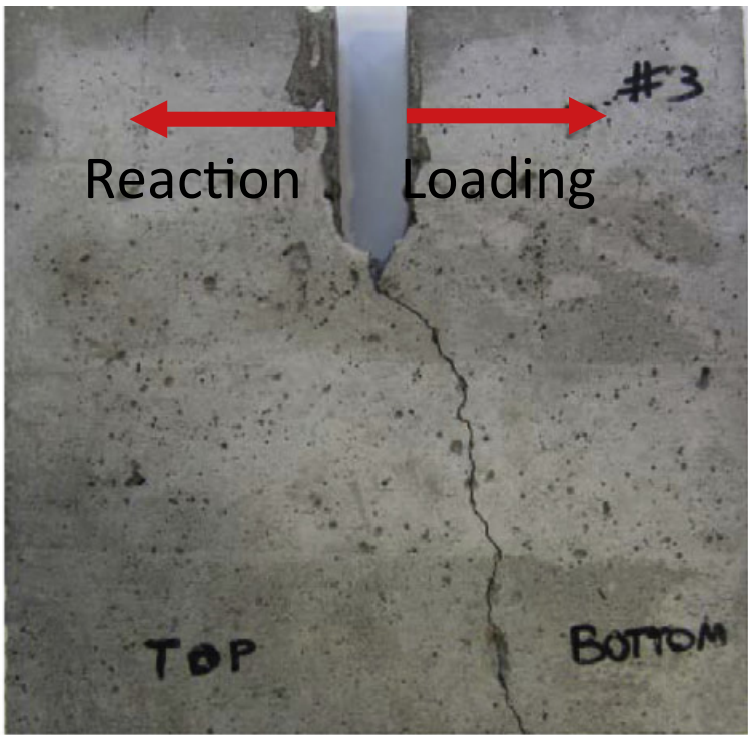}
            \end{center}
            \begin{center}
            (a)
            \end{center}
        \end{minipage}
        \begin{minipage}{0.3\linewidth}
            \begin{center}
            \includegraphics[height=1.3in]{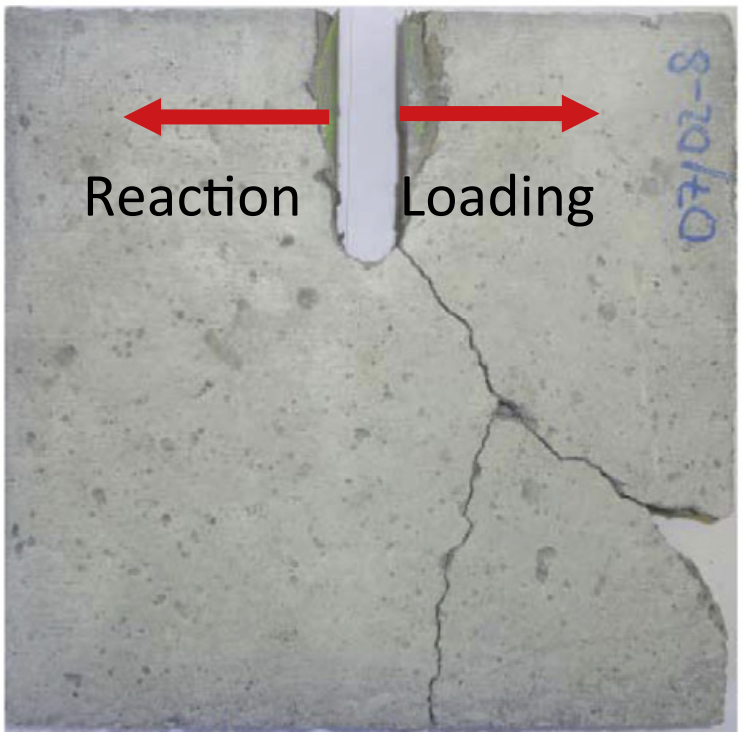}
            \end{center}
            \begin{center}
            (b)
            \end{center}
        \end{minipage}   
        \begin{minipage}{0.3\linewidth}
            \begin{center}
            \includegraphics[height=1.3in]{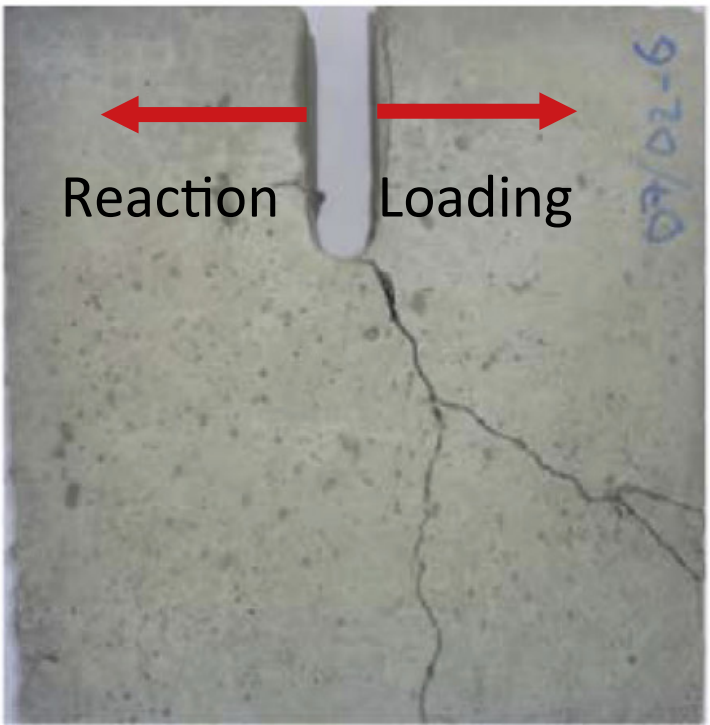}
            \end{center}
            \begin{center}
            (c)
            \end{center}
        \end{minipage}
        \begin{minipage}{0.3\linewidth}
            \begin{center}
            \includegraphics[height=1.8in]{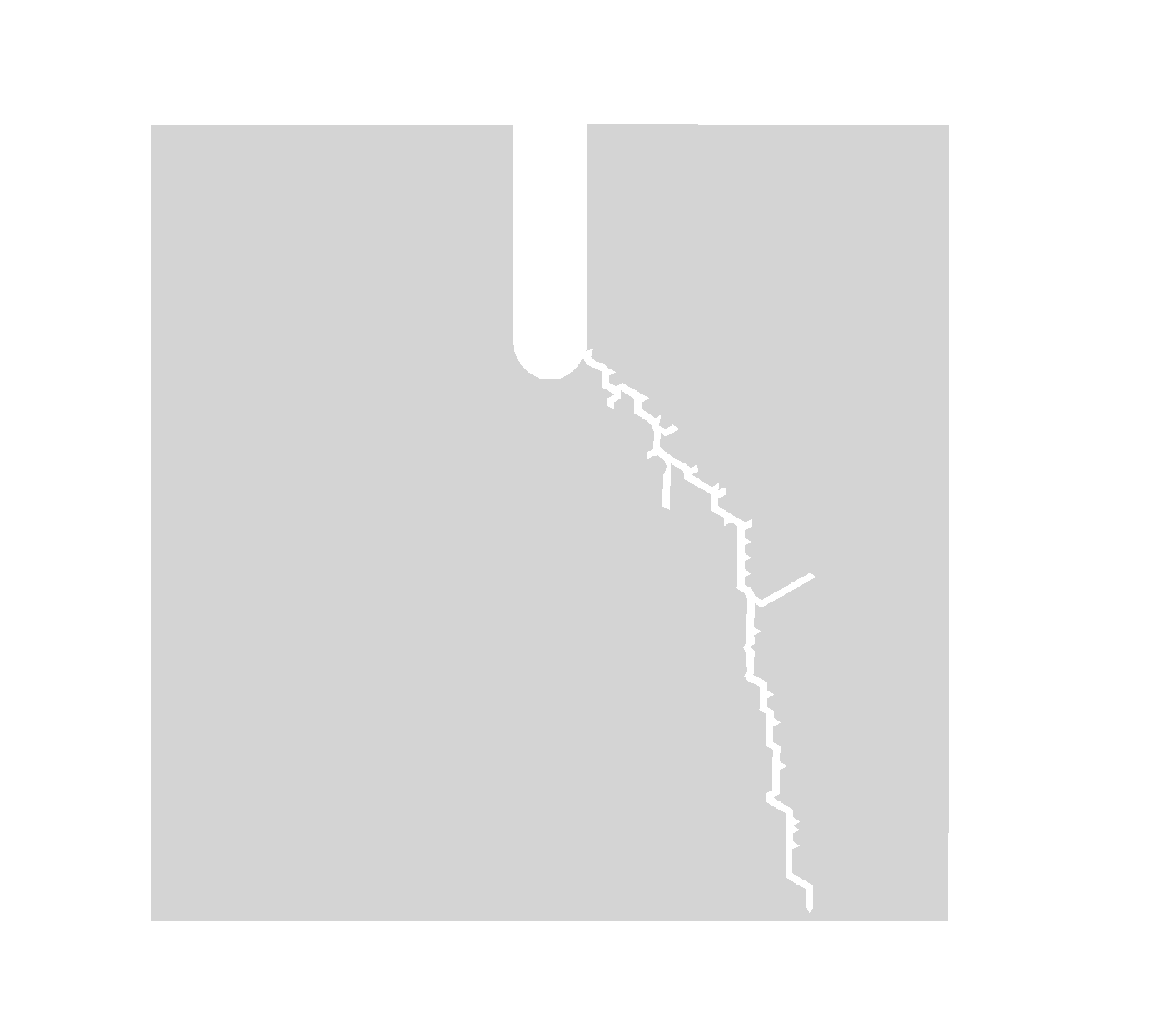}
            \end{center}
            \begin{center}
            (d)
            \end{center}
        \end{minipage}
        \begin{minipage}{0.3\linewidth}
            \begin{center}
            \includegraphics[height=1.8in]{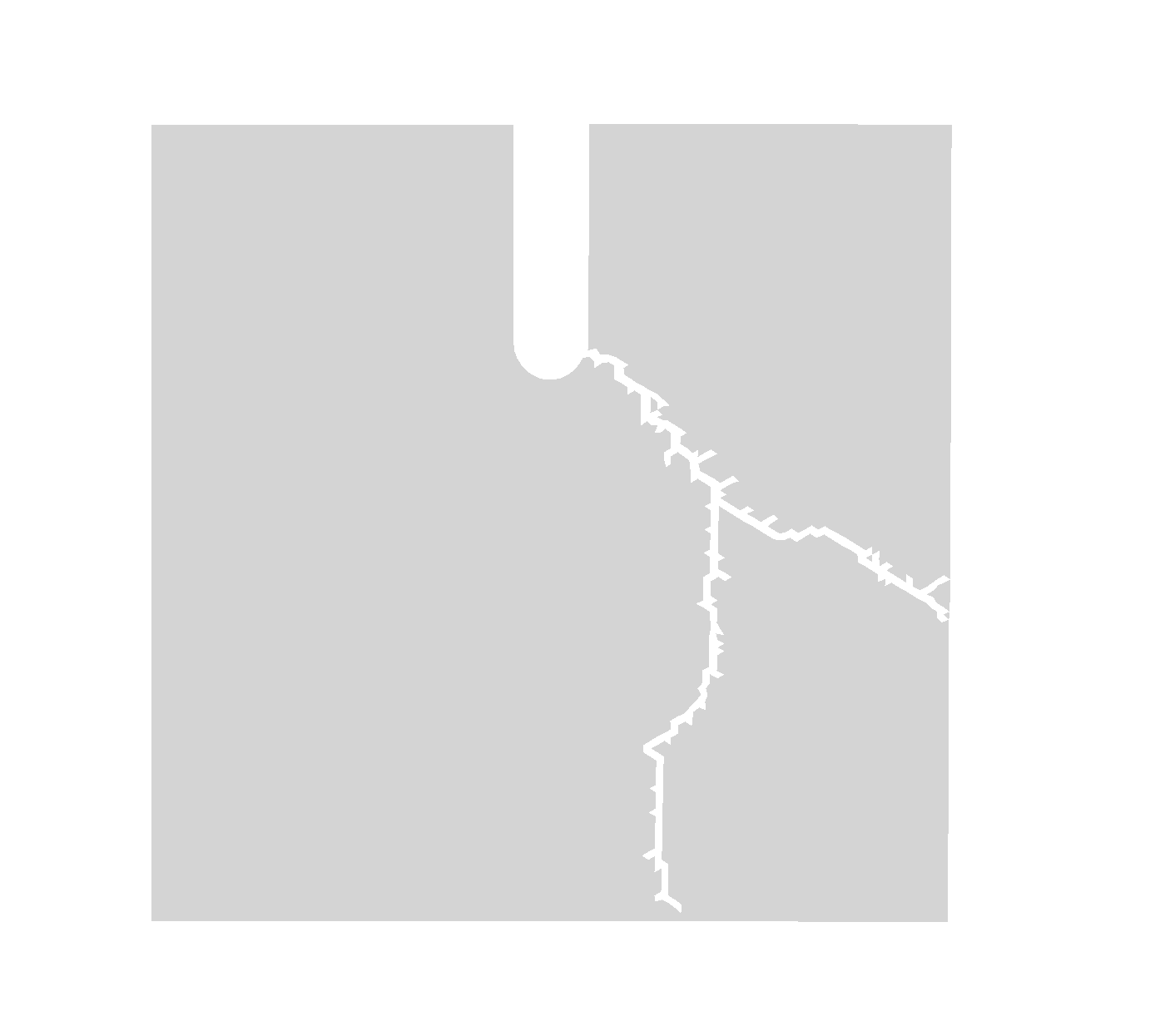}
            \end{center}
            \begin{center}
            (e)
            \end{center}
        \end{minipage}   
        \begin{minipage}{0.3\linewidth}
            \begin{center}
            \includegraphics[height=1.8in]{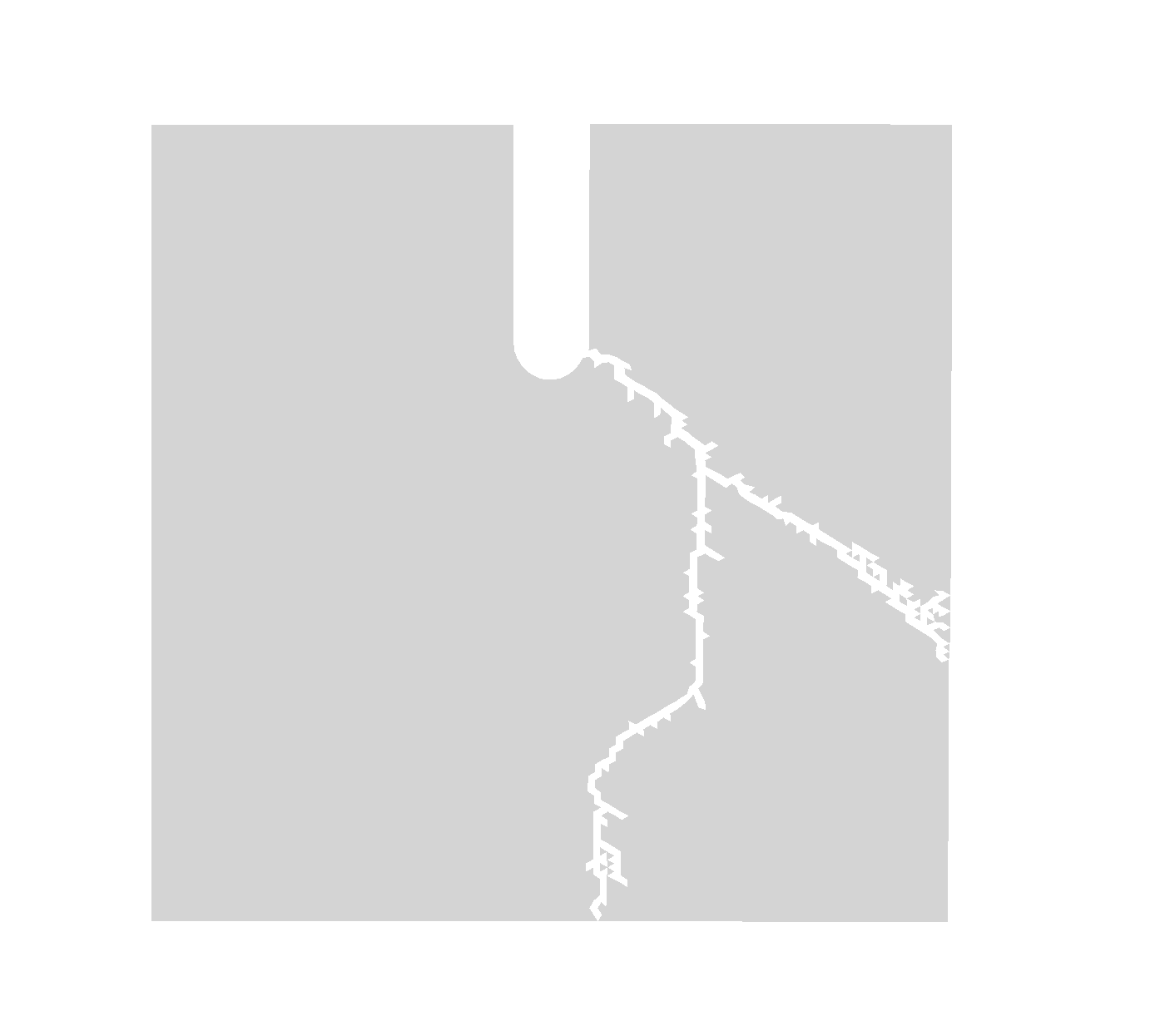}
            \end{center}
            \begin{center}
            (f)
            \end{center}
        \end{minipage}
        \begin{minipage}{0.3\linewidth}
            \begin{center}
            \includegraphics[height=1.8in]{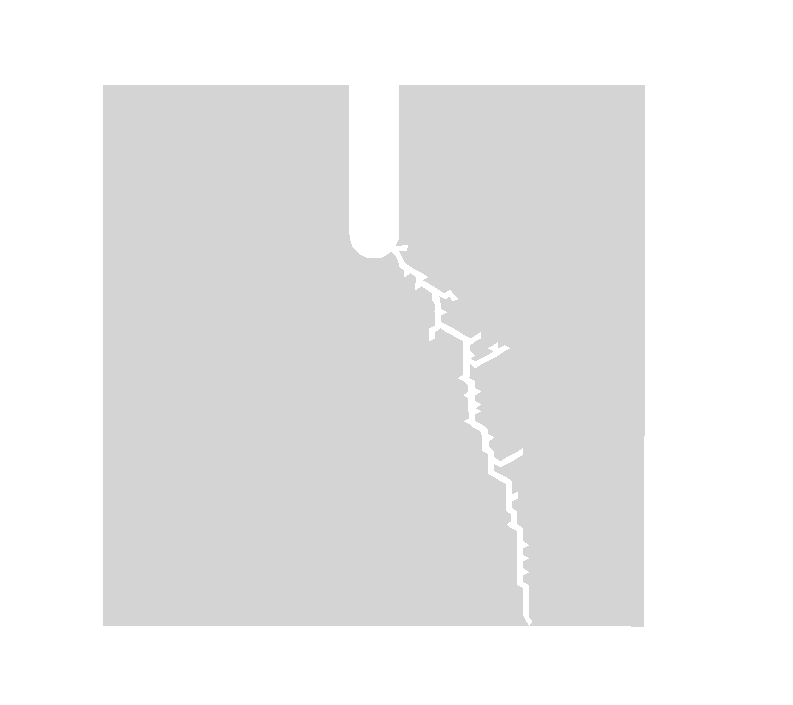}
            \end{center}
            \begin{center}
            (g)
            \end{center}
        \end{minipage}
        \begin{minipage}{0.3\linewidth}
            \begin{center}
            \includegraphics[height=1.8in]{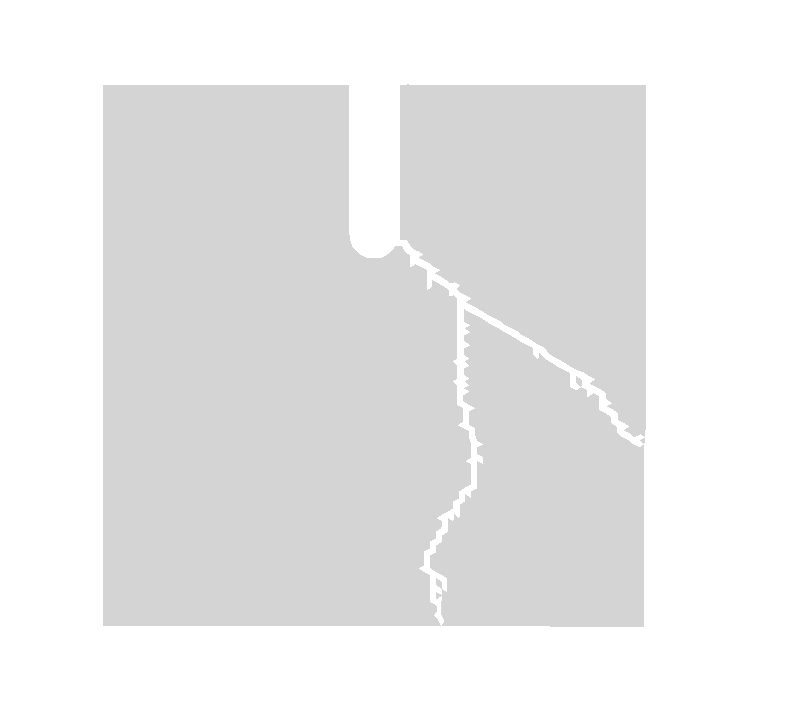}
            \end{center}
            \begin{center}
            (h)
            \end{center}
        \end{minipage}   
        \begin{minipage}{0.3\linewidth}
            \begin{center}
            \includegraphics[height=1.8in]{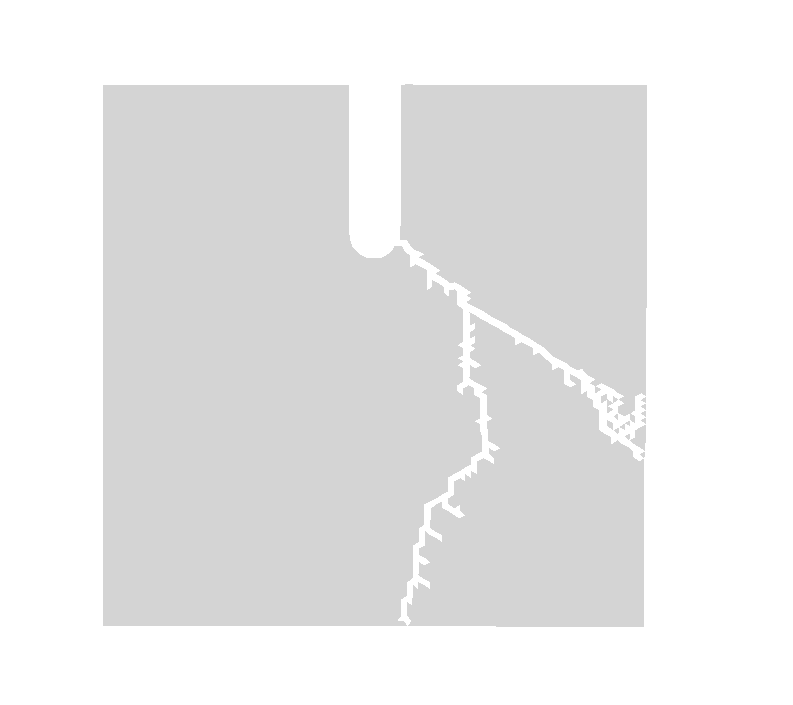}
            \end{center}
            \begin{center}
            (i)
            \end{center}
        \end{minipage}
        \begin{minipage}{0.3\linewidth}
            \begin{center}
            \includegraphics[height=1.8in]{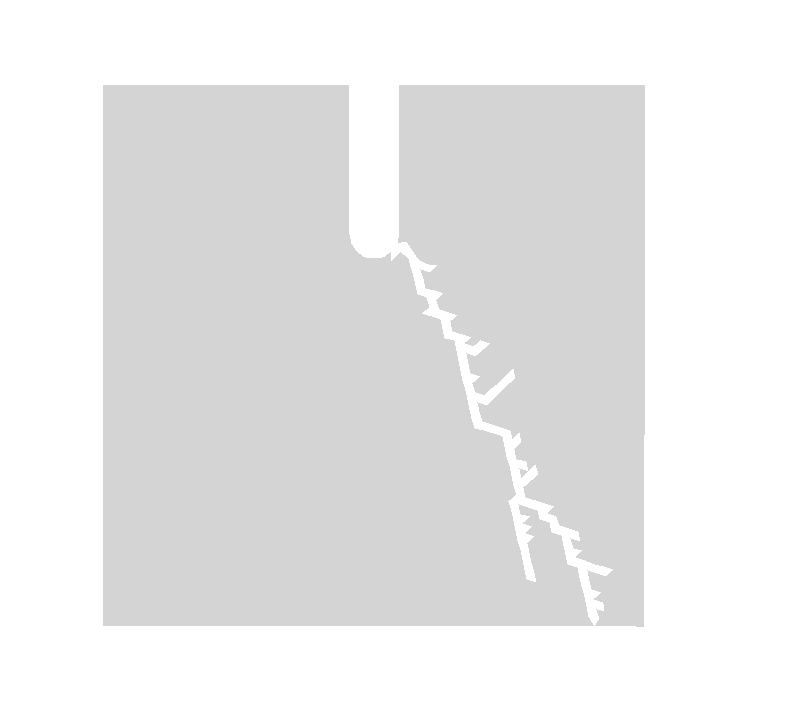}
            \end{center}
            \begin{center}
            (j)
            \end{center}
        \end{minipage}
        \begin{minipage}{0.3\linewidth}
            \begin{center}
            \includegraphics[height=1.8in]{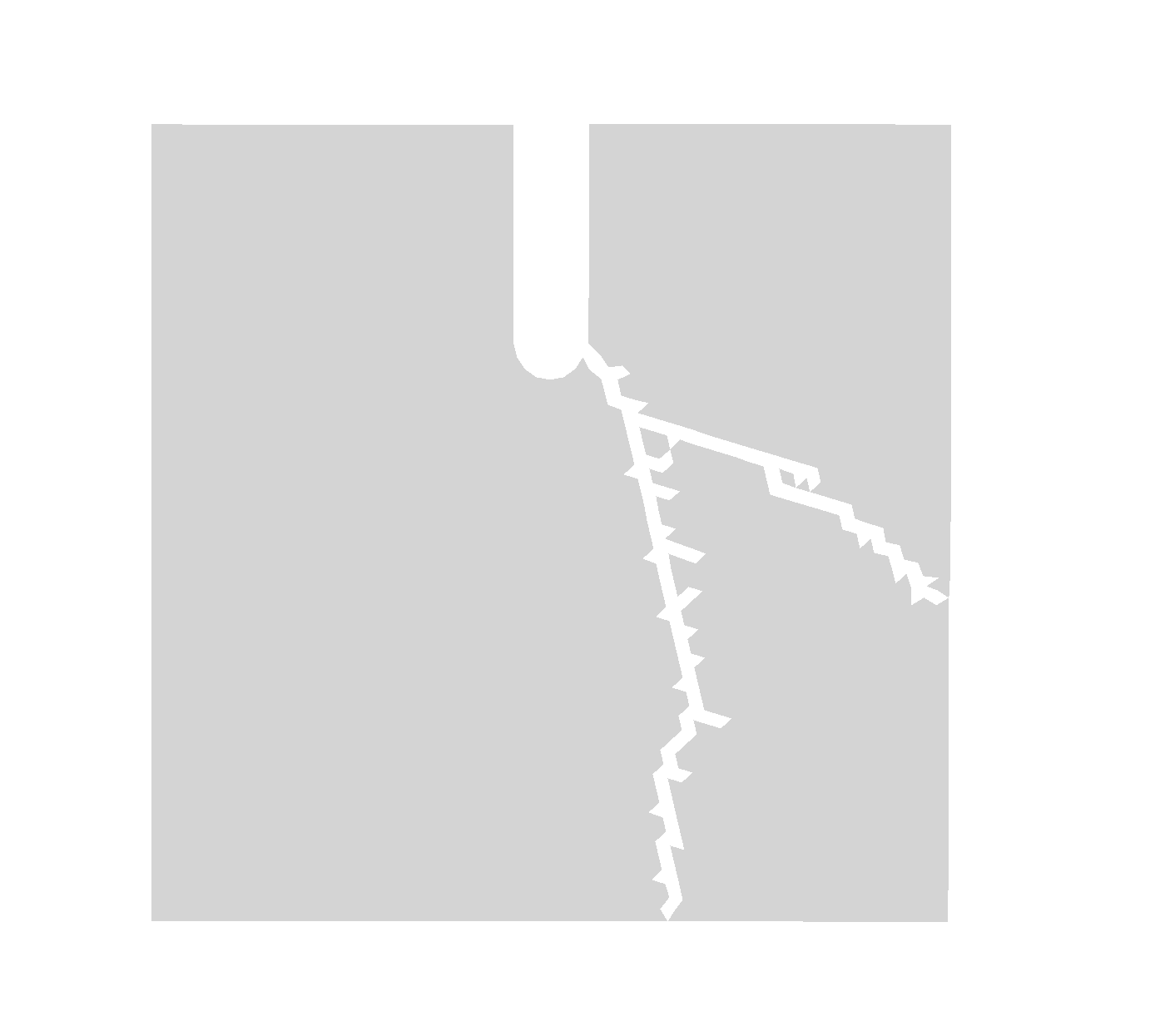}
            \end{center}
            \begin{center}
            (k)
            \end{center}
        \end{minipage}   
        \begin{minipage}{0.3\linewidth}
            \begin{center}
            \includegraphics[height=1.8in]{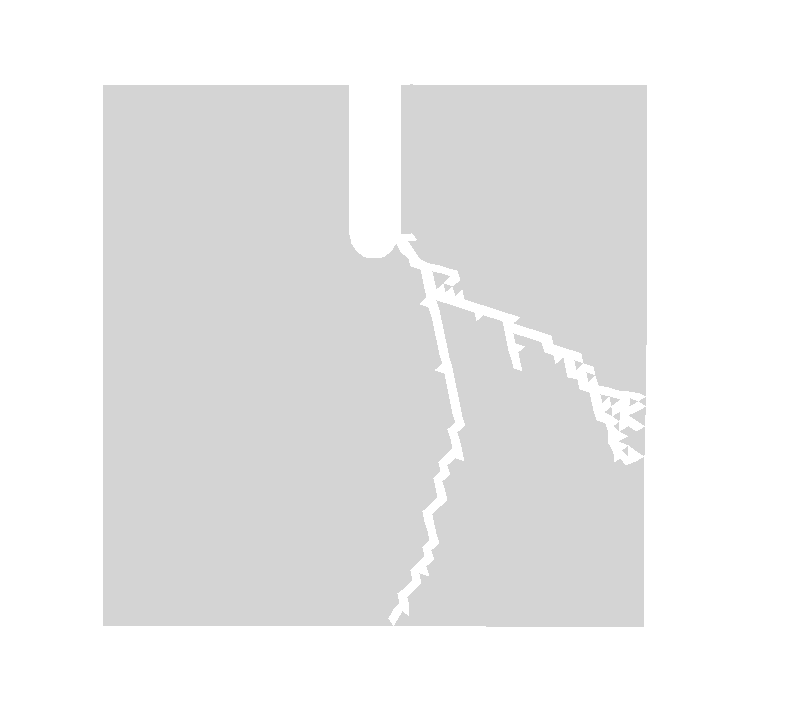}
            \end{center}
            \begin{center}
            (l)
            \end{center}
        \end{minipage}
        \caption{(a-c). experimental crack patterns under applied velocity of $1.375\ m/s$, $3.318\ m/s$ and $3.993\ m/s$ (\cite{ovzbolt2013dynamic}); (d-f). numerical crack patterns of $N=11939,\ E=23413$ model under applied velocity of $1.375\ m/s$, $3.318\ m/s$ and $3.993\ m/s$; (g-i). numerical crack patterns of $N=7643,\ E=14914$ model under applied velocity of $1.375\ m/s$, $3.318\ m/s$ and $3.993\ m/s$; (j-l). numerical crack patterns of $N=3563,\ E=6874$ model under applied velocity of $1.375\ m/s$, $3.318\ m/s$ and $3.993\ m/s$. }
        \label{fig5-16: cb-dirichlet-crack-patterns}
\end{figure}

The final crack patterns of the three meshes with different various applied velocities are compared to corresponding experimental results in Figure.\ref{fig5-16: cb-dirichlet-crack-patterns}.

It can be concluded that the proposed Multiple Crack-tips Tracking algorithm in two-dimensional CEM is capable to model crack branching under Dirichlet boundary condition. Because when loading velocity exceeds $3.0\ m/s$ the crack branching pattern observed experimentally can be reproduced even though using a relative coarse grid with $3563$ nodes and $6874$ elements. At a lower velocity of $1.375\ m/s$, the same set of proposed algorithm also accurately reflects formation of a single crack, aligning well with experimental findings. These outcomes suggest that the Multiple Crack-tips Tracking algorithm in CEM is well-suited for simulating two-dimensional, transient-dynamic crack branching under Dirichlet boundary condition.

\begin{figure}[htp]
	\centering
        \begin{minipage}{0.24\linewidth}
            \begin{center}
            \includegraphics[height=1.5in]{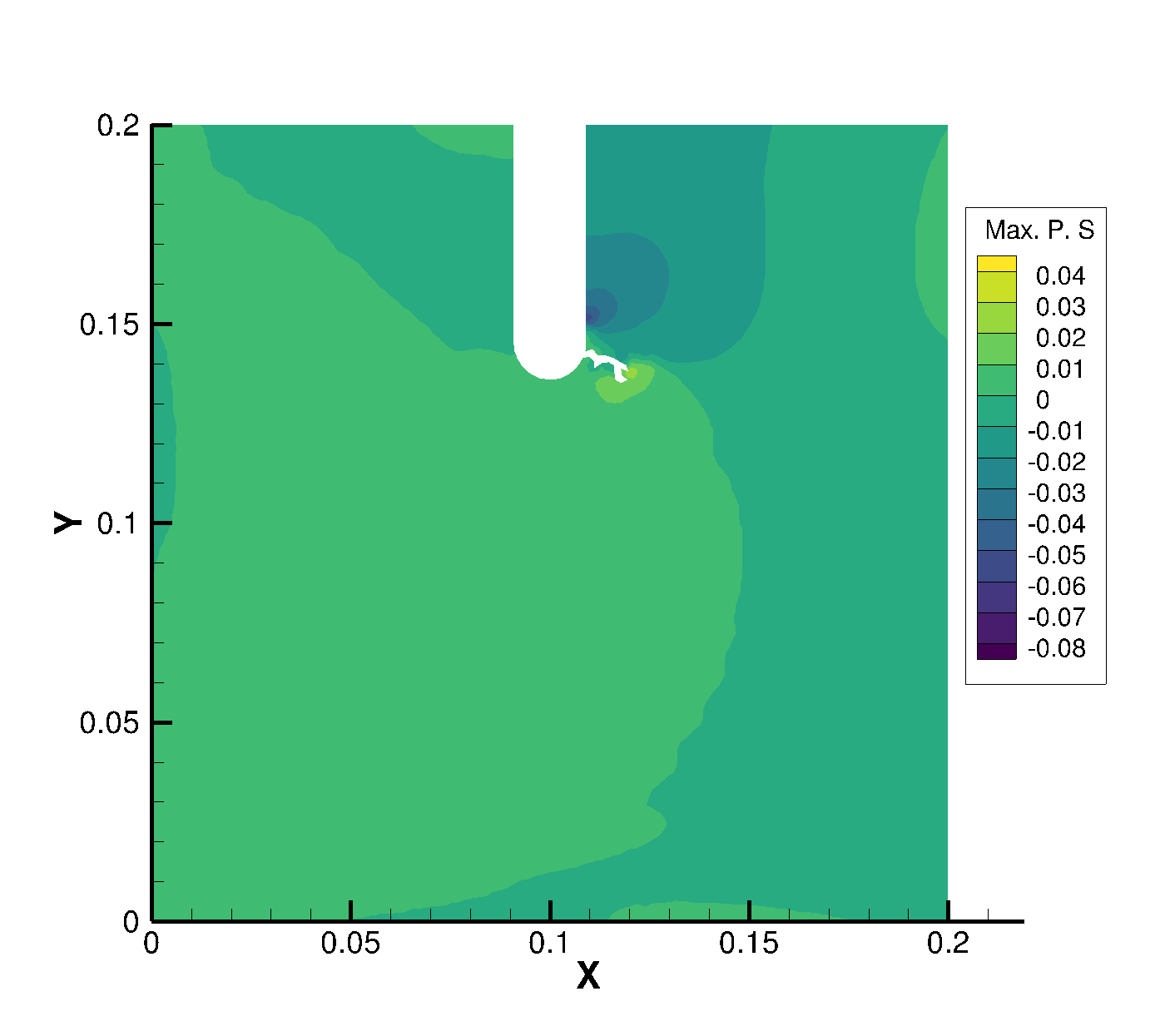}
            (a)
            \end{center}
        \end{minipage}
        \hfill
        \begin{minipage}{0.24\linewidth}
            \begin{center}
            \includegraphics[height=1.5in]{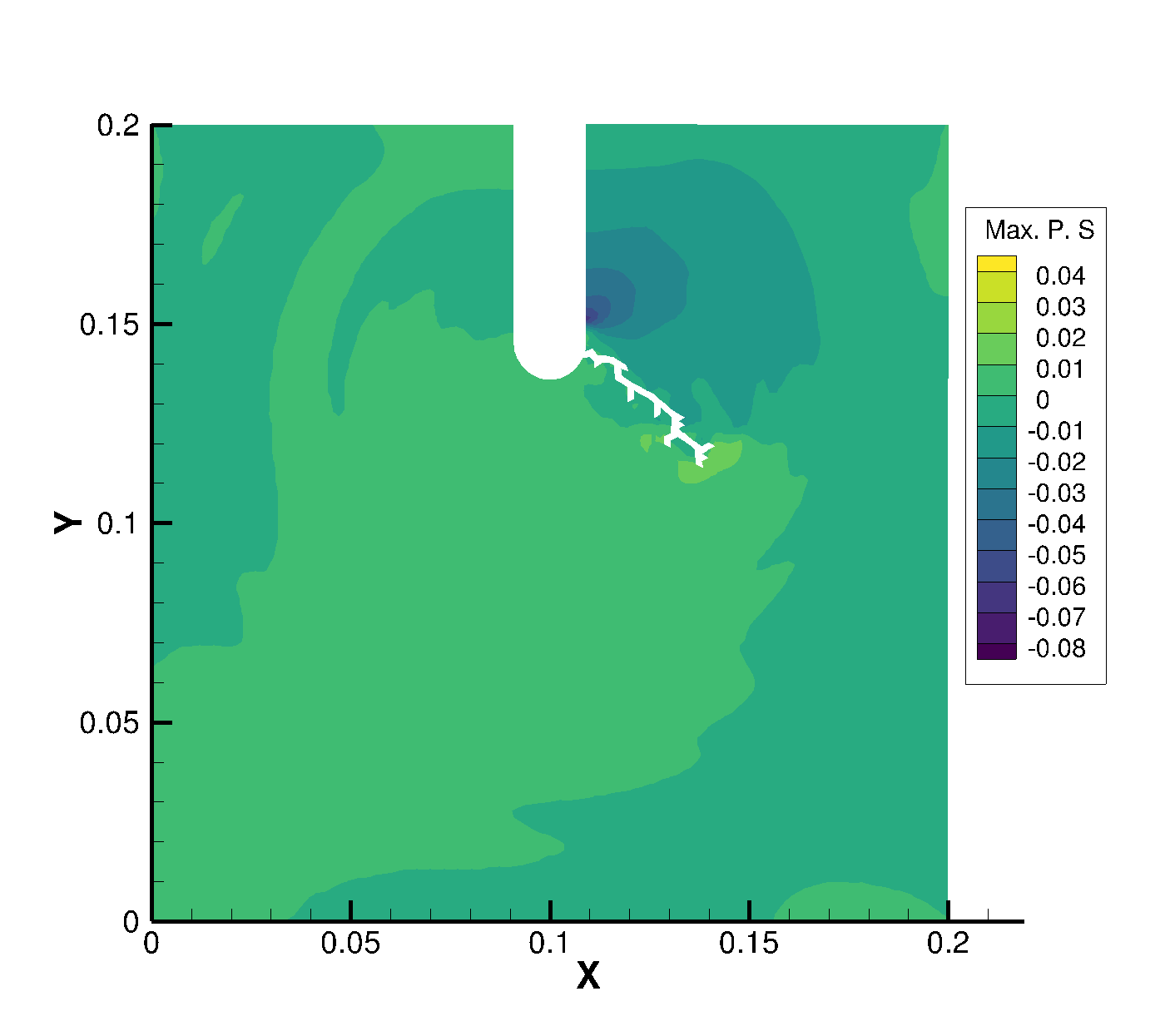}
            (b)
            \end{center}
        \end{minipage}   
        \hfill
        \begin{minipage}{0.24\linewidth}
            \begin{center}
            \includegraphics[height=1.5in]{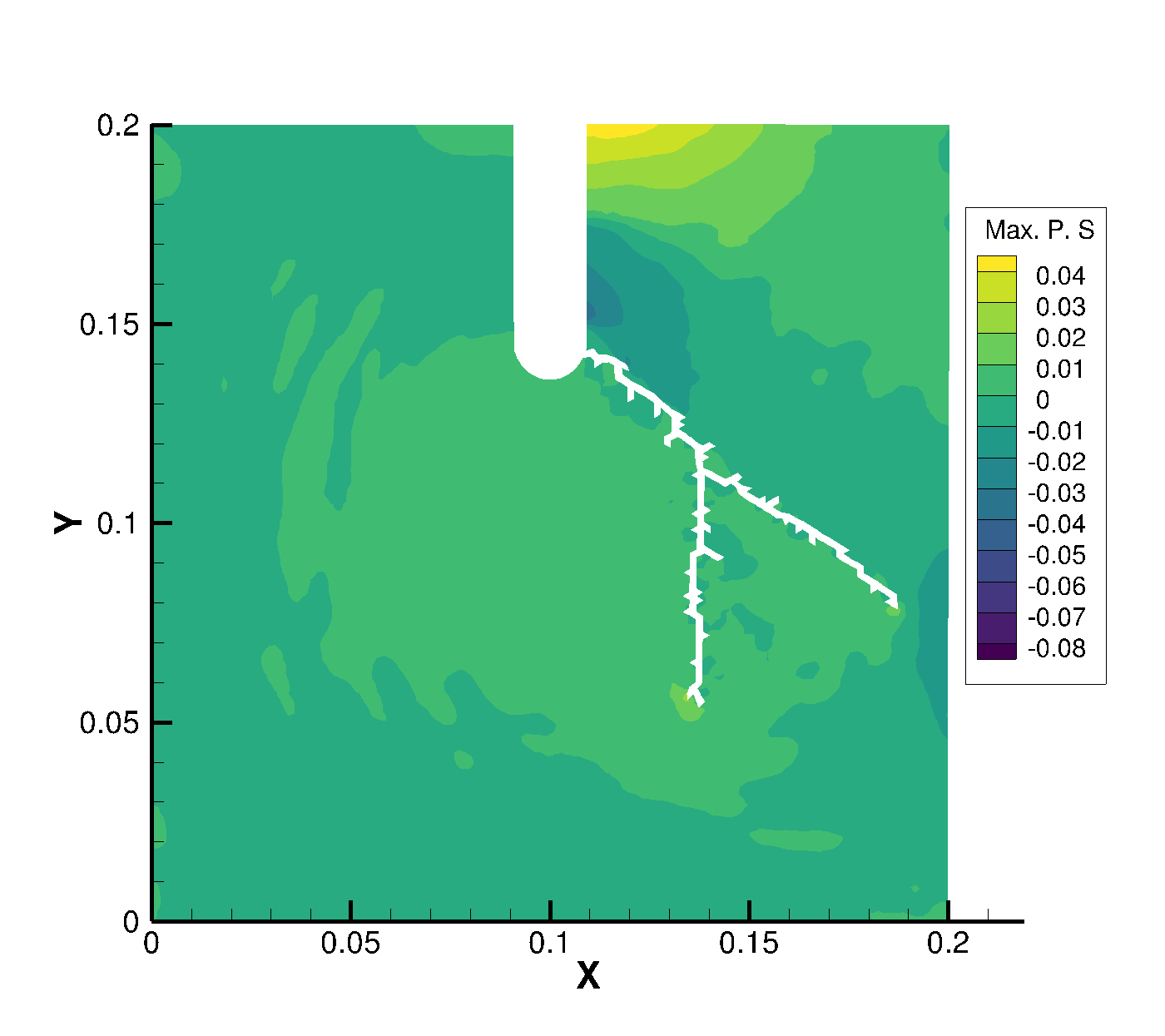}
            (c)
            \end{center}
        \end{minipage}
        \hfill
        \begin{minipage}{0.24\linewidth}
            \begin{center}
            \includegraphics[height=1.5in]{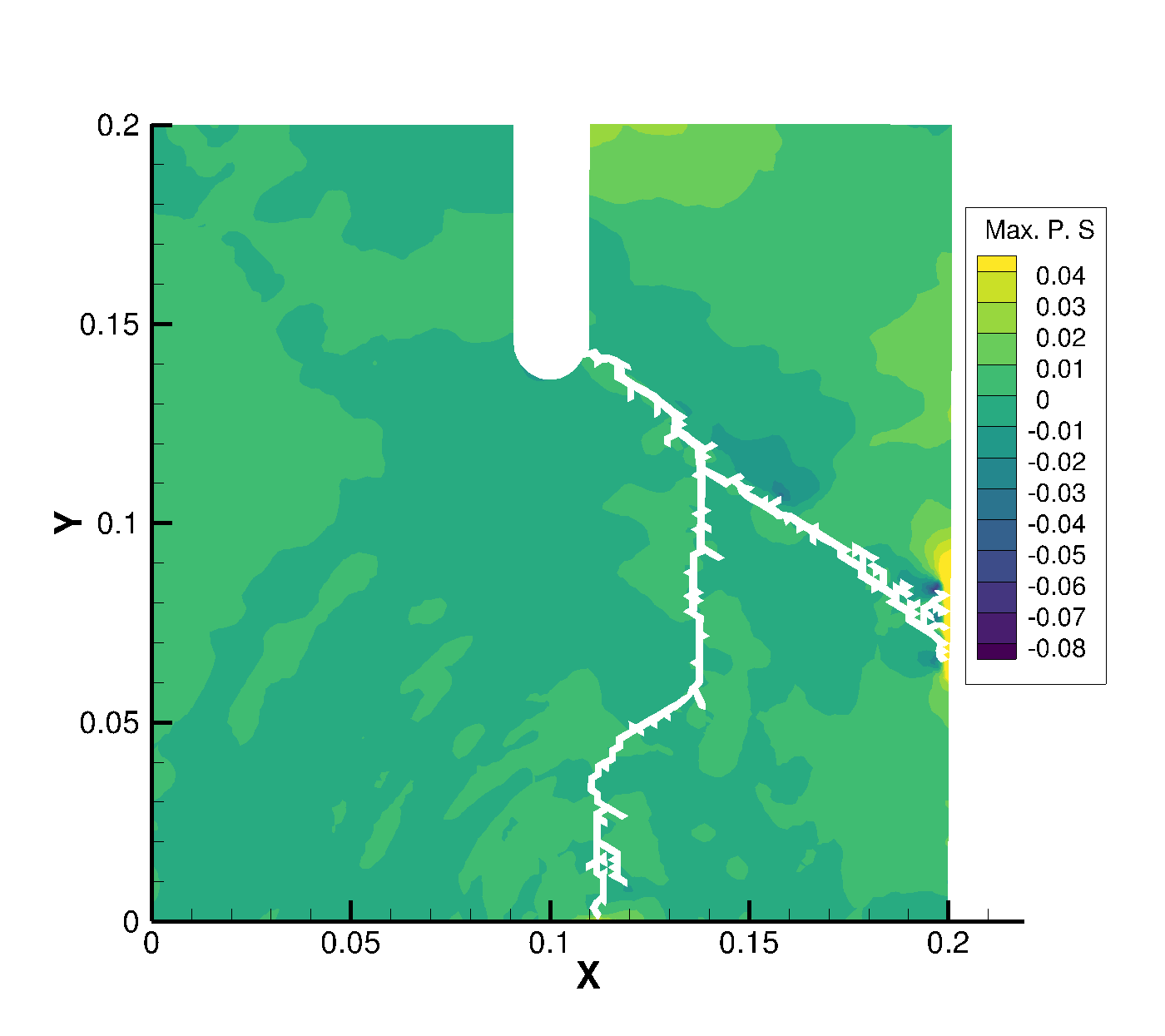}
            (d)
            \end{center}
        \end{minipage}
        \caption{Max. principle stress contour and crack patterns evolution of fine mesh $N=11939$, $E=23413$ with $v_0=3.993\ m/s$ at times at (a). $t=60\ \mu s$, (b). $t=84\ \mu s$, (c). $t=144\ \mu s$, (d). $t=300\ \mu s$.}
        \label{fig5-17: cb-dirichlet-crack-evolution-1}
\end{figure}

\begin{figure}[htp]
	\centering
        \begin{minipage}{0.24\linewidth}
            \begin{center}
            \includegraphics[height=1.5in]{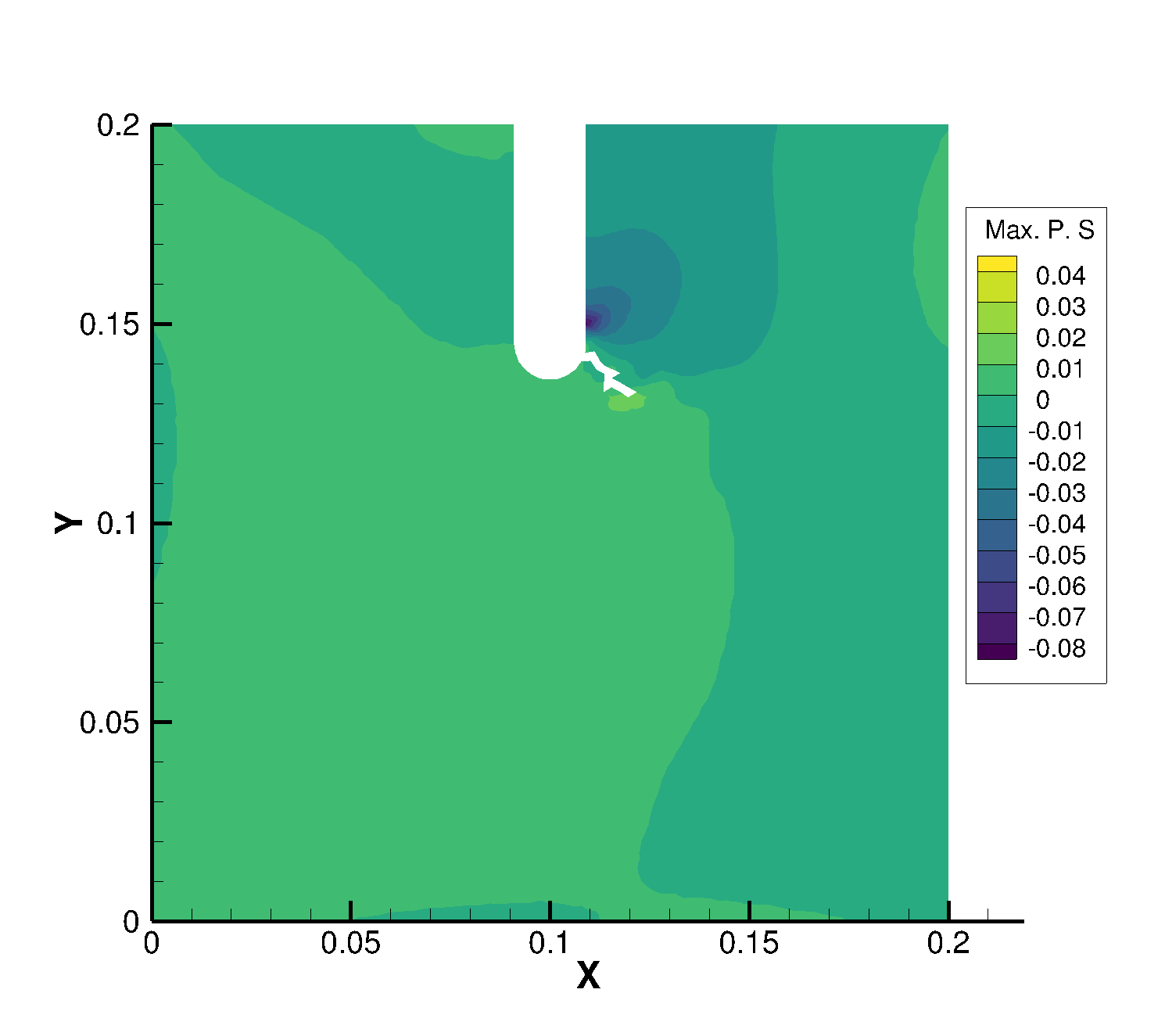}
            (a)
            \end{center}
        \end{minipage}
        \hfill
        \begin{minipage}{0.24\linewidth}
            \begin{center}
            \includegraphics[height=1.5in]{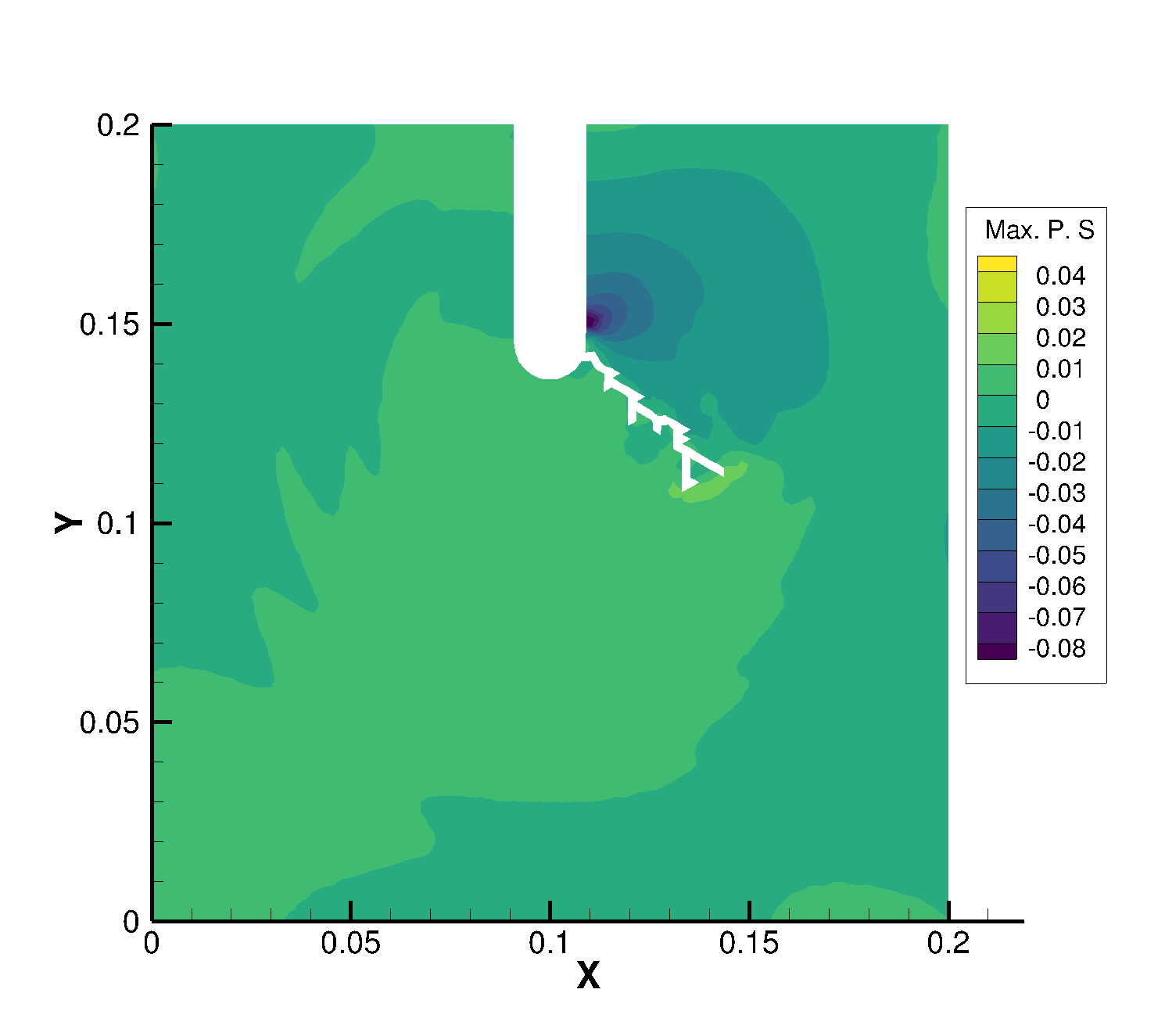}
            (b)
            \end{center}
        \end{minipage}   
        \hfill
        \begin{minipage}{0.24\linewidth}
            \begin{center}
            \includegraphics[height=1.5in]{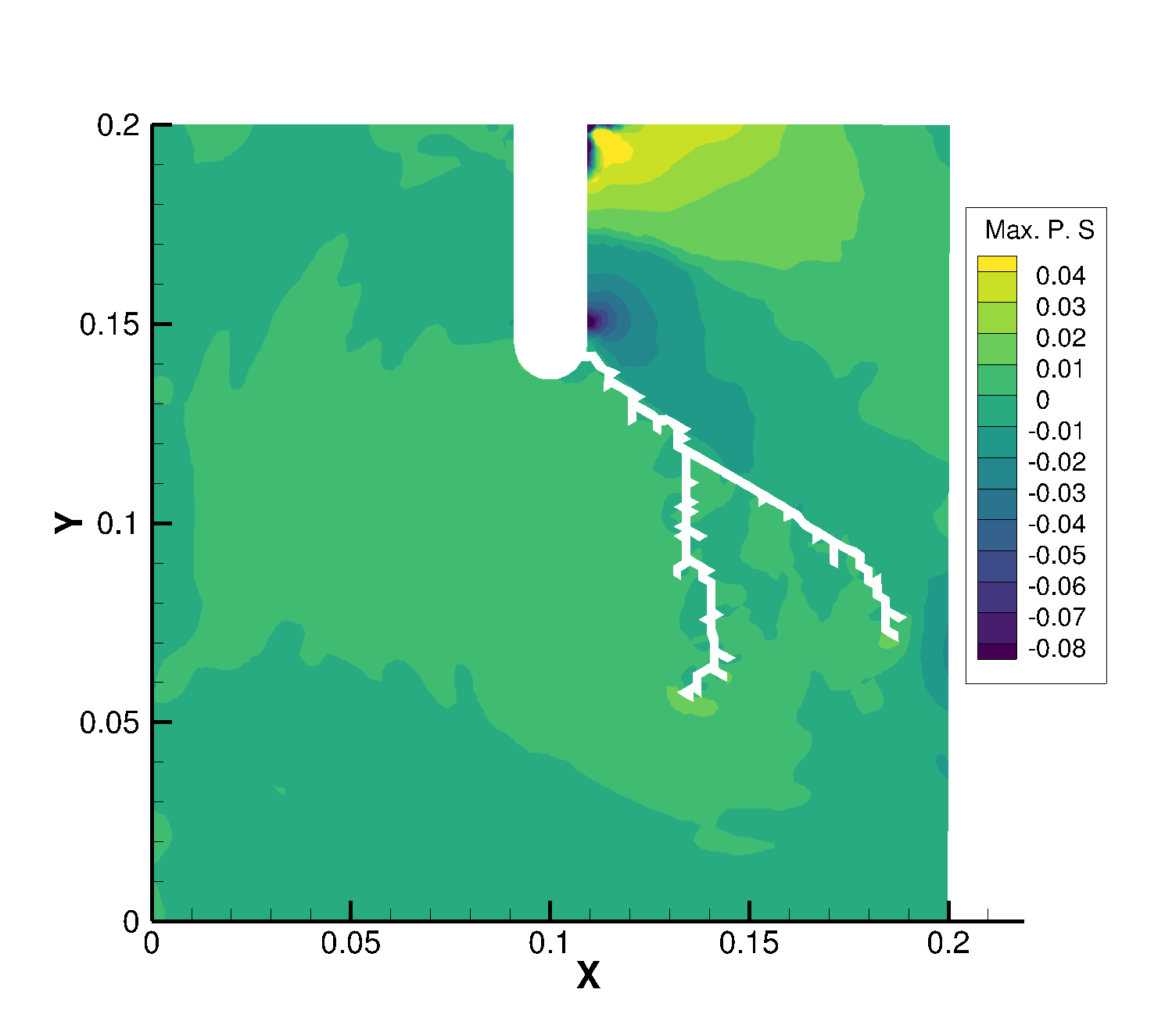}
            (c)
            \end{center}
        \end{minipage}
        \hfill
        \begin{minipage}{0.24\linewidth}
            \begin{center}
            \includegraphics[height=1.5in]{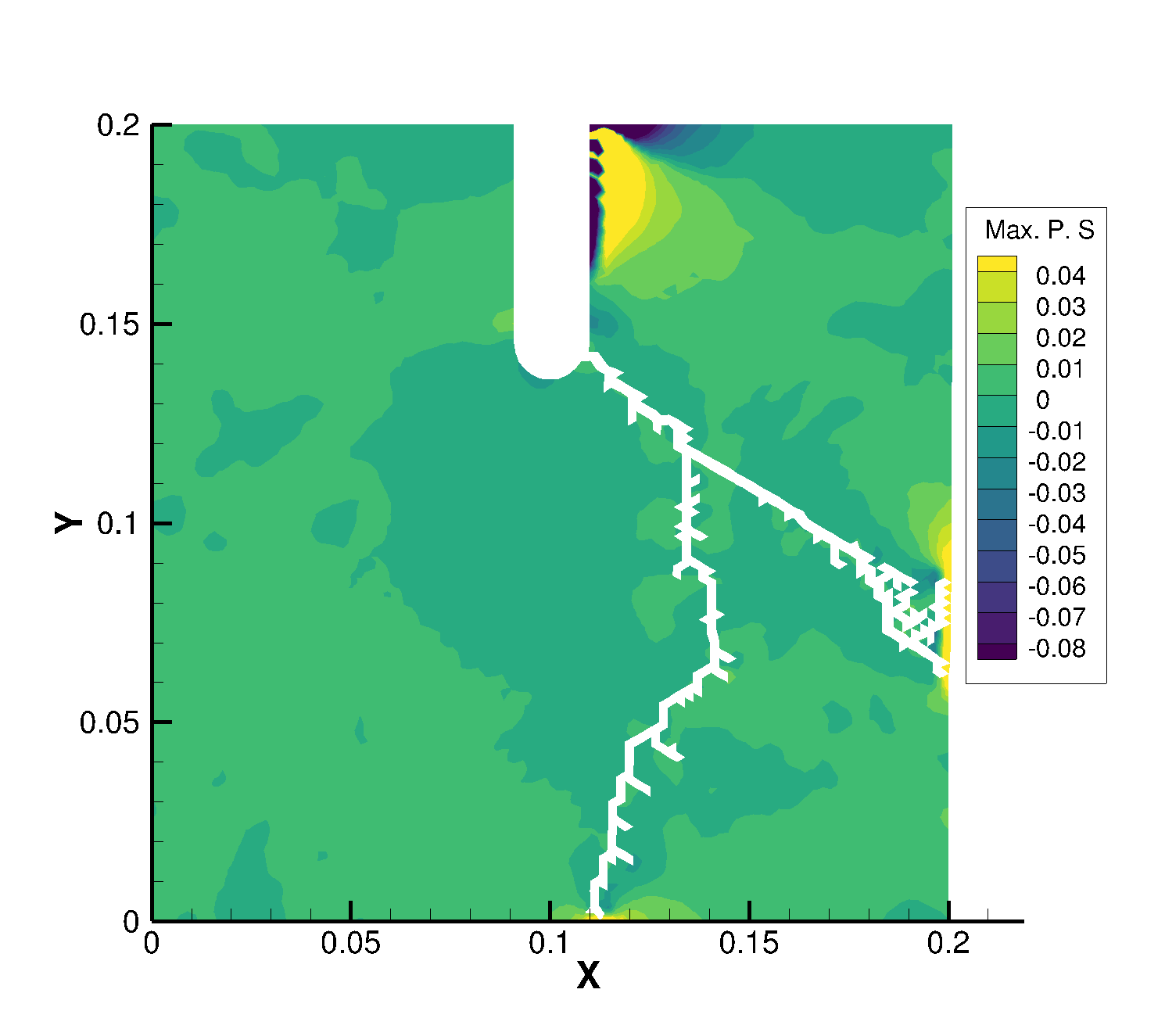}
            (d)
            \end{center}
        \end{minipage}
        \caption{Max. principle stress contour and crack patterns evolution of medium mesh $N=7643$, $E=14914$ with $v_0=3.993\ m/s$ at times at (a). $t=60\ \mu s$, (b). $t=84\ \mu s$, (c). $t=144\ \mu s$, (d). $t=300\ \mu s$.}
        \label{fig5-18: cb-dirichlet-crack-evolution-2}
\end{figure}

\begin{figure}[htp]
	\centering
        \begin{minipage}{0.24\linewidth}
            \begin{center}
            \includegraphics[height=1.5in]{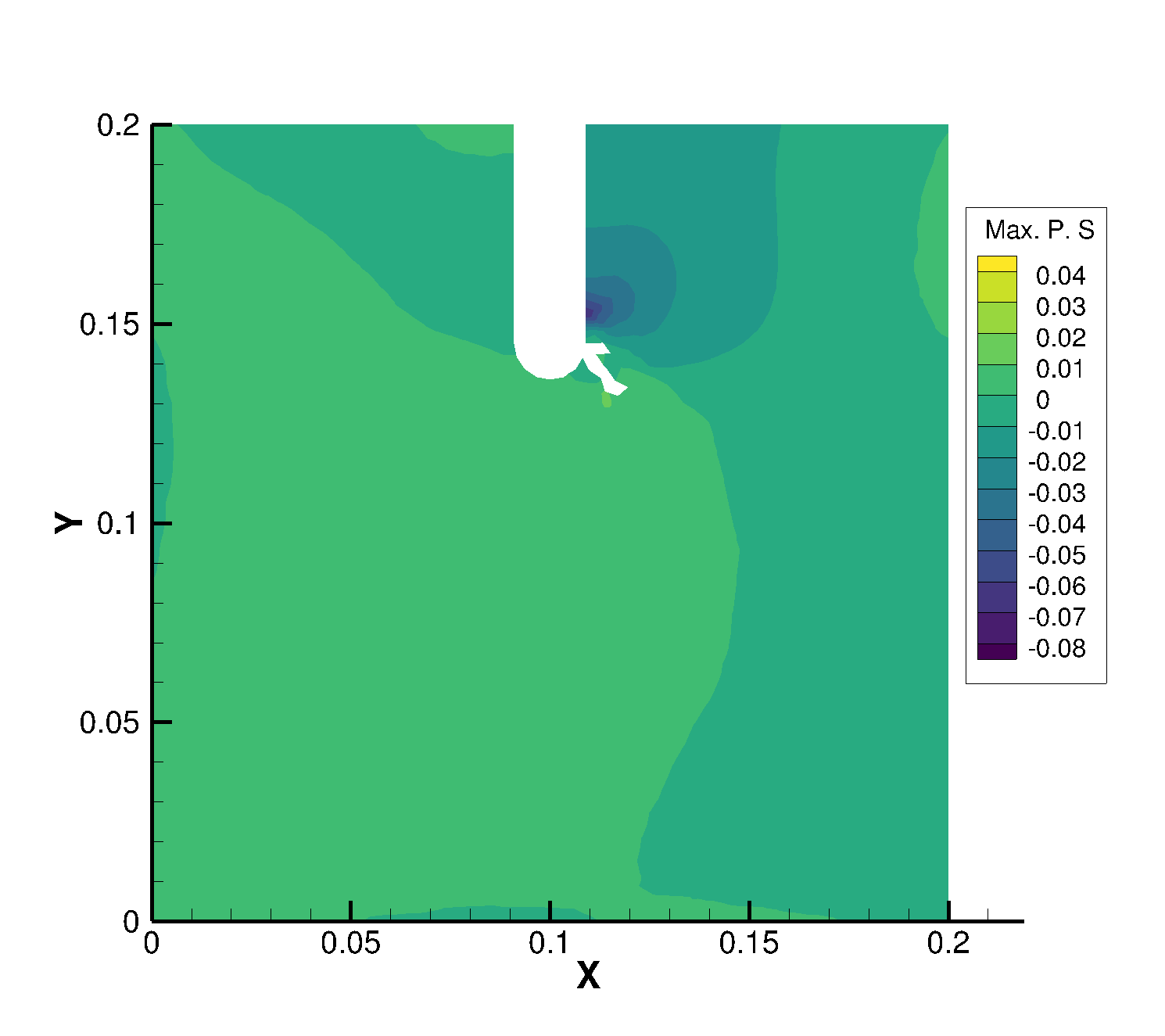}
            (a)
            \end{center}
        \end{minipage}
        \hfill
        \begin{minipage}{0.24\linewidth}
            \begin{center}
            \includegraphics[height=1.5in]{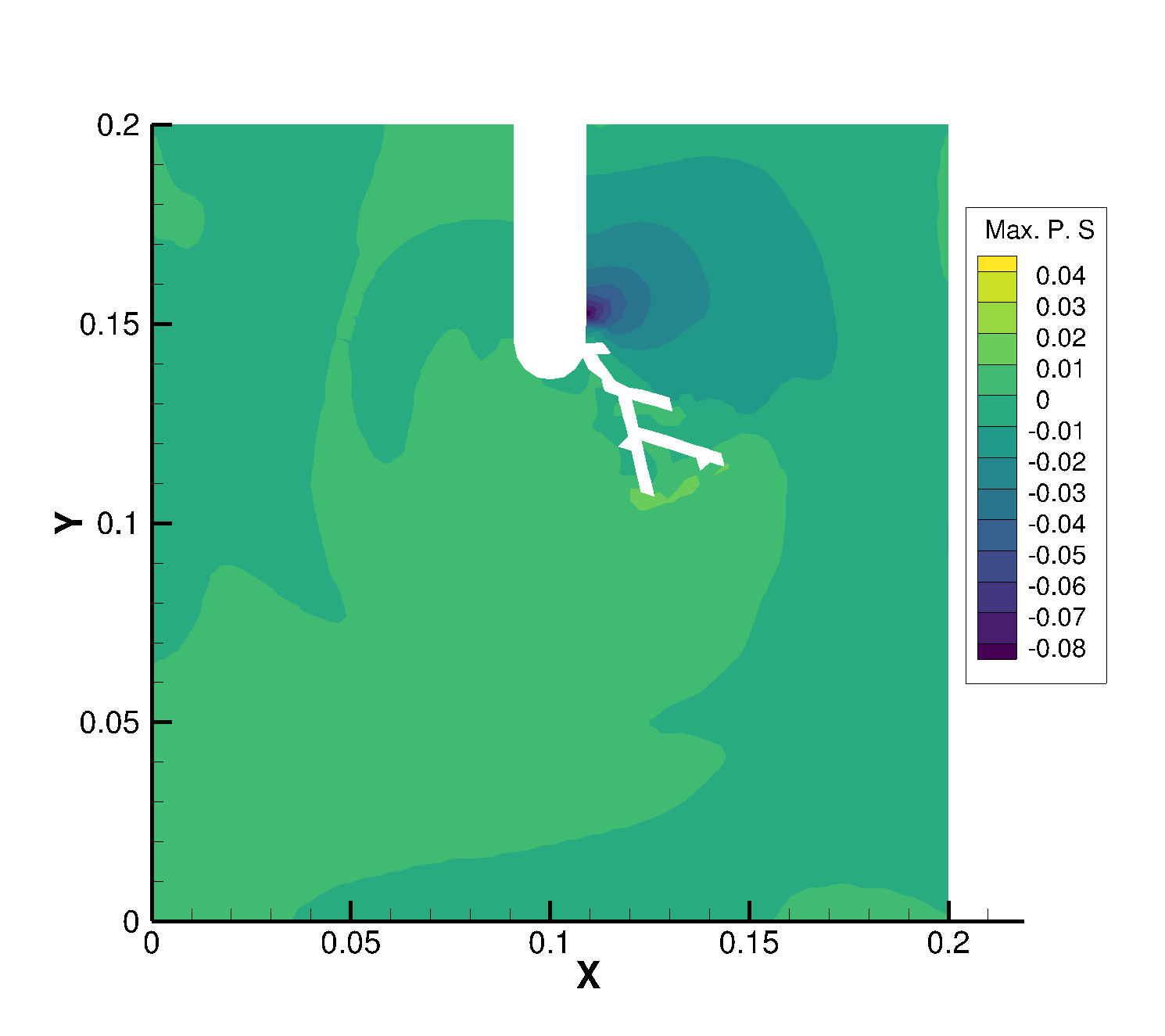}
            (b)
            \end{center}
        \end{minipage}   
        \hfill
        \begin{minipage}{0.24\linewidth}
            \begin{center}
            \includegraphics[height=1.5in]{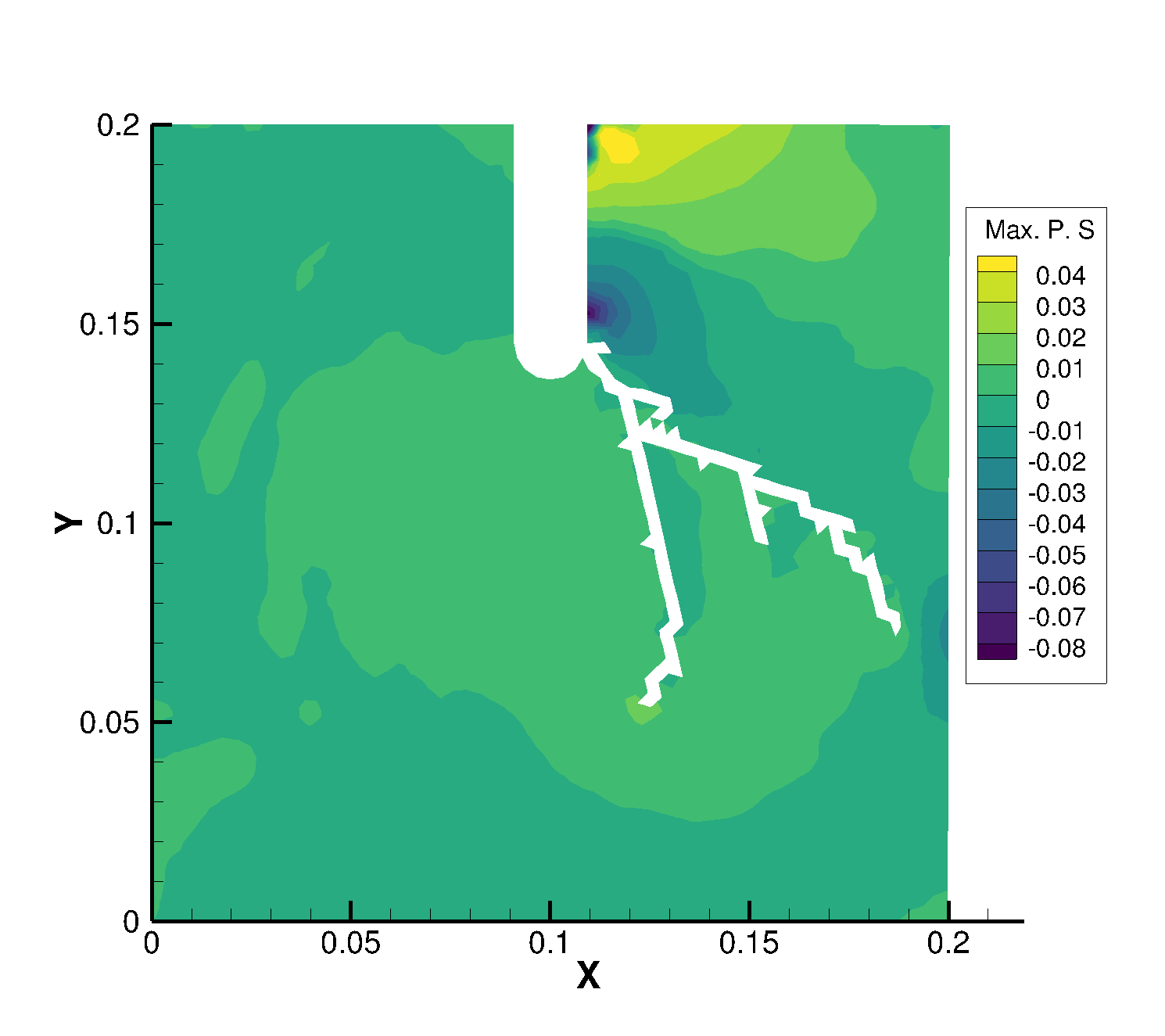}
            (c)
            \end{center}
        \end{minipage}
        \hfill
        \begin{minipage}{0.24\linewidth}
            \begin{center}
            \includegraphics[height=1.5in]{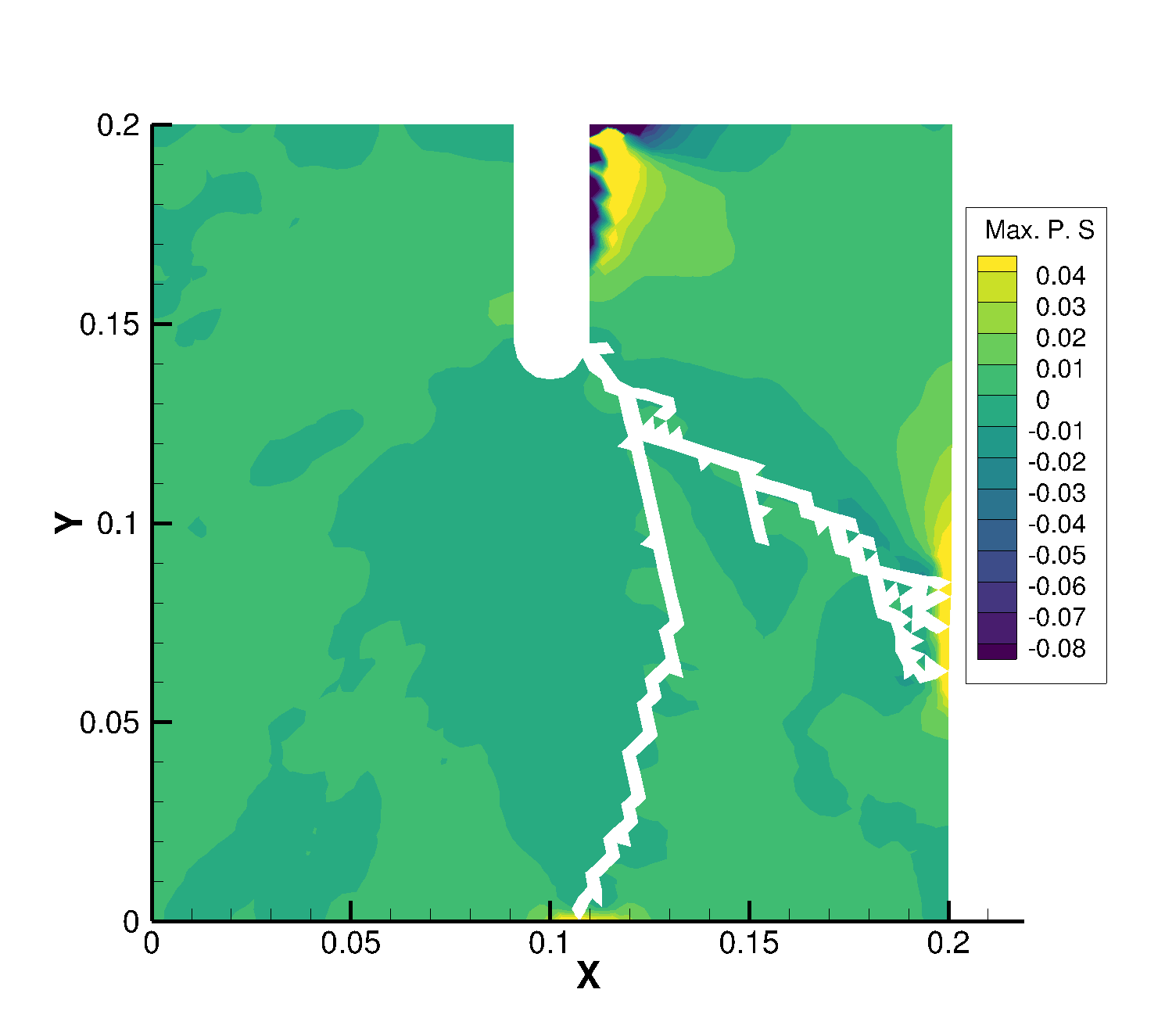}
            (d)
            \end{center}
        \end{minipage}
        \caption{Max. principle stress contour and crack patterns evolution of coarse mesh $N=3563$, $E=6874$ with $v_0=3.993\ m/s$ at times at (a). $t=54\ \mu s$, (b). $t=78\ \mu s$, (c). $t=144\ \mu s$, (d). $t=300\ \mu s$.}
        \label{fig5-19: cb-dirichlet-crack-evolution-3}
\end{figure}

In addition, the progression of the crack pattern of the mesh with $11939$ nodes and $23413$ elements under applied velocity $v_0 = 3.993\ m/s$ at different time steps is demonstrated in Figure.\ref{fig5-17: cb-dirichlet-crack-evolution-1},
the crack pattern progression of the mesh with $7643$ nodes and $14914$ elements under applied velocity $v_0=3.993\ m/s$ at different time steps is demonstrated in Figure.\ref{fig5-18: cb-dirichlet-crack-evolution-2},
and the crack pattern progression of mesh with $3563$ nodes and $6874$ elements under applied velocity $v_0=3.993\ m/s$ is demonstrated in Figure.\ref{fig5-19: cb-dirichlet-crack-evolution-3}.

\begin{figure}[htp]
	\centering
            \begin{center}
            \includegraphics[height=2.4in]{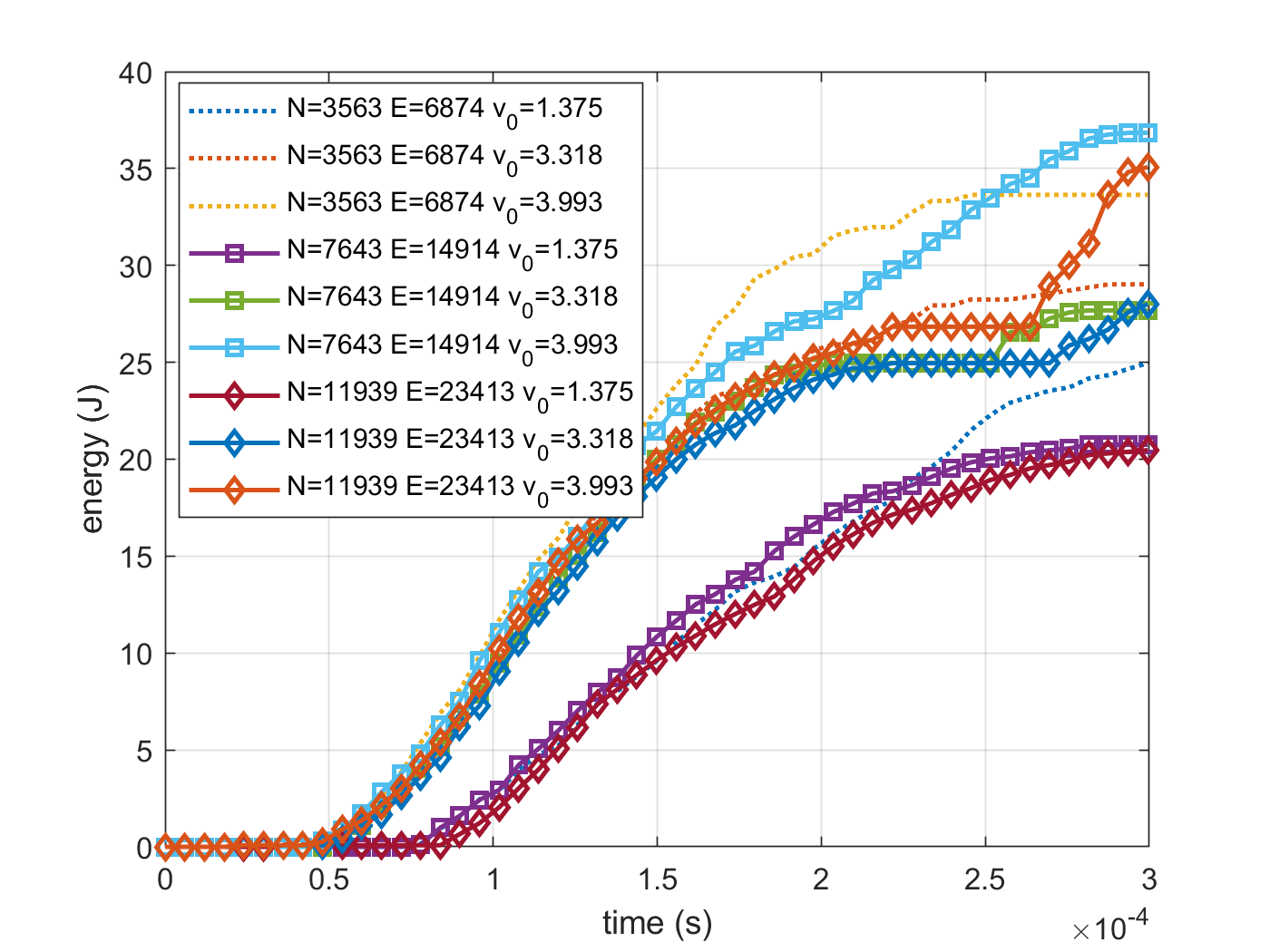}
            \end{center}
        \caption{Comparison of dissipated energy $U_d$ among three representative meshes under three different applied velocities.}
        \label{fig5-20: cb-dirichlet-Ud}
\end{figure}

Beyond the qualitative comparisons of crack patterns discussed above, Figure.\ref{fig5-20: cb-dirichlet-Ud} presents a quantitative analysis of the dissipated energy for three meshes resolutions: a fine mesh with $11939$ nodes and $23413$ elements, a medium mesh with $7643$ nodes and $14914$ elements, and a coarse mesh with $3563$ nodes and $6874$ elements. These results are shown under three different applied velocities: $v_0 = 1.375\ m/s$, $v_0=3.318\ m/s$ and $v_0=3.993\ m/s$. It is observed that all three meshes resolutions exhibit lower energy dissipation under the smaller applied velocity of $v_0 = 1.375\ m/s$, aligning with both experimental and numerical findings that indicate the formation of a single crack at this loading rate. Except for the coarse mesh model, the final dissipated energy for the fine and medium meshes at the lower velcity remains consistently around $U_d = 20\sim 25\ J/m$, indicating that the resulting single crack patterns are largely similar across different mesh resolutions. Then all three meshes under other two applied velocities demonstrate convergent and stable dissipated energy at final stage. For example, with applied velocity of $v_0 = 3.318\ m/s$ the final dissipated energy converges around $U_d = 25 \sim 30\ J/m$ and with applied velocity of $v_0 = 3.993\ m/s$ the final dissipated energy converges around $U_d = 35\ J/m$. These results underscore the intrinsic challenges of accurately modeling crack branching behavior in two-dimensional fracture simulations.

Through the comparative evaluation of the proposed Multiple Crack-tips Tracking algorithm in CEM in simulating two-dimensional crack branching under both Dirichlet and Neumann boundary conditions, it becomes evident that the method yields more reliable and physically consistent crack patterns when Dirichlet boundary conditions are applied. This improved performance is likely due to the relatively stable stress distribution maintained under prescribed displacements, which enhances the method’s robustness in transient dynamic simulations. In contrast, under Neumann boundary conditions, where external forces are directly applied, the resulting fluctuations in the stress field may hinder the accuracy of crack propagation predictions. This issue is further compounded by the way surface tractions are imposed, through transformation into equivalent nodal forces, which introduces additional numerical imbalance. Since surface elements often vary in area, this approach can lead to uneven force distribution across the boundary, thereby exacerbating the instability and reducing the fidelity of the simulation results.

\subsection{Fragmentation in pressured cylinder}
Crack fragmentation is the process by which a material under high loading rates or dynamic conditions develops multiple cracks that propagate and interact, ultimately breaking the material into several pieces. Unlike single-crack growth, fragmentation involves a complex interplay of rapidly evolving cracks influenced by the stress field, material heterogeneity, and loading conditions. This phenomenon is especially important in dynamic fracture problems, such as blast loading or high-velocity impacts, where strain rates are high and energy release is abrupt.

\begin{figure}[htp]
        \centering
        \begin{minipage}{0.45\linewidth}
            \begin{center}
            \includegraphics[height=2.4in]{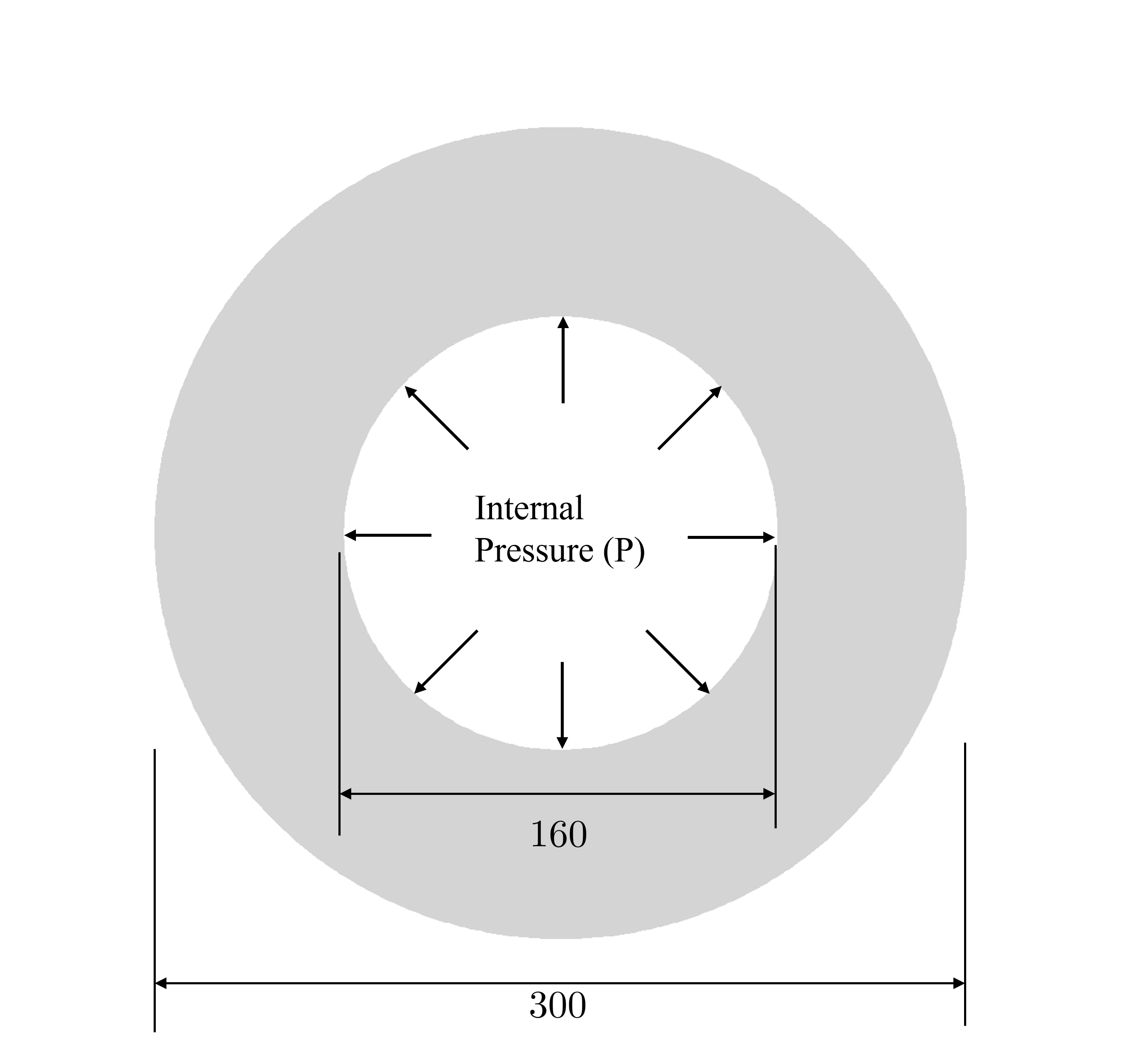}
            \end{center}
            \begin{center}
            (a)
            \end{center}
        \end{minipage}
        \begin{minipage}{0.45\linewidth}
            \begin{center}
            \includegraphics[height=2.4in]{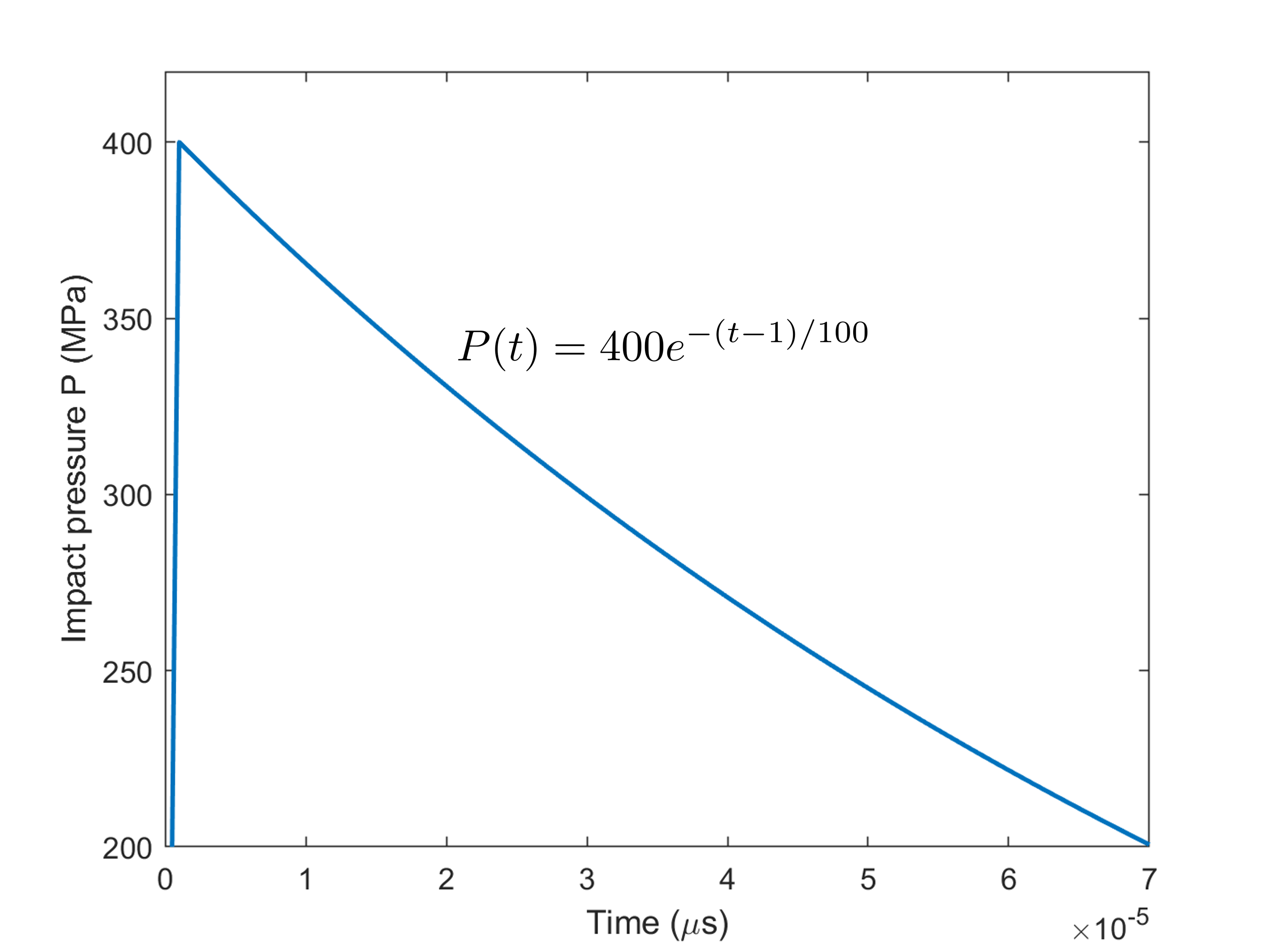}
            \end{center}
            \begin{center}
            (b)
            \end{center}
        \end{minipage}   
        \caption{(a). The whole geometry, boundary condition of pressure-loaded cylinder; (b). internal impact pressure variation with respect to time.}
        \label{fig5-21: fragmentation-cylinder}
\end{figure}

A well-known benchmark for studying crack fragmentation is the fragmentation of a pressure-loaded hollow cylinder, as shown in Figure.\ref{fig5-21: fragmentation-cylinder}(a). In this setup, a cylindrical shell is subjected to a sudden increase in internal pressure, commonly produced by an explosive charge, causing it to expand rapidly. The loaded internal pressure varies with respect to time (see Figure.\ref{fig5-21: fragmentation-cylinder}(b)), as formulated as following
\begin{equation}
p =
\begin{cases} 
\frac{t}{t_0}p_0,  & \text{if }\ t \le t_0, \\
400\times e^{-(t-t_0)/100}, & \text{if } \ t > t_0.
\end{cases}
\end{equation}
in which, $p_0 = 400\ MPa$, $t_0 = 1\mu s$.

As the internal pressure builds up, the hoop stress within the cylinder wall increases uniformly, leading to the nearly simultaneous initiation of multiple radial cracks. These cracks propagate from the inner surface toward the outer wall, causing the cylinder to fragment into wedge-like pieces. The number and distribution of these fragments are determined by the strain rate, material toughness and geometry of the cylinder. The material parameters of the benchmark are as follow: Young's Modulus $E = 210\ GPa$, Poisson ratio $v=0.3$, density $\rho=7850\ kg/m^3$ and critical fracture release rate $\mathcal{G}_c = 2000\ N/m$.

A series of studies on the benchmark introduce perturbation of elastic modulus to avoid mesh-dependent symmetry in their models (\cite{song2009cracking}, \cite{leon2014reduction}, \cite{paulino2010adaptive}, \cite{rabczuk2004cracking}, \cite{bui2022simulation}). In present work, the two-dimensional cylinder is discretized into two different Constant Strain Triangle grids: a fine mesh of $52916$ elements and $26942$ nodes, and a coarse mesh of $29598$ elements and $15161$ nodes. It is worthy to emphasize that NO artificial perturbation on elastic modulus to avoid structure symmetry since the two grids are essentially arbitrary. 

\begin{figure}[htp]
	\centering
        \begin{minipage}{0.3\linewidth}
            \begin{center}
            \includegraphics[height=1.8in]{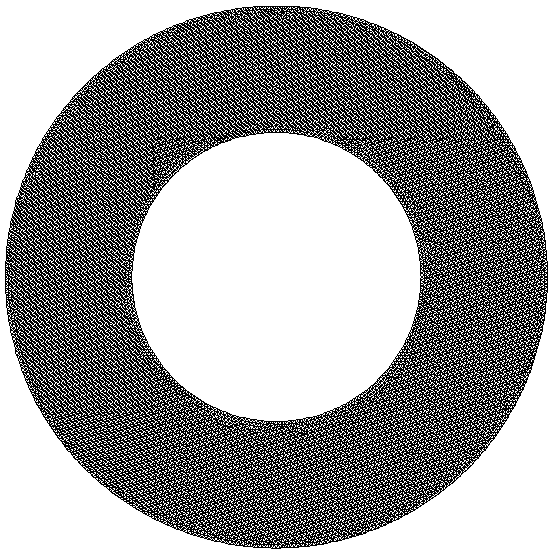}
            \end{center}
            \begin{center}
            (a)
            \end{center}
        \end{minipage}
        \begin{minipage}{0.3\linewidth}
            \begin{center}
            \includegraphics[height=1.8in]{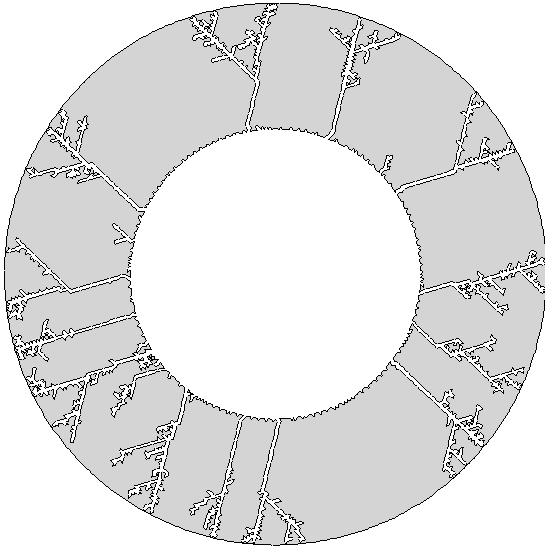}
            \end{center}
            \begin{center}
            (b)
            \end{center}
        \end{minipage}   
        \begin{minipage}{0.3\linewidth}
            \begin{center}
            \includegraphics[height=1.8in]{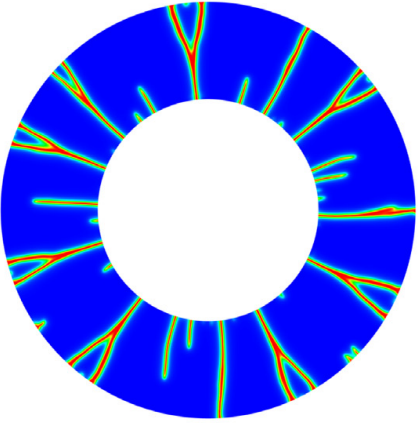}
            \end{center}
            \begin{center}
            (c)
            \end{center}
        \end{minipage}
        \begin{minipage}{0.3\linewidth}
            \begin{center}
            \includegraphics[height=1.8in]{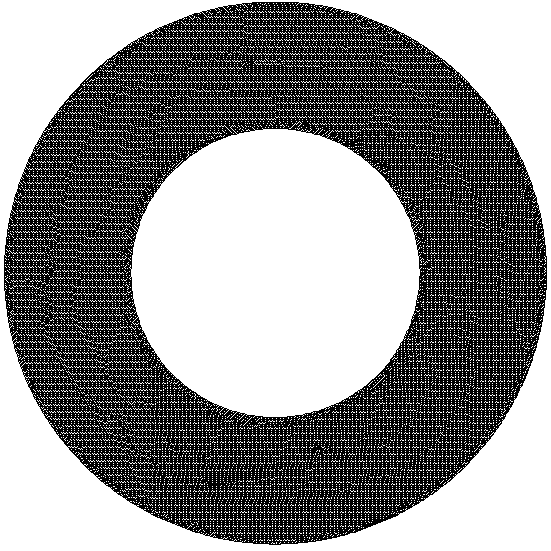}
            \end{center}
            \begin{center}
            (d)
            \end{center}
        \end{minipage}
        \begin{minipage}{0.3\linewidth}
            \begin{center}
            \includegraphics[height=1.8in]{fragmentation-N26942_E52916-crack-pattern-N26942.png}
            \end{center}
            \begin{center}
            (e)
            \end{center}
        \end{minipage}
        \begin{minipage}{0.3\linewidth}
            \begin{center}
            \includegraphics[height=1.8in]{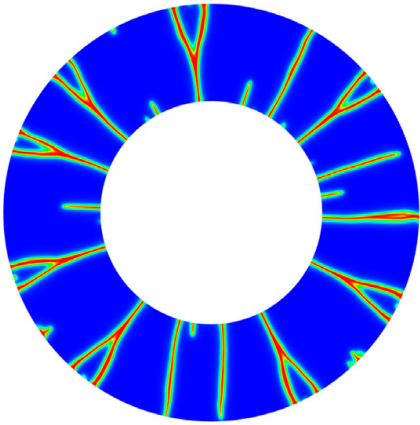}
            \end{center}
            \begin{center}
            (f)
            \end{center}
        \end{minipage}
        \caption{The discretization and final crack patterns of two representative Constant Strain Triangle meshes are illustrated: (a-b). $29598$ elements and $15161$ nodes; (d-e). $52916$ elements and $26942$ nodes. And (c,f). reference crack patterns of two meshes (\cite{geelen2019phase}).}
        \label{fig5-22: fragmentation-crack-pattern}
\end{figure}

\begin{figure}[htp]
	\centering
        \begin{minipage}{0.45\linewidth}
            \begin{center}
            \includegraphics[height=2.4in]{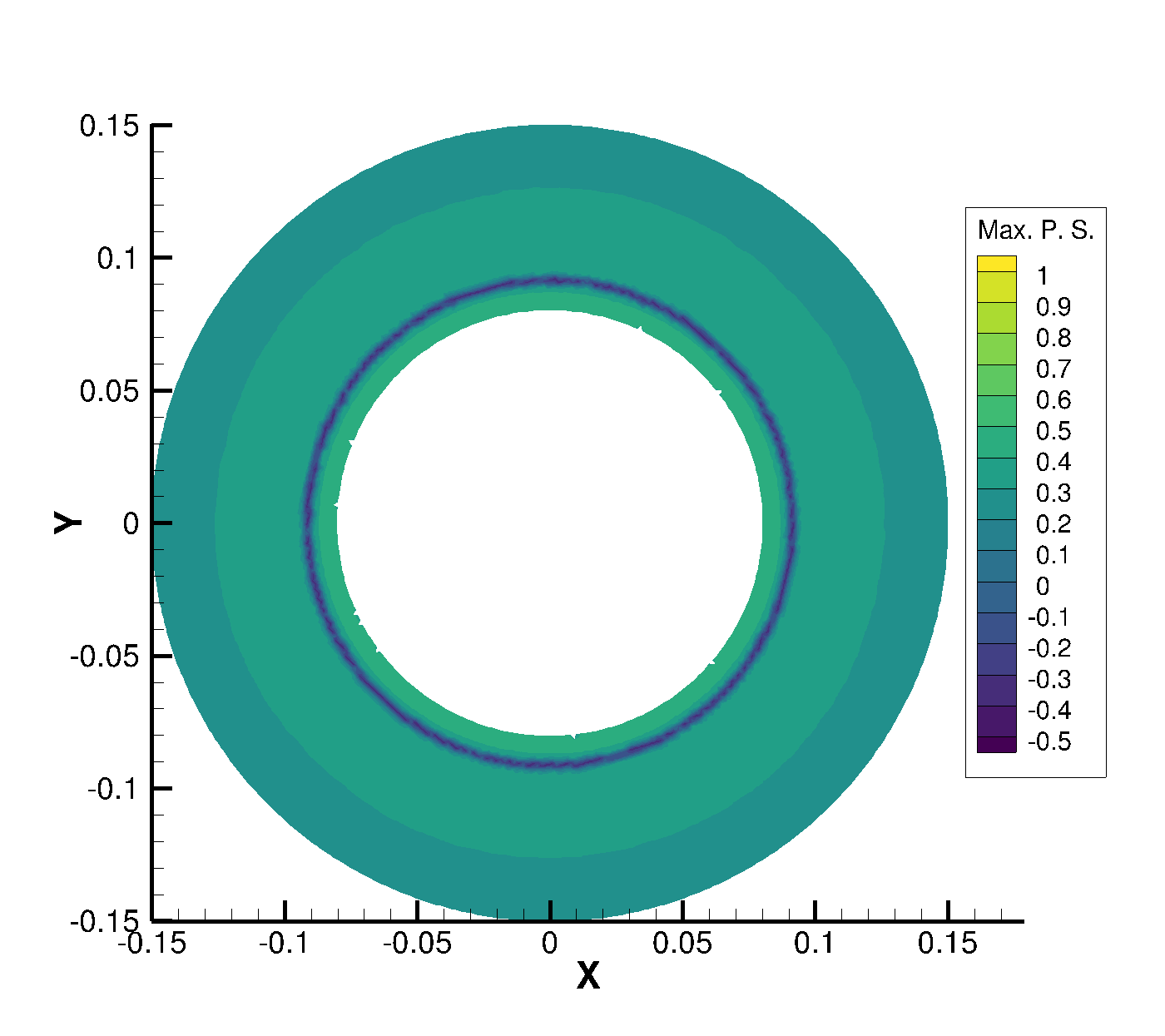}
            \end{center}
            \begin{center}
            (a)
            \end{center}
        \end{minipage}
        \begin{minipage}{0.45\linewidth}
            \begin{center}
            \includegraphics[height=2.4in]{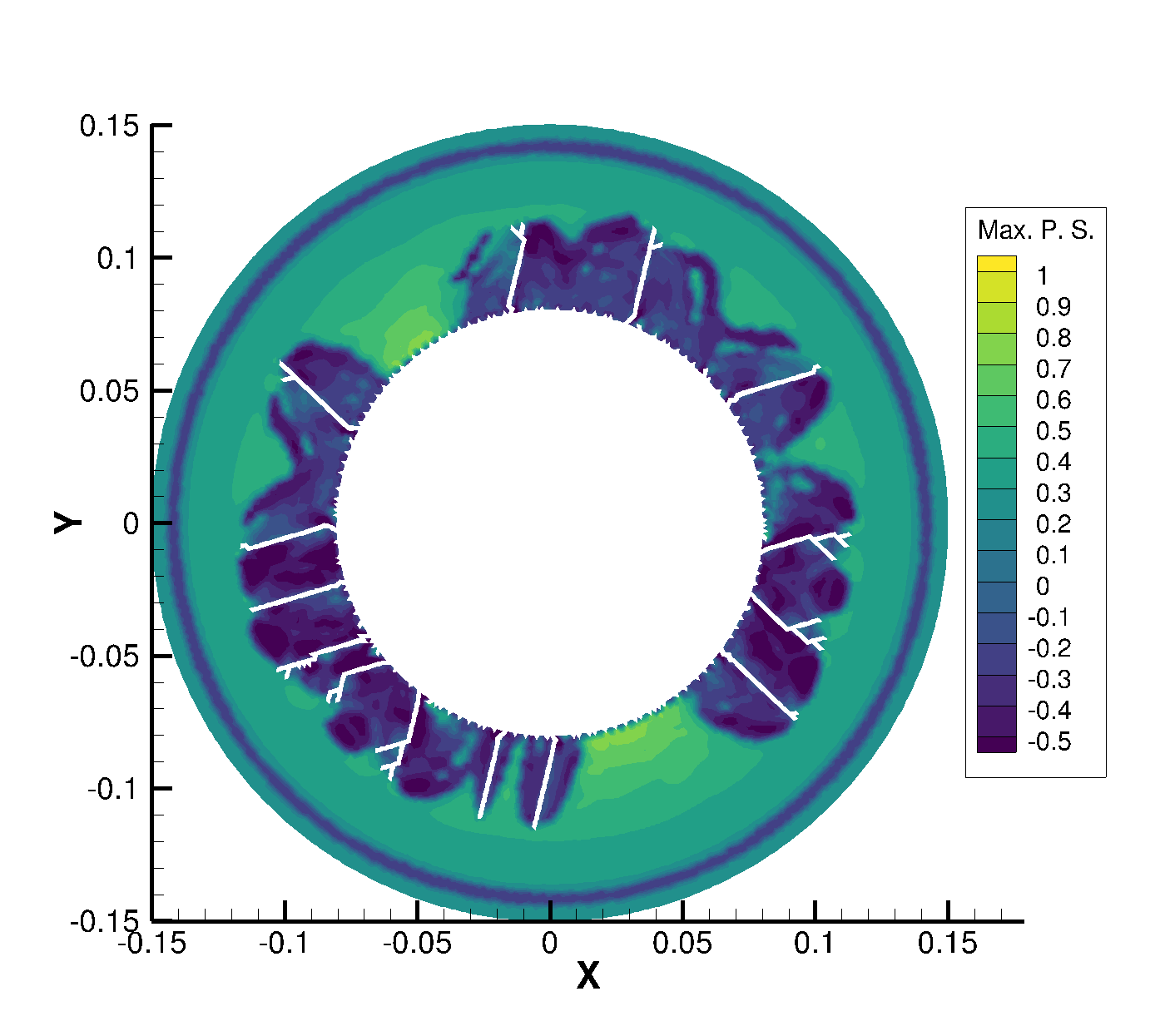}
            \end{center}
            \begin{center}
            (b)
            \end{center}
        \end{minipage}   
        \begin{minipage}{0.45\linewidth}
            \begin{center}
            \includegraphics[height=2.4in]{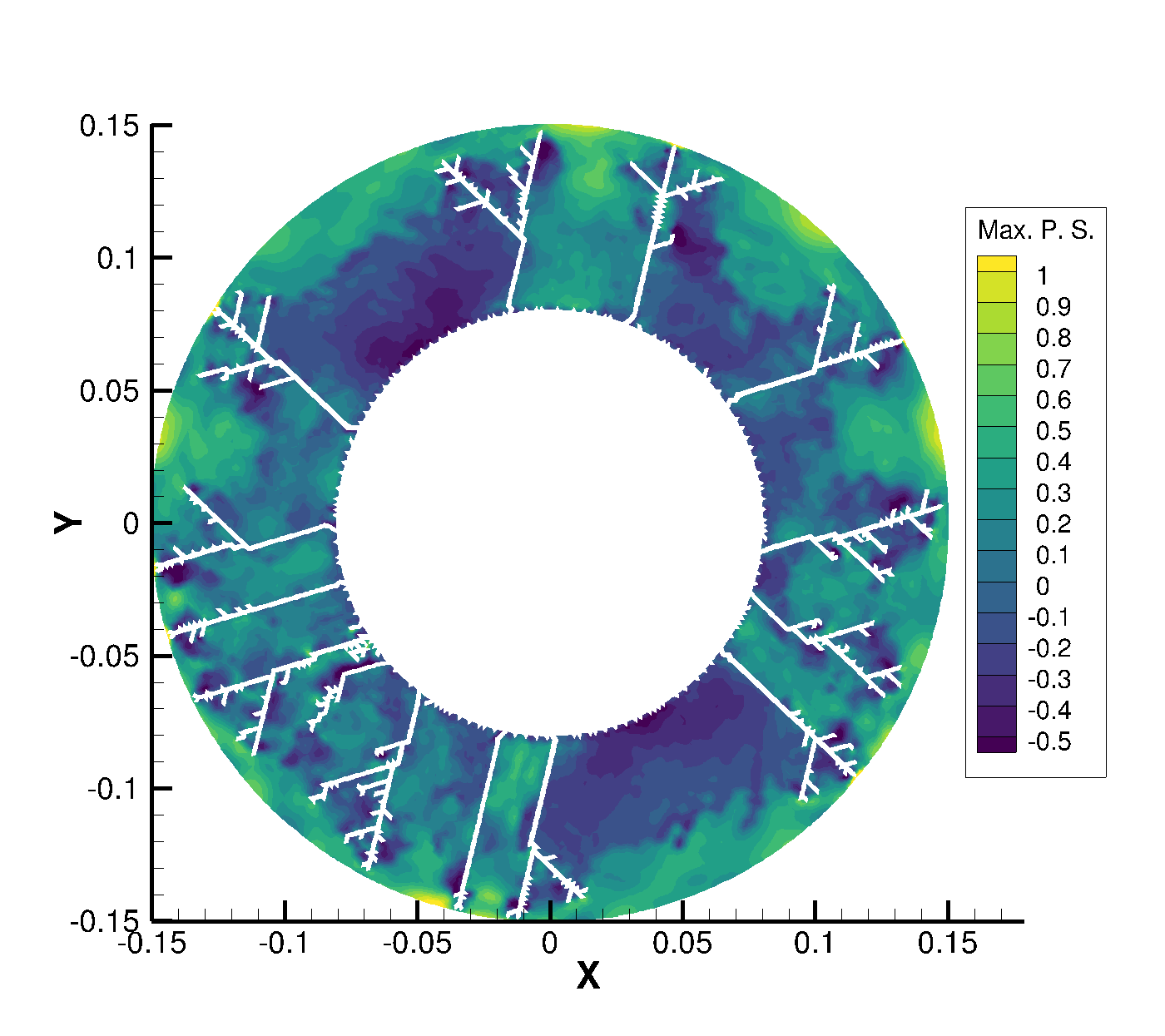}
            \end{center}
            \begin{center}
            (c)
            \end{center}
        \end{minipage}
        \begin{minipage}{0.45\linewidth}
            \begin{center}
            \includegraphics[height=2.4in]{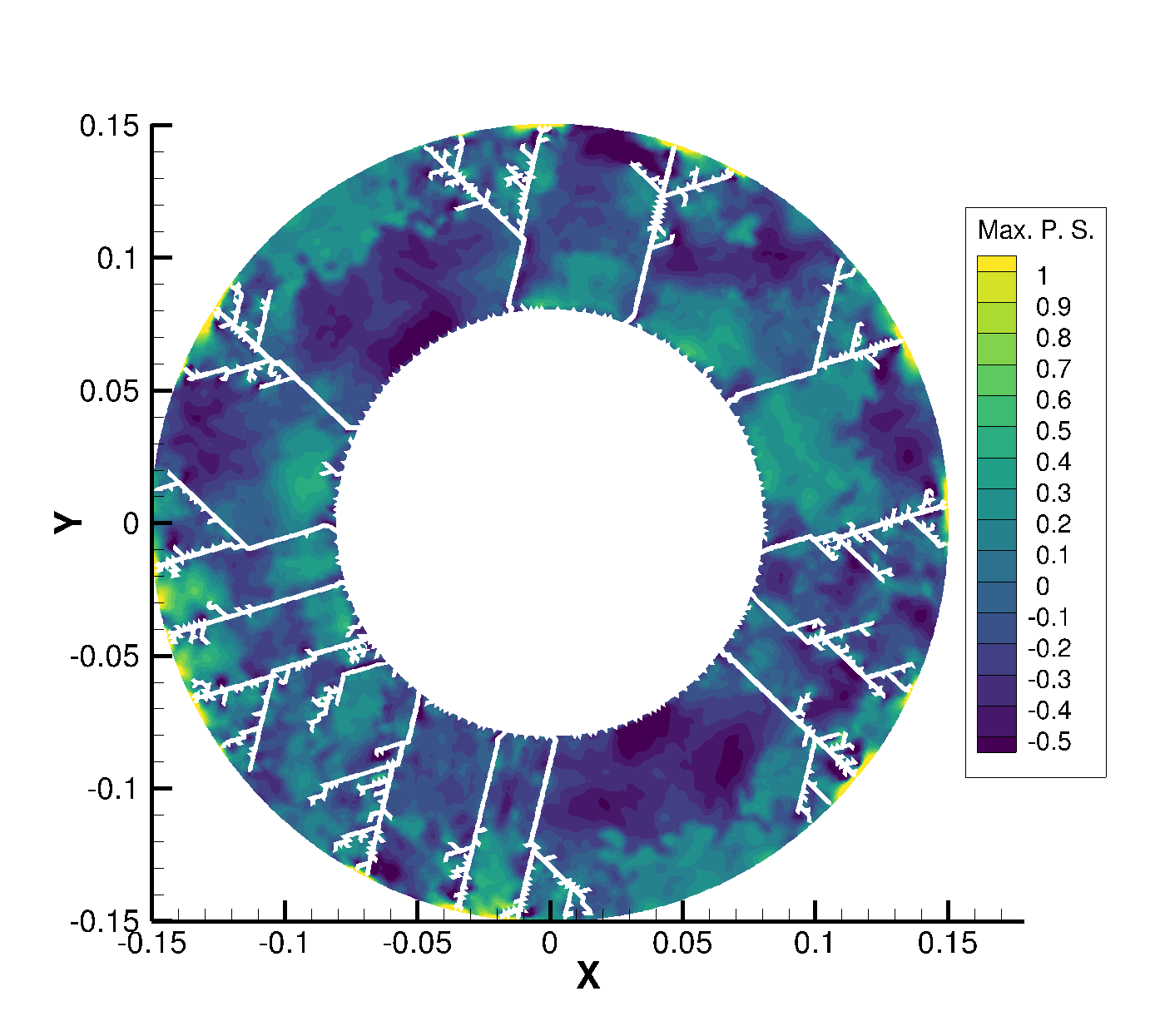}
            \end{center}
            \begin{center}
            (d)
            \end{center}
        \end{minipage}
        \caption{Development of maximum principal stress through crack propagation process in the mesh of $52916$ elements and $26942$ nodes, at times: (a). $30\ \mu s$; (b). $40\ \mu s$; (c). $54\ \mu s$; (d). $60\ \mu s$.}
        \label{fig5-23: fragmentation-evolution}
\end{figure}

The final crack patterns comparison between the present work and Geelen et al. work (\cite{geelen2019phase}) for different meshes are shown in Figure.\ref{fig5-22: fragmentation-crack-pattern}, 
and the development of the maximum principal stress as crack propagation for the fine mesh of $52916$ elements and $26942$ nodes is shown in Figure.\ref{fig5-23: fragmentation-evolution}.
Besides, we also compare the numbers of major fragments among various numerical methods and summarize in Table-\ref{tbl-1:fragmentation}.
\begin{table}[width=.9\linewidth,cols=4,pos=h]
\caption{The comparison on numbers of large fragments among various numerical methods}\label{tbl1}
\begin{tabular*}{\tblwidth}{@{} LLLL@{} }
\toprule
Approaches & No. of major fragments (No. of elements) \\
\midrule
Phase-field theory (\cite{geelen2019phase}) & $13$($218400$), $11$($400000$), $12$($728000$)  \\
Cohesive fracture model (\cite{hirmand2019block}) & $14$($20434$), $15$($30258$), $16$($40654$)  \\
Cracking node model (\cite{song2009cracking}) & $18$($10443$), $19$($32383$), $20$($75202$)  \\
Local damage model (\cite{bui2022simulation}) & $19$($9325$), $19$($14550$), $18$($25839$) \\
The Crack Element Model (CEM)  & $13$($29598$), $11$($52916$)  \\
\bottomrule
\end{tabular*}
\label{tbl-1:fragmentation}
\end{table}

In terms of final crack pattern comparison (see Figure.\ref{fig5-22: fragmentation-crack-pattern}), an agreement is observed between the results of the proposed Multiple Crack-tips Tracking algorithm in CEM and those reported in reference studies, especially the phenomenon that some crack branchings occurring at outer boundary of the cylinder is captured by both the proposed method and the referenced phase-field model. Furthermore, the number of major fragments obtained in this study, along with reference data extracted from the literature, is presented in Table-\ref{tbl-1:fragmentation} for comparison. Obvious discrepancies in the reported number of fragments can be observed among various studies in the Table-\ref{tbl-1:fragmentation}. Our results are consistent to that reported in both Geelen et al. (\cite{geelen2019phase}) and Hirmand et al. (\cite{hirmand2019block}). From the Figure.\ref{fig5-23: fragmentation-evolution} describing evolution of maximum principal stress along with the development of sample fragmentation, we find out that multiple cracks start to initiate at around $30\ \mu s$. Through the whole process, some cracks may get arrested while others keep propagating even branching until reaching the outer boundary of the cylinder.

Summarily, the benchmark of internally pressured cylinder is a standard test in dynamic fracture studies since it captures essential features of fragmentation such as crack initiation, branching, and interaction under transient, high-energy conditions. Through this example, it is validated that the proposed Multiple Crack-tips Tracking algorithm in two-dimensional CEM is capable to replicate the complex behavior observed in real scenarios.

\section{Conclusion and Prospects}
Crack branching remains a fascinating and unresolved challenge in the field of fracture mechanics. Its occurrence, particularly in brittle solids and stress-corroded metals, marks a critical shift in fracture behavior—one that often signals the onset of catastrophic failure. While conventional theories grounded in linear elastic fracture mechanics and energy-based criteria such as the $\mathcal{J}$-integral provide a foundational understanding of crack propagation, they fall short of capturing the intricacies of crack branching, especially under dynamic or non-uniform loading conditions. Early attempts to explain branching largely revolved around critical crack-tip velocities, proposing that branching is triggered once a certain speed threshold is crossed. However, empirical observations have consistently contradicted this simplification, revealing branching at much lower velocities than predicted. This discrepancy has led to a growing recognition that branching is not governed by a single mechanism, but rather emerges from a confluence of factors—local stress field perturbations, microstructure interactions, and dynamic instabilities that develop near the crack front.

In response to the need for more accurate modeling tools, recent developments in numerical fracture simulation have introduced the Crack Element Model (CEM) as a powerful framework. While the original two-dimensional CEM was designed for single-crack propagation, its extension into three dimensions yielded unexpected success in simulating complex branching behaviors without the need for artificial branching criteria. This natural emergence of branching in three dimensions suggests that the model inherently captures critical aspects of fracture physics that traditional approaches overlook. Building on this foundation, the present work revisits and refines the two-dimensional CEM, introducing an energy-based Multiple Crack-Tips Tracking algorithm that enables the simulation of not just single cracks, but also branching and fragmentation within a unified computational scheme. For example, in Kalthoff-Winkler plate example the Multiple Crack-tips Tracking algorithm can reproduce an approximated crack pattern as experimental results but some micro-cracks ignored by single crack-tip tracking algorithm are introduced. The comparison between crack branching simulation with Dirichlet boundary condition and Neumann boundary condition demonstrates the Multiple Crack-tips Tracking algorithm in two-dimensional Crack Element Model (MCT-2D-CEM) obtain a more physically reasonable crack pattern in Dirichlet boundary condition than in Neumann boundary condition. And the capability of MCT-2D-CME in simulating complex fragmentation is validated in the benchmark example of internally pressured cylinder.

Unlike earlier methods that rely on externally imposed conditions, this approach lets the physics of the problem dictate the crack evolution. As a result, the model offers a more realistic and flexible tool for simulating fracture processes across a broad range of scenarios. By enhancing the two-dimensional CEM with this tracking algorithm, we take a step closer to bridging the gap between theoretical fracture mechanics and the unpredictable reality of material failure. This work not only deepens our understanding of the mechanics behind crack branching but also opens the door to more reliable predictive modeling in structural analysis and material design.

On the other hand, as the crack paths propagate toward the boundaries of the specimen, they exhibit unexpected and complex (sub-)branching behaviors. This phenomenon is likely attributed to intensified stress fluctuations and stress field distortions occurring near the boundaries, which are not sufficiently controlled by the current modeling framework. These irregularities in the stress distribution may act as perturbations that promote unstable crack growth and deviation from the main path. To better capture and mitigate these effects, it is essential to develop a more robust and effective stress regularization scheme. The implementation and validation of such a scheme will be the focus of future investigations aimed at improving the accuracy and physical fidelity of crack propagation simulations, particularly in regions influenced by boundary effects.

\newpage




\bibliographystyle{cas-model2-names}

\bibliography{cas-refs}






\end{document}